\documentclass[twocolumn]{aastex62}
\usepackage[demo]{rotating}
\usepackage{appendix}

\usepackage{siunitx} 
\usepackage{float}
\graphicspath{{./}}

\received{TBD}
\revised{TBD}
\accepted{TBD}

\shorttitle{Fundamental Properties of mid-to-late type M Dwarfs}
\shortauthors{Iyer et al}

\begin{document}


\title{The {\tt SPHINX} M dwarf Spectral Grid. II. New Model Atmospheres and Spectra to Derive Fundamental Properties of mid-to-late type M-dwarfs}

\correspondingauthor{Aishwarya Iyer}
\email{aishwarya.iyer@nasa.gov}

\author[0000-0003-0971-1709]{Aishwarya R. Iyer}
\affil{NASA Goddard Space Flight Center, 8800 Greenbelt Rd, Greenbelt, MD 20771}
\affil{NASA Postdoctoral Fellowship through Oak Ridge Associated Universities}

\author[0000-0002-2338-476X]{Michael R. Line}
\affil{School of Earth and Space Exploration,
Arizona State University,
525 E. University Dr., Tempe AZ 85281}

\author[0000-0002-0638-8822]{Philip S. Muirhead}
\affil{Department of Astronomy and Institute for Astrophysical Research,
Boston University,
725 Commonwealth Ave., Boston, Massachusetts, 02215}

\author[0000-0002-9843-4354]{Jonathan J. Fortney}
\affil{Department of Astronomy and Astrophysics,
University of California, Santa Cruz,
1156 High St, Santa Cruz, CA 95064}

\author[0000-0001-6251-0573]{Jacqueline K. Faherty}
\affil{Department of Astrophysics, American Museum of Natural History, Central Park West at 79th St., New York, NY 10024, USA}

\begin{abstract}
M-dwarfs are the most dominant stars in the Galaxy. Their interiors and atmospheres exhibit complex processes including dust condensation, convective feedback, and magnetic activity-driven heterogeneity. Standard stellar characterization methods often struggle to capture these coupled effects. Part I of this series introduced \texttt{SPHINX I}, a validated grid of self-consistent radiative–convective model atmospheres and spectra for M-dwarfs with up-to-date molecular opacities suitable for early-to-mid M-dwarfs. Here, we present \texttt{SPHINX II}, which extends the model grid to cover mid-to-late type M-dwarfs, including both gray and physically motivated condensate cloud treatments and shorter convective mixing lengths. We validate \texttt{SPHINX II} using 39 benchmark FGK+M binary systems observed with SpeX/IRTF \citep{mann2014prospecting} and apply it to 32 mid-to-late-type M-dwarfs from the SpeX Prism Library. \texttt{SPHINX II} yields improved fits that are statistically consistent with empirical benchmarks, achieving precisions of 0.078 dex in metallicity and 0.13 dex in C/O. Across the model grid, condensate cloud mass peaks between 2100–2400 K, decreasing sharply toward both cooler and hotter temperatures. We find the onset of the cloud-free regime around $\sim$2900 K, and below 2100 K, we see formation of deep/buried clouds. As a case study, we also model Trappist-1 and show that even mass-limited silicate grains subtly modify its emergent spectrum, suppressing near-infrared flux and reddening the mid-infrared slope via shallow cloud formation near 10$^{-2}$ bar. In sum, \texttt{SPHINX II} provides an improved framework for constraining the fundamental properties of mid-to-late M-dwarfs.
\end{abstract}


\section{INTRODUCTION}\label{paper2intro}
Main-sequence M-dwarf stars are small, spanning a wide range of masses from 0.08 to 0.6 solar masses and effective temperatures between 2300 and 4000 K. They dominate the solar neighborhood, the Milky Way, and the Universe, comprising over 75\% of all stars by number and over 50\% of baryonic mass \citep{tarter2007reappraisal,woolf2020m}. Despite their prevalence, there remain many outstanding questions concerning their fundamental properties. A key challenge is that current stellar atmosphere models struggle to properly characterize low-mass stars, especially those with effective temperatures below 3000 K, as they fail to fully capture the effects of stellar activity, cloud formation, and convection \citep{freytag2010role}. Discrepancies between model predictions and observational constraints in inferred fundamental properties such as metallicity can significantly impact several areas of astrophysics, including stellar evolution, exoplanet characterization, and galactic chemical enrichment \citep{allard2001}. Given that M-dwarfs host a substantial fraction of the detected terrestrial exoplanet population \citep{bochanski2010}, refining their atmospheric models is critical for improving our understanding of both the stars themselves and the planets they host.

Atmospheric models play a crucial role in determining elemental abundances, which are a fundamental ingredient for constraining stellar formation histories and planet compositions \citep{fortney2012carbon}. For Sun-like stars, precise stellar abundances have driven major advances in exoplanet characterization, including the observed correlation between planetary occurrence rates and host-star metallicities, which supports the core-accretion model of  planet formation \citep{fischer2005planet,osborn2019}. Additionally, the detailed chemical composition of Sun-like stars has provided key insights into the bulk composition of their exoplanets. For example, studies of nearby solar analogs have shown that the Sun is deficient in certain refractory elements, suggesting that these materials were sequestered in planetary formation rather than incorporated into the star itself \citep{melendez2009peculiar}. These considerations become even more critical in the M-dwarf regime, where the high frequency of rocky planets \citep{dressing2015occurrence} and the amplified signal-to-noise ratios for both radial velocity and transit observations due to their small sizes \citep{kopparapu2013habitable} make accurate stellar characterization essential for interpreting exoplanet properties.

However, inferring metallicities for M dwarfs, particularly for mid-to-late spectral types, has remained an ongoing challenge using both empirical calibration methods \citep{mann2014prospecting,neves2012metallicity,rojas2010metal,newton2013near} and theoretical atmospheric models \citep{allard1997,husser2013}. The broad diversity across the entire M-dwarf spectral type is enhanced by molecule-rich environments that create broadband spectral features that obscure traditional atomic absorption lines used for abundance determinations. Additionally, the spectral energy distributions of late-type M-dwarfs in particular, are plagued by a complex interplay of dust and cloud condensation \citep{tsuji1996evolution,hurt2024uniform}, molecular opacities, and pseudo-continuum effects, all of which further complicate standard spectroscopic methods for inferring metallicities and other fundamental parameters \citep{rajpurohit2018exploring}.

M-dwarfs also exhibit a diverse range of internal and atmospheric processes—magnetic activity, convective overshoot, and chromospheric variability—that introduce significant degeneracies in retrieved stellar properties \citep{charbonneau2014solar,shkolnik2014hazmat,schneider2018hazmat,loyd2018hazmat}. The influence of these processes is not limited to the outer stellar atmosphere but extends to the deeper photosphere, where magnetic fields can alter molecular equilibrium chemistry and impact inferred stellar properties \citep{muirhead2020magnetic}. Understanding how these physical processes affect observed spectra is crucial for improving empirical calibrations, particularly in benchmark binary systems where M-dwarfs are compared to F-, G-, and K-type primary stars. 

\begin{figure}[t]
    \centering
    \includegraphics[width=\columnwidth]{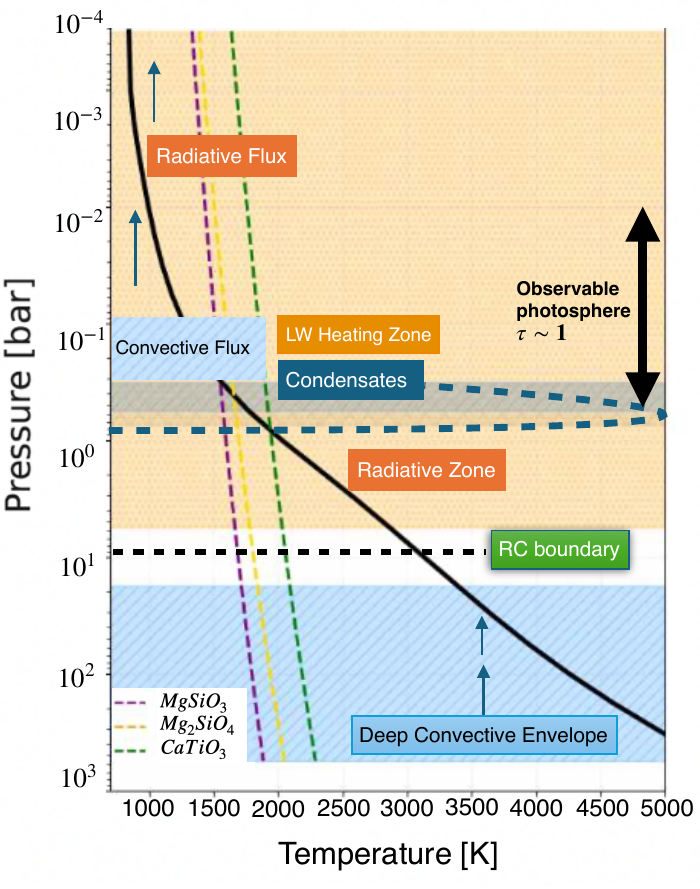}
    \caption{Schematic illustration of clouds, convection, and radiative transfer in a mid-to-late-type (T$_\mathrm{eff}$ $\sim$ 2000–2900 K) M-dwarf atmosphere. The solid black curve shows a representative temperature–pressure (T-P) profile. Orange shading marks the radiative zone, blue shading the convective envelope, and the dashed black line denotes the radiative–convective (R-C) boundary. The solid black double-headed arrow indicates the approximate observable photosphere ($\tau \sim$ 1), from which emergent flux across optical (0.4-0.9$\mu$m), NIR (1-2.5 $\mu$m), and MIR (3-8 $\mu$m) arises. Purple (MgSiO$_3$), yellow (Mg$_2$SiO$_4$), and green (CaTiO$_3$) dashed curves show condensation boundaries for major silicates and titanates, with the shaded ``Condensates'' region marking the expected cloud-forming zone near 10$^{-1}$ bar. Upward arrows denote convective and radiative energy transport, while the orange bar (``LW Heating Zone'') indicates layers influenced by long-wavelength radiative heating. Even when the cloud deck lies just below the $\tau \sim$ 1 surface, its additional opacity raises the effective photosphere to cooler layers and smooths the near- and mid-infrared continuum, producing the flux suppression and reddening characteristic of late-M dwarfs.}
    \label{fig:cloud_schematic}
\end{figure}

Several recent works have emphasized that uncertainties in stellar models propagate into exoplanet atmospheric retrievals \citep{iyer2020influence,rackham2024stellar,lim2023effects,fauchez2025impact}, particularly highlighting that inaccuracies in M-dwarf spectra can introduce biases in both transmission and emission spectra, affecting interpretations of planetary atmospheres observed with $JWST$. The increasing role of M-dwarfs exoplanet surveys—such as CARMENES \citep{reiners2018carmenes}, the Habitable Zone Planet Finder \citep{mahadevan2012habitable}, and JWST exoplanet programs \citep{greene2019characterizing,kanodia2024searching}, including the Hot Rocks Survey \citep{diamondlowe2024hotrocks}—all implore the need for an improved framework for deriving accurate stellar parameters. Upcoming missions such as \textit{Pandora} \citep{quintana2021pandora} will also enable complementary efforts to understand M-dwarf variability, offering multi-wavelength long-baseline strategies to further insights into time-dependent processes such as spot evolution and magnetic activity, which can refine the applicability of stellar models for exoplanetary research.

In part I of this paper series \citep{iyer2023sphinx}, we introduced \texttt{SPHINX I}, a new 1D self-consistent radiative-convective thermochemical equilibrium model grid designed for early- to mid-type M-dwarfs. This model grid was validated against observations of benchmark G+M widely separated binary systems \citep{mann2013metal,mann2015constrain} and M-dwarfs with available interferometric angular diameter measurements \citep{boyajian2012stellar}. However, the previous model grid did not fully address spectroscopic degeneracies arising from convection, and cloud opacity, particularly in cooler M-dwarfs. In this work, we introduce \texttt{SPHINX II}, an updated model grid that expands upon these previous efforts by incorporating: (1) a gray cloud opacity treatment to better capture dust-driven flux redistribution, (2) a revised convective mixing length prescription to improve atmospheric structures for mid-to-late M-dwarfs, and (3) a condensate cloud treatment based on the Ackerman \& Marley sedimentation framework \citep{ackerman2001precipitating}, which parameterizes the vertical distribution of condensates as a balance between upward mixing and downward rainout, with feedback between cloud opacity and convective stability.

This paper is organized as follows. Section \S\ref{SPHINX-IImodel} reviews the \texttt{SPHINX I} framework from Part I and introduces the upgrades implemented in \texttt{SPHINX II}. Section \S\ref{interpolation} outlines our interpolation routine and likelihood evaluation used in the grid-model retrieval (grid-trieval) approach. Section \S\ref{paper2results} presents results, including fits to low-resolution SpeX NIR spectra from the SpeX Prism Library \citep{burgasser2014spex} for 32 late-type M-dwarfs, as well as empirical validation using 39 mid-to-late type M-dwarfs with FGK primaries \citep{mann2014prospecting}. Section \S\ref{paper2discussion} discusses broader implications of these results. In \S\ref{mdwarfprop}, we show how incorporating cloud opacity and reduced convective efficiency improves metallicity estimates for mid-to-late M-dwarfs. \S\ref{jwstsection} places our findings in the context of JWST observations and exoplanet host-star characterization, and  \S\ref{Trappist} then focuses on Trappist-1 as an example late-type M-dwarf, where the inclusion of condensates, cloud–convection feedback, and photospheric heterogeneities further improves spectral fits. \S\ref{condensates} explores condensate opacity trends across the M-dwarf sequence, highlighting how cloud base migration and silicate absorption features shape near- and mid-infrared spectra. Finally, Section \S\ref{paper2summary} summarizes the key conclusions and outlines directions for future work. Please note, that throughout this paper, we refer to condensates in M-dwarf atmospheres as ``clouds'', following the convention in exoplanet and recent brown dwarf modeling literature \citep[e.g.,][]{ackerman2001precipitating, marley2013}. However, in much of the ultracool dwarf and stellar atmosphere literature, these condensates are also frequently referred to as ``dust'' 
\citep[e.g.,][]{tsuji1996dust, allard2001, allard2012}. For clarity, we treat ``clouds'' and ``dust'' as synonymous in this work, with both terms denoting the opacity contribution from silicate and refractory condensates (e.g., Mg$_2$SiO$_4$, MgSiO$_3$, CaTiO$_3$) in the photospheres of mid-to-late M-dwarfs.

\section{Model Description and Data Selection}\label{SPHINX-IImodel}

We use \texttt{SPHINX I}—a validated 1D self-consistent radiative-convective equilibrium chemistry model—built using the \texttt{ScCHIMERA} framework, originally developed for modeling extrasolar planet atmospheres \citep{bonnefoy2018,piskorz2018,Arcangeli2018,Kreidberg2018,gharib2019influence,gharib2021exoplines,mansfield2021unique}. The code iteratively solves for vertical profiles of volume mixing ratios, cloud and condensate properties, temperature structure, and the top-of-atmosphere disk-integrated stellar spectrum. A full description of the core \texttt{SPHINX I} modeling framework is presented in part I of this series \citep{iyer2023sphinx}.

\texttt{SPHINX I} model atmospheres include the following components:
\begin{enumerate}
    \item \textbf{Radiative transfer:} We adopt a two-stream approximation following \citet{toon1989rapid}, assuming plane-parallel geometry and hydrostatic equilibrium. We solve for temperature and opacity across all atmospheric layers.
    
    \item \textbf{Equilibrium chemistry:} We use the NASA CEA package \citep{gordon1996nasa} for the equation of state and to compute chemical abundances for all species in the model for a given elemental composition, temperature and pressure grid.
    
    \item \textbf{Gas opacities:} We include updated molecular opacities relevant to cool stellar atmospheres from the \texttt{EXOPLINES} database \citep{gharib2021exoplines}. A full list of both molecular and atomic line lists and sources is provided in part I.
    
\end{enumerate}

Part I focused on early-type M-dwarfs that are too hot for significant dust formation. Here, we extend the framework to mid- to late-type M-dwarfs, where condensates strongly influence the photospheric temperature structure and emergent spectra \citep{patience2012spectroscopy,allard1997,allard2001,allard2012}. These extensions collectively define \texttt{SPHINX II}.

\begin{itemize}
    \item \textbf{Clouds:} We implement both parametric and physically motivated cloud treatments. 
    
    \textbf{Gray Cloud Model}: Condensates (or “dust,” in the M-dwarf literature; e.g., \citealt{allard2001,tsuji2002water}) are first represented through a vertically uniform, wavelength-independent opacity term $\kappa$ (cm$^2$ g$^{-1}$). This simple parameterization captures the bulk reddening and muted molecular features characteristic of dusty late-M atmospheres \citep{faherty2016population,patience2012spectroscopy,dupuy2010studying}. Figure \ref{cloudplot} illustrates the effect of varying $\kappa$ on the spectral energy distribution of a 2700 K model, with significant differences throughout optical, NIR and MIR bandpasses.

    \begin{figure*}[!tbp]
    \centering
    \includegraphics[width=\textwidth]{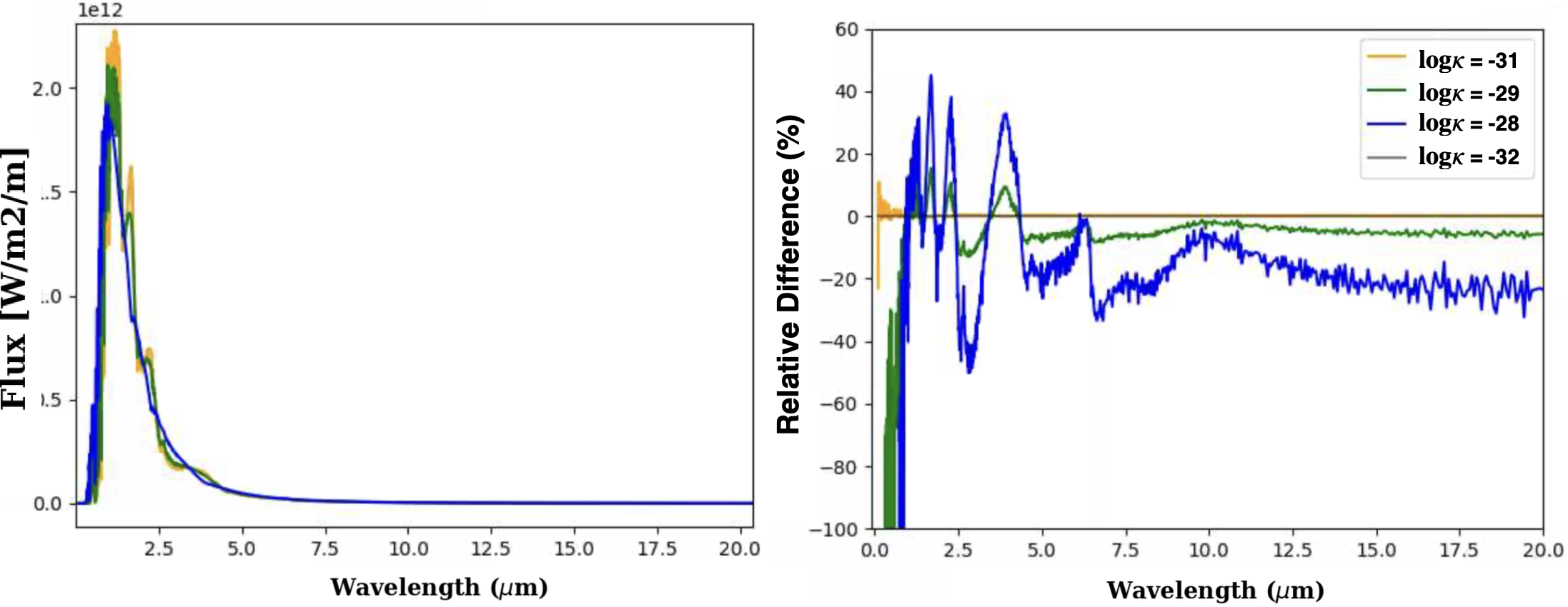}
    \caption{Spectra  \textbf{(Left)} and relative differences in ``spectral shape'' \textbf{(Right)} with varying levels of gray cloud opacity $\kappa$ (cm$^2$ g$^{-1}$). Assumes $T_{\mathrm{eff}} = 2700$ K, $\log g = 5.0$, and solar metallicity and C/O. The cloud-free model (gray), minimal cloud (green), and optically thick cloud (blue) cases correspond to $\log(\kappa)$ = -32, -29, and -28, respectively. The model corresponding to $\log(\kappa)$ = -31 (yellow) and -30 (not shown in the plot) do not predict dramatic spectral differences relative to the cloud-free case. The flat gray line at 0 in the right panel represents the cloud-free reference spectrum. \label{cloudplot}}
    \end{figure*}

    \textbf{Physical Condensate Model:}
    Beyond the gray parameterization, \texttt{SPHINX II} includes a physical condensate treatment following the steady-state sedimentation balance of \citep{ackerman2001precipitating}. Condensate mass fractions evolve under the competition between upward turbulent mixing and downward sedimentation: 

    \begin{equation}\label{am01}
        K_{zz} \frac{dq_t}{dz} + f_{sed}v_{fall}q_c = 0
    \end{equation}
    where $q_t = q_v + q_c$ is the total vapor plus condensate mass fraction, $v_{\mathrm{fall}}$ is the particle terminal velocity, $f_{\mathrm{sed}}$ is the sedimentation efficiency, and $K_{zz}$ is the eddy diffusion coefficient. The terminal velocity $v_{\mathrm{fall}}$ depends directly on the local effective particle radius $r_{\mathrm{eff}}$, thereby linking grain growth and sedimentation strength \citep{ackerman2001precipitating,helling2008comparison}. Larger $r_{\mathrm{eff}}$ increases settling efficiency, while smaller grains remain lofted longer by turbulent mixing.

   In most implementations of the Ackerman–Marley (2001) cloud model (e.g., \citealt{mukherjee2022}), the eddy diffusion coefficient K$_{zz}$ is parameterized from the convective flux at the radiative–convective (R–C) boundary and extended upward to represent turbulent mixing. In \texttt{SPHINX II}, we adopt a single physically motivated prescription relevant for M-dwarfs: (1) mass transport is computed self-consistently via mixing-length theory (MLT), which directly yields convective fluxes and velocities that govern mixing throughout the convective zone, and (2) a small fixed K$_{zz}$ = 1 cm$^2$ s$^{-1}$ is applied only within the radiative region to maintain numerical stability in the Ackerman–Marley condensation scheme. This fixed value does not represent additional mixing but simply provides a diffusion floor where convection ceases.
   
   The assumption is well supported by 3D radiation-hydrodynamic simulations of M-dwarfs, which show that convective overturning extends up to optical depths of $\tau \sim 10^{-3}$, while overshoot and wave-driven mixing above the R–C boundary are orders of magnitude weaker \citep{freytag2010,allard2012,Chabrier2000,tremblin2015cloud}. Thus, convective mixing overwhelmingly dominates vertical transport across all observable pressures in late-type M-dwarfs \citep{Rajpurohit2013}, making the use of a small radiative K$_{zz}$ both physically reasonable and numerically stable. In future work, we plan to test the impact of enhanced non-convective K$_{zz}$ values on condensate lofting and particle growth in more weakly convective or partially radiative regimes.
   
   The sedimentation efficiency is fixed at f${sed}$ = 2, matching empirical calibrations from brown-dwarf studies \citep{morley2012neglected,morley2024sonora}. Particle radii follow a log-normal distribution, with the mean r$_{eff}$ varying with local thermodynamic conditions following Eq.~17 of \citet{ackerman2001precipitating}. r$_{eff}$ depends primarily on local supersaturation and terminal velocity, ensuring that grain sizes evolve naturally with altitude even when K$_{zz}$ is fixed \citep{ackerman2001precipitating,morley2012neglected,Morley2014water,tremblin2015cloud,lefevre2022cloud}. K$_{zz}$ primarily modulates the amount of condensate lofted ($q_c$), rather than prescribing a single fixed r$_{eff}$ \citep{ackerman2001precipitating}. Optical properties (absorption, scattering, and extinction) are treated using
    wavelength-dependent Mie theory \citep{Mie1908,Bohren1983}. We adopt published Mie optical-property tables 
    (e.g., \citealt{Wakeford2015,Kitzmann2018}), which provide $Q_{\rm ext}(\lambda,r)$, $Q_{\rm sca}(\lambda,r)$, $\omega_0(\lambda,r)$, and $g(\lambda,r)$ as a function of grain radius and wavelength. These are interpolated onto our spectral grid and integrated over a log-normal size distribution to obtain mass extinction coefficients for each condensate. The resulting cloud opacities are coupled directly into the radiative-transfer solver, modifying both the temperature gradient and emergent spectrum self-consistently.

    \item \textbf{Convection:} Convection in \texttt{SPHINX II} is modeled using 
    standard mixing–length theory (MLT; \citealt{bohm1958wasserstoffkonvektionszone,hubeny2017model}), 
    which provides a self–consistent prescription for energy transport in regions 
    where the radiative temperature gradient exceeds the adiabatic gradient. 
    Following Equation 34 of \citet{hubeny2017model}, the convective heat flux is
    
    \begin{equation}\label{mlteq}
    F_{\mathrm{conv}} 
    = \left( \frac{g\,\delta\,H_{P}}{32} \right)^{1/2}
    (\rho c_{p} T)\,(\nabla - \nabla_{\mathrm{el}})^{3/2}\,\alpha^2,
    \end{equation}
    
    where $g$ is the surface gravity, $H_{P}$ is the pressure scale height, 
    $\rho$ is the mass density, $c_{p}$ is the specific heat at constant pressure, 
    and $\alpha = \ell/H_{P}$ is the dimensionless mixing–length parameter. 
    Here $\nabla = d\ln T / d\ln P$ is the actual temperature gradient and 
    $\nabla_{\mathrm{el}}$ is the temperature gradient of a convective element. 
    The quantity $\delta \equiv -(\partial \ln \rho/\partial \ln T)_{P,\mu}$ 
    is the dimensionless thermodynamic derivative (often denoted $Q$ in the MLT literature), 
    ensuring that Eq.~\ref{mlteq} has units of 
    erg\,cm$^{-2}$\,s$^{-1}$ for a heat flux.

    In the MLT framework, buoyant elements transport energy and mass over a distance $l = \alpha H_p$, with characteristic convective velocities scaling as $v_{\mathrm{conv}} \propto [g Q H_p (\nabla - \nabla_{\mathrm{el}})]^{1/2}$. These velocities naturally set the efficiency of both energy and material transport, eliminating the need for an additional diffusive mixing term. By allowing MLT to govern both convective flux and mass transport directly, \texttt{SPHINX II} maintains physical self-consistency while avoiding redundancy with the cloud model’s fixed radiative $K_{zz}$ floor.
    
    The mixing-length parameter $\alpha$ remains a tunable control on convective efficiency: smaller $\alpha$ yields steeper temperature gradients and cooler photospheres, while larger $\alpha$ produces deeper convection zones and warmer emergent spectra. Figure~\ref{convectionplot} illustrates how variations in $\alpha$ modify the convective flux profile and the emergent spectrum for a representative 2700~K M-dwarf model. A smaller mixing-length parameter ($\alpha$ = 0.5) produces shallower convective fluxes and a deeper radiative–convective boundary, reflecting less efficient energy transport. Although counterintuitive at first, this behavior follows directly from mixing-length theory: reducing $\alpha$ decreases the size and efficiency of convective elements, steepening the temperature gradient and yielding cooler photospheres. Such reduced surface-convection efficiency has also been inferred from 3D radiation-hydrodynamic simulations of late-M dwarfs, which favor $\alpha$ $\sim$ 0.5–0.7 near the photosphere \citep[e.g.,][]{,Magic2019}.

    We treat the mixing length parameter $\alpha_{MLT}$ as an effective convective efficiency parameter. Empirical constraints on M-dwarf radii and temperatures indicate the need to decrease $\alpha_{MLT}$ toward lower masses, with $\alpha_{MLT} \sim 1$ at $\sim 0.5 M_\odot$ and $\sim 0.5$ at $\sim 0.3 M_\odot$ (\cite{mann2015constrain}, see subsection 8.5.3). Consistent with these results and expectations for molecule-rich convective envelopes, we adopt $\alpha_{MLT} \in [0.5,1.0]$ to bracket plausible convective behavior in late-type M-dwarfs. Exploring lower values associated with magnetic inhibition of convection (e.g. \cite{Cox1981,Chabrier2007}) is deferred to future work.

    \begin{figure*}[!tbp]
    \centering
    \includegraphics[width=\textwidth]{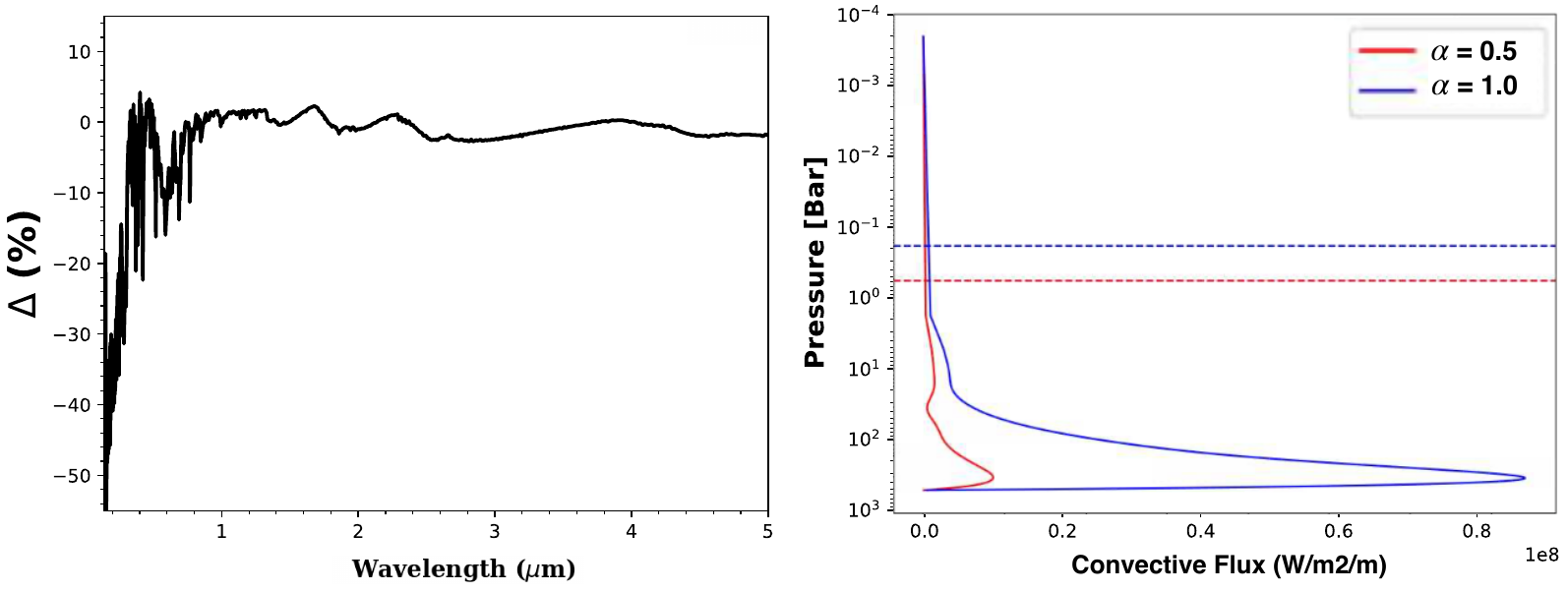}
    \caption{Relative spectral differences \textbf{(Left)} and convective-flux profiles \textbf{(Right)} for a mid-type M-dwarf with varying mixing-length parameters ($\alpha$ = 1.0, 0.5). The model assumes $T_{\mathrm{eff}}$ = 2700 K, $\log g$ = 5.0, and solar composition. The right panel shows deeper radiative–convective (R–C) boundaries (dashed lines) in the lower-$\alpha$ case, where convection is less efficient and the radiative zone extends deeper. This reduced convective heat flux steepens the temperature gradient and cools the photosphere, producing up to 55 $\%$ lower near-IR flux below 1 $\mu$m—consistent with the weaker convective efficiency expected in late-M dwarfs \citep{Cox1981,Chabrier2007}. \label{convectionplot}}
    \end{figure*}


 \item \textbf{Cloud--Convection Feedback:} In addition to including cloud opacity in the radiative transfer, we implement a feedback mechanism that links condensates to convective stability, following the approach outlined for substellar objects in \citet{lefevre2022cloud}. Cloud opacity modifies the local radiative gradient $\nabla_{\mathrm{rad}}$, which determines convective instability. As condensates increase the optical depth, regions that would otherwise remain radiative can become convective, altering the depth and extent of convection zones. Following \cite{lefevre2022cloud}, we implement this through a modified local gradient:
 \begin{equation}
    \nabla_{\mathrm{rad}}' \;=\; \nabla_{\mathrm{rad}}
    \left[\,1 + f_{\mathrm{cloud}}(\tau_{\mathrm{cond}})\,\right],
    \end{equation}
    where $\tau_{\mathrm{cond}}$ is the condensate optical depth in each layer and $f_{\mathrm{cloud}}$ is a scaling function that steepens $\nabla_{\mathrm{rad}}$ proportionally to the additional cloud opacity. This modified gradient, $\nabla_{\mathrm{rad}}'$, is passed directly into the MLT solver, ensuring that both radiative and convective transport adjust self-consistently in response to condensate loading.
    This convection feedback refers purely to the radiative--dynamical coupling between clouds and convective stability, not to non-equilibrium chemistry or convective quenching. While transport-induced chemistry is expected to operate in the coolest M dwarfs and substellar objects (e.g., CO/CH$_4$ and NH$_3$ quenching), SPHINX-II presently assumes chemical equilibrium and does not include explicit kinetic quenching. Incorporating non-equilibrium chemistry and dynamical mixing timescales is an extension for even cooler substellar objects that we will address in forthcoming work.

\end{itemize}

We use two primary datasets for our analysis to test and validate the models:
\begin{enumerate}
    \item A sample of 32 M-dwarfs observed using the SpeX spectrograph \citep{rayner2003} on the NASA Infrared Telescope Facility (IRTF), compiled from the SpeX Prism Library \citep{burgasser2014spex}. See Table~\ref{spexdatatable} for target details.

    \item A set of 39 mid-to-late M-dwarfs (M4.5-9.5) with FGK and early-M-type primaries from \citet{mann2014prospecting}, also observed with SpeX/IRTF. These spectra were obtained in cross-dispersed mode with a 0\farcs3 slit and cover 0.8–2.4~$\mu$m at $R\sim2000$. Data extraction and reduction procedures are described in Section 3.2 of \citet{mann2014prospecting}. See Table~\ref{manndatatable}. This sample is explicitly used for model validation due to the empirically constrained metallicities of M-dwarfs by \citep{mann2014prospecting}.
\end{enumerate}

As in part I, we restrict our analysis to low-resolution spectra in order to leverage broadband molecular features in M-dwarf atmospheres while avoiding the need to model complex processes such as microturbulence. The \texttt{SPHINX II} grid is computed at a native resolution of $R\sim250$. All spectra are therefore smoothed to match this resolution and flux-calibrated using Mauna Kea Observatories (MKO) H-band photometry.

Finally, since the SpeX instrument samples the instrumental line spread function across approximately 2.5 pixels, we follow the procedure of \citet{line2015uniform,line2017,zalesky2019,zalesky2022uniform} and smooth the data by averaging every third spectral point before our grid-model retrieval analysis.

\section{Grid-Model Interpolation Routine, Priors, and Likelihood Evaluation}\label{interpolation}

Following the analysis laid out in paper I \citep{iyer2023sphinx}, we again use grid-model retrievals (grid-trievals) technique to solve for five fundamental stellar properties: effective temperature ($T_{\rm eff}$), surface gravity (log $g$), metallicity ([M/H]), carbon-to-oxygen ratio (C/O), and stellar radius ($R_*$), which is constrained through the quantity $(R_*/D)^2$, where $D$ is the distance taken from the Gaia DR3 catalog \citep{gaiadr3} (see Tables~\ref{spexdatatable} and \ref{manndatatable}).

We perform linear interpolation across the model grid using a Delaunay triangulation method \citep{delaunay1934bulletin}. Posterior exploration and evidence-based model comparison are performed using the \texttt{MultiNest} nested sampling algorithm\citep{feroz2009multinest,feroz2019multinest}, integrated into the interpolation framework to ensure robust convergence across multimodal likelihood surfaces and efficient sampling of degenerate parameter spaces.

The parameter priors are uniform across the \texttt{SPHINX$~$II} grid ranges:
\begin{itemize}
    \item $2000 \leq T_{\rm eff} \leq 4000$ K, $\Delta T = 100K$
    \item $4.0 \leq \log g \leq 5.0$, $\Delta \log g = 0.25$
    \item $-1.0 \leq$ [M/H] $\leq 1.0$, $\Delta [M/H] = 0.25$
    \item $0.3 \leq$ C/O $\leq 0.7$, $\Delta C/O = 0.2$
    \item $-32  \leq log  \ \kappa \leq -28$, $\Delta \log \kappa = 1$
    \item $0.5 \leq \alpha_{mlt} \leq 1.0 $, $\Delta \alpha_{mlt} = 0.5$

\end{itemize}

To simplify and speed up the fitting process, we only restrict our analysis with the simpler gray cloud opacity model grid. We also assume uncorrelated residuals between model and data, and do not include Gaussian Process-based correlated noise modeling as implemented in part I. Instead, we account for potential model imperfections and observational systematics by introducing an error inflation term during the likelihood evaluation.

Our log-likelihood function is defined as:
\begin{equation}\label{nonstarfishlikelihood}
    \ln \, p(D|M) = -\frac{1}{2} \sum_{i=1}^{n} \frac{(y_i - F_i(x))^2}{s_i^2} - \frac{1}{2} \ln(2\pi s_i^2),
\end{equation}
where $y_i$ is the observed flux at the $i$th wavelength point, $F_i(x)$ is the corresponding model flux, and $s_i$ is the total uncertainty.

The total uncertainty $s_i$ includes both the observational error $\sigma_i$ and an error inflation term defined as:
\begin{equation}\label{errorinflate}
    s_i^2 = \sigma_i^2 + 10^b,
\end{equation}
where $b$ is a free parameter sampled during the fitting process.

\section{Results}\label{paper2results}

\begin{figure*}[t]
    \centering
    \includegraphics[width=0.9\textwidth]{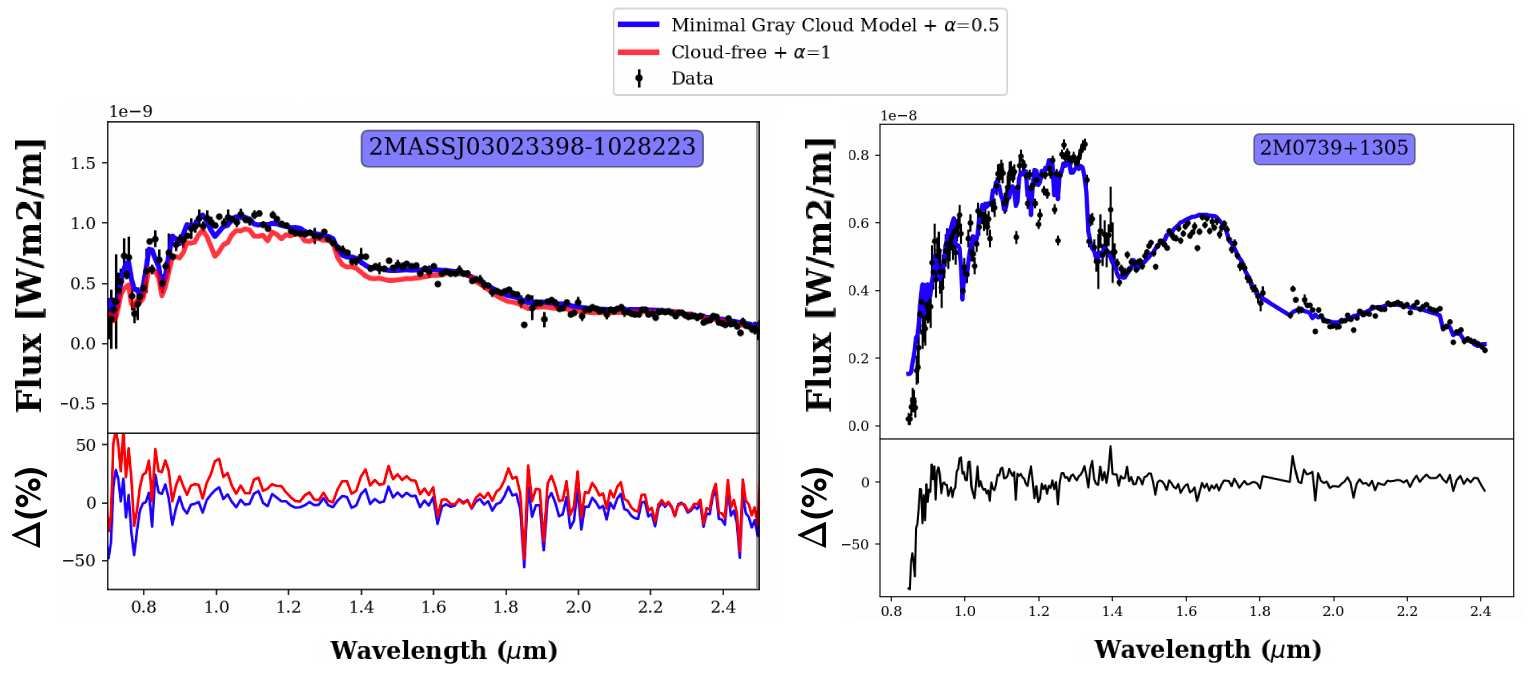}
    \caption{Representative spectral fits comparing \texttt{SPHINX II} cloudy and \texttt{SPHINX I} cloud-free models. 
    \textbf{(Left)} A SpeX Prism Library late-type M-dwarf (2MASS J03023398--1028223). 
    \textbf{(Right)} A benchmark FGK+M companion from \citet{mann2014prospecting} (2M0739+1305). 
    Black points show observed spectra, with error bars where visible. 
    Red curves denote cloud-free \texttt{SPHINX I} models with $\alpha = 1$, 
    while blue curves include a minimal gray cloud opacity \texttt{SPHINX II} models ($\log \kappa = -29$) and reduced mixing length $\alpha = 0.5$. 
    Residuals are plotted below each spectrum. 
    The cloudy, low-$\alpha$ model improves near-infrared slopes and yields metallicities consistent with FGK primaries, 
    whereas the cloud-free case tends to bias metallicity high.}
    \label{fig:rep_fits}
\end{figure*}

We begin by applying our cloud-free \texttt{SPHINX I} model grid (default with mixing-length parameter $\alpha = 1$), to our first sample of 32 M-dwarf near-infrared (NIR) spectra from the SpeX Prism Library Database \citep{burgasser2014spex}. For each target, we fit for five fundamental parameters: effective temperature ($T_{\rm eff}$), surface gravity (log $g$), radius ($R_*$), metallicity ([M/H]), and carbon-to-oxygen ratio (C/O). These best-fit models are shown in red in Figure~\ref{fig:rep_fits} left panel, and Figure \ref{spexfits}.

We then repeat the fits using the upgraded \texttt{SPHINX$~$II} model, assuming a minimal gray cloud opacity (log $\kappa = -29$) and a reduced convective mixing-length parameter ($\alpha = 0.5$), appropriate for cooler M-dwarfs \citep{xuan2024validation,mann2015constrain,Cox1981,Chabrier2007}. The choice of fixing gray cloud opacity and mixing length parameter was done to reduce the dimensionality of the grid and reduce computational burden on the interpolation routine. These results are shown in blue in Figure~\ref{fig:rep_fits} left panel, and Figure \ref{spexfits}. Across all fits, the relative residuals between data and model remain within 20\% across the 0.8–2.3 $\mu$m SpeX bandpass—comparable to those achieved with higher precision spectra such as from JWST \citep{rackham2023towards}.

\begin{figure*}[!tbp]
    \includegraphics[width=\textwidth,trim={0 1.2cm 0 0},clip]{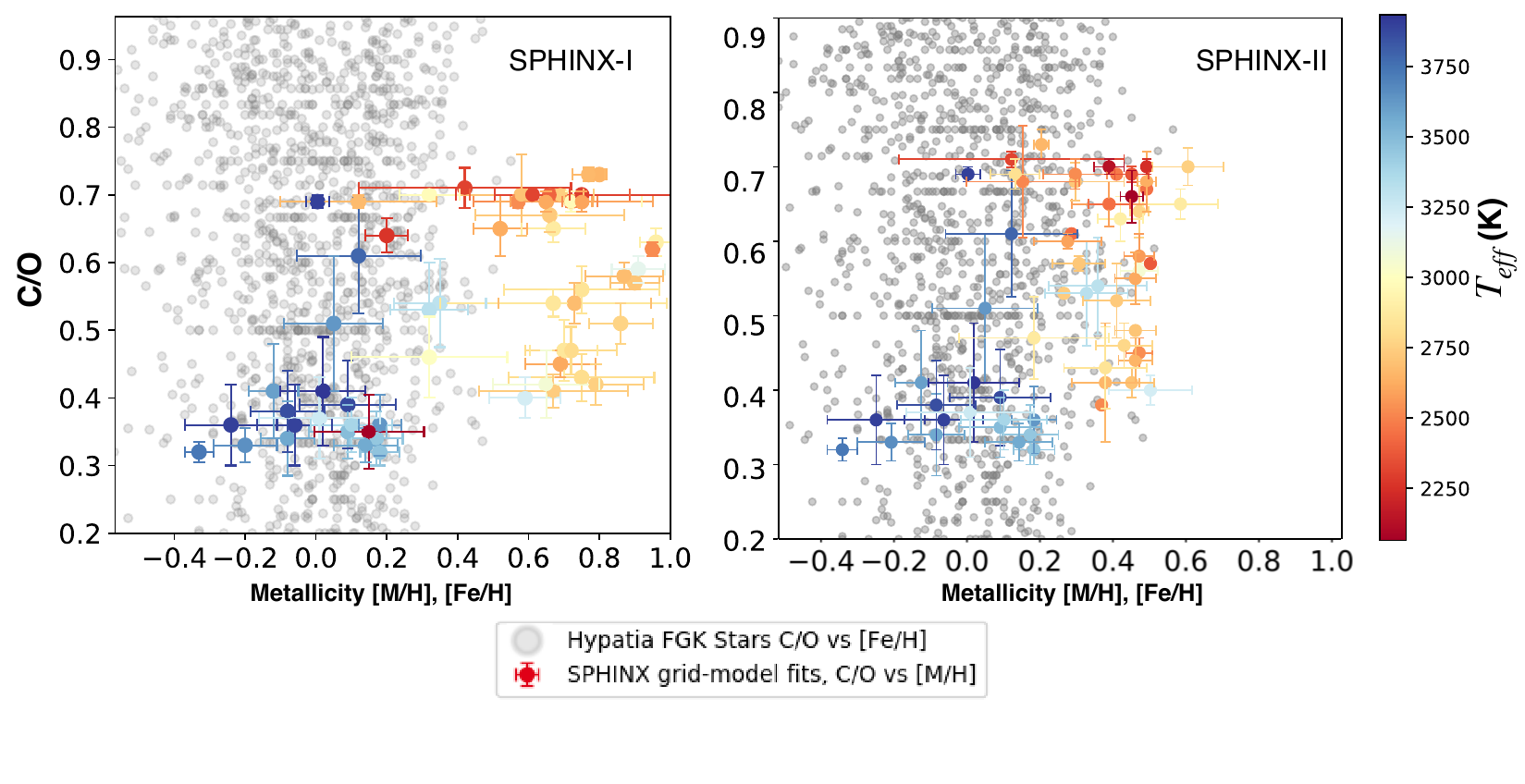}
    \caption{Model Inferred metallicities and C/O values for all M-dwarfs in the SpeX Database \citep{burgasser2014spex} sample. We also overlay our results from part I \citep{iyer2023sphinx} with values inferred for WBS + IS early-M-type targets. \textbf{(Left)} all targets were fitted using our fiducial \texttt{SPHINX I} model including cloud-free atmospheres and mixing-length value $\alpha$=1.  \textbf{(Right)} we show fits for the same using the upgraded \texttt{SPHINX II} models with a fixed minimal gray cloud opacity (log$\kappa$=-29) and lower mixing-length value of convection ($\alpha$=0.5). The right panel shows how the model upgrades yield comparable [M/H] values that are more consistent with neighborhood FGK-type stars as taken from the Hypatia Catalog \citep{hinkel2014} shown in gray.
        \label{fig:met_vs_ctoo_beforeandafter}
    }
\end{figure*}

Model comparison using the Bayesian Information Criterion (BIC) favors \texttt{SPHINX II} for 29 out of 32 targets (90.6\%), indicating that including cloud opacity and reduced convective efficiency improves spectral fits. Furthermore, metallicities and C/O ratios inferred from \texttt{SPHINX II} are more consistent with expectations based on nearby FGK-type stars, as cataloged in the Hypatia Catalog \citep{hinkel2014}, compared to values derived from the cloud-free, $\alpha=1$ \texttt{SPHINX I} fits (see Figure~\ref{fig:met_vs_ctoo_beforeandafter}).

We extend our analysis to 39 benchmark mid-to-late type M-dwarfs with FGKM primaries from \citet{mann2014prospecting}, again assuming log $\kappa = -29$ and $\alpha = 0.5$. These systems allow for empirical validation of \texttt{SPHINX II} by comparing model-derived metallicities to those inferred from their primary stars. Best-fit spectra for these companion targets are shown in the right panel of Figure~\ref{fig:rep_fits}, and Figure \ref{fig:mannfits_app}, with median relative residuals below 20\%.

\begin{figure}[!tbp]
    \includegraphics[width=\columnwidth]{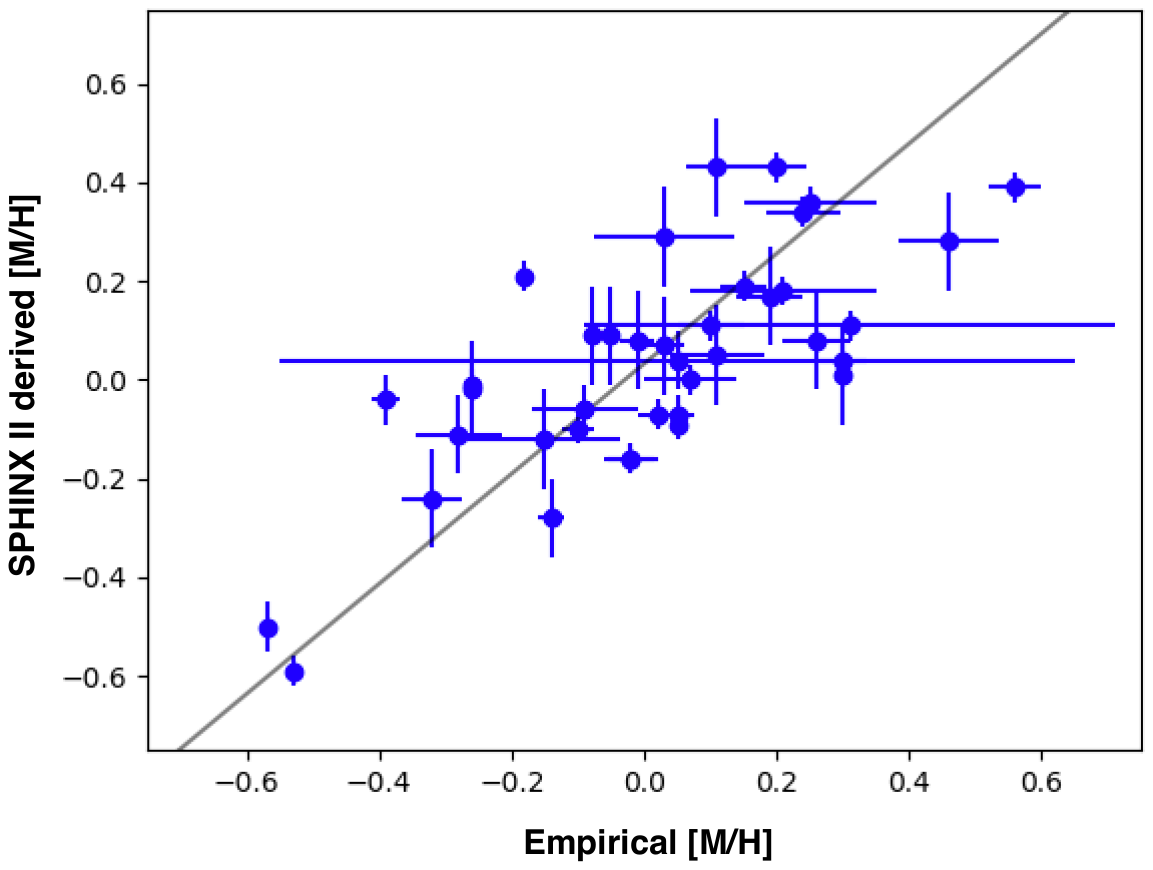}
    \caption{Model Inferred metallicities compared to empirically derived [M/H] values from \cite{mann2014prospecting} for FGKM+companion mid-to-late M-dwarfs from table \ref{manndatatable}. Here we show that both empirical- and model-derived values are strongly consistent with a linear function with a slope of 1.1 and intercept of 0.03 with a p-value of 1e-9. Therefore, we validate our \texttt{SPHINX II} model against these benchmark mid-to-late type M-dwarfs with a scatter of 0.078 dex in \textit{SPHINX II} derived [M/H], consistent with the empirically calibrated values.
        \label{fig:met_empirical_check}
    }
\end{figure}

In Figure~\ref{fig:met_empirical_check}, we compare model-derived metallicities to empirical values from \citet{mann2014prospecting}. A linear regression accounting for uncertainties in both axes yields a slope of 1.1 and an intercept of 0.03, with a p-value of $1 \times 10^{-9}$, indicating a statistically significant correlation. This rejects the null hypothesis at the 95\% confidence level. The overall scatter in \texttt{SPHINX II}–derived metallicities (0.078 dex) is also consistent with the empirical dispersion reported by \citet{mann2014prospecting}.

Best-fit parameters for all targets in both datasets are listed in Tables~\ref{bestfitspexvalues} and \ref{manndatabestfitvalues}. All posterior distributions are provided in the supplementary figures.

\section{Discussion}\label{paper2discussion}

\subsection{Effect of Clouds on mid-to-late type M-dwarf Fundamental Properties}\label{mdwarfprop}
\begin{figure}[!tbp]
    \includegraphics[width=\columnwidth]{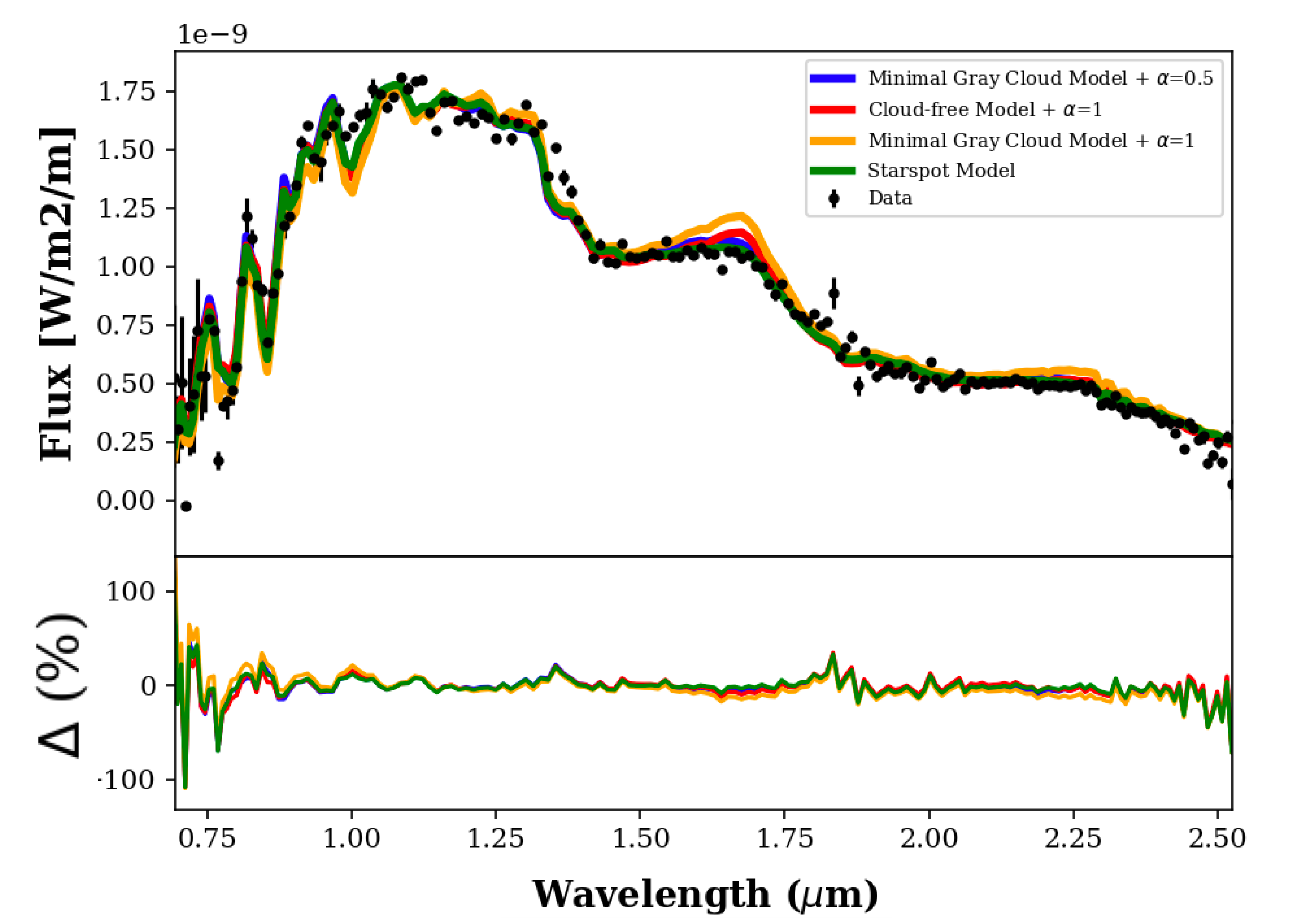}
    \caption{Best fits to potentially young SpeX target 2MASSJ00013044+1010146 using four models: cloud-free + $\alpha$=1 (red), gray cloud with log $\kappa$=-29 + $\alpha$=1 (yellow), gray cloud + $\alpha$=0.5 (blue), and starspot-parameterized cloud-free + $\alpha$=1 (green). While all achieve residuals below 20\% in the 0.8–2.4 $\mu$m range, the BIC strongly favors the blue (\texttt{SPHINX II}) and green (\texttt{SPHINX II} with spots) models (by $\sim$3 dex).}
    \label{fig:allmodcompare_tospex}
\end{figure}

Our updated \texttt{SPHINX II} grid demonstrates meaningful improvements over cloud-free or high-convective-efficiency models, particularly for late-type M-dwarfs. By including a fixed gray cloud opacity and reduced mixing-length parameter ($\alpha=0.5$), we see broad improvements in both spectral fits and physical inferences. These enhancements are most evident in comparisons between cloud-free \texttt{SPHINX I} vs cloud-inclusive \texttt{SPHINX$~$II} fits for SpeX target 2MASS J00013044+1010146 as shown in Figure~\ref{fig:allmodcompare_tospex}, where models including clouds significantly outperform the baseline. We additionally also include an analysis with a stellar spot parameterization to capture photospheric heterogeneity as done in Figure 11 of part I \citep{iyer2023sphinx} and the resultant BIC comparisons favor both the \texttt{SPHINX II} (blue) and the starspot model (green) by more than 3 dex relative to the cloud-free baseline. This result is aligned with a current hypothesis of being a potentially young object with low surface gravity and dust. \citep{gagne2014banyan}.

The posterior histograms (See Appendix Figure \ref{cornerplot}) highlight how degeneracies play out: although both \texttt{SPHINX II} and the spot model perform comparably in BIC, the spot parameters push against prior boundaries, suggesting the cloud model offers a more physically plausible explanation. Notably, only \texttt{SPHINX II} (log $\kappa = -29$, $\alpha=0.5$) yields metallicities in agreement with expectations from nearby FGK-type stars (Figure \ref{fig:met_vs_ctoo_beforeandafter}).

Interestingly, a subset of spectra (e.g.,
2MASS J18112466+3748513) show poor fits regardless of model choice, suggesting additional physics may be needed, such as higher-order cloud structures or magnetic effects. For some targets, even including stellar spots yields no improvement (see supplementary figures), reinforcing the limitations of simple heterogeneity or cloud prescriptions.

\begin{figure}[!tbp]
    \includegraphics[width=\columnwidth]{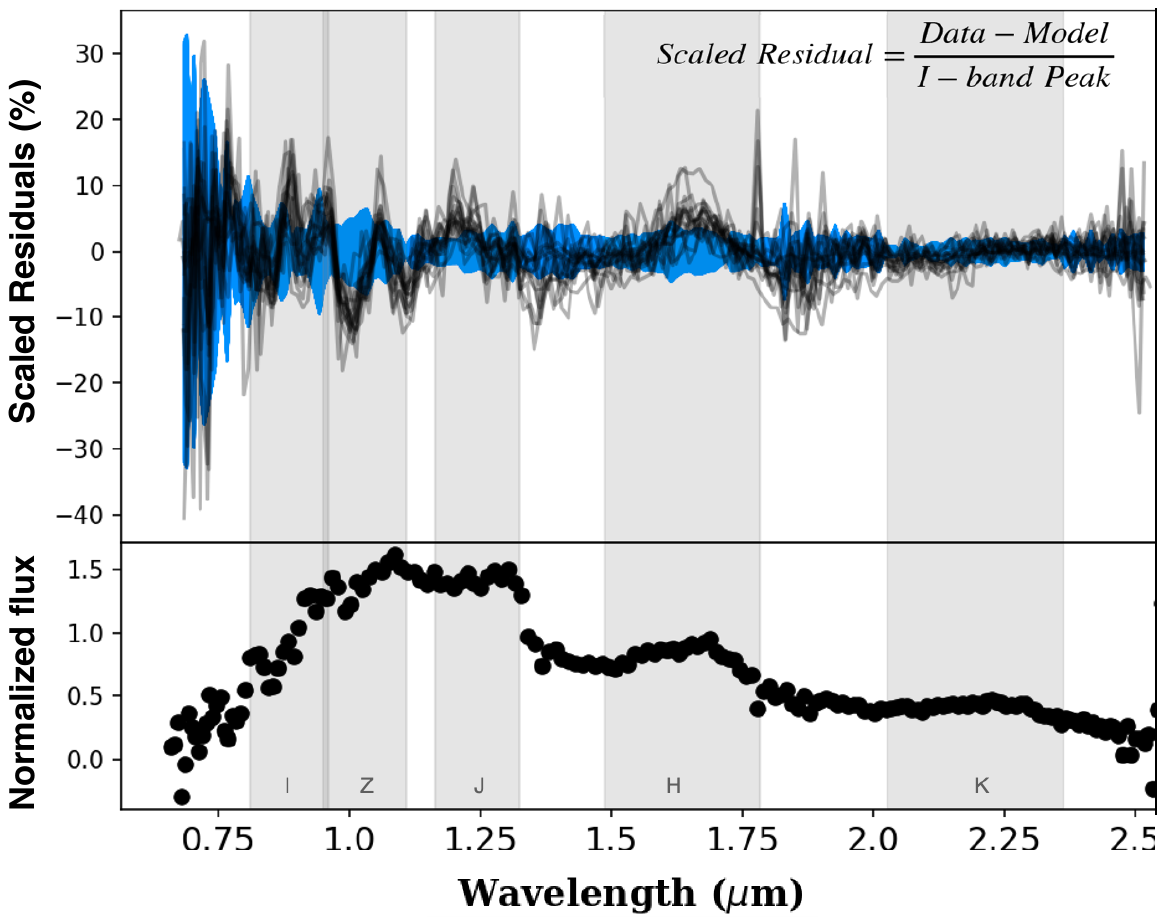}
    \caption{Top: Stacked residuals from fiducial (cloud-free, $\alpha$=1) \texttt{SPHINX I} fits to all SpeX targets. Blue shading shows the 1$\sigma$/2$\sigma$ scatter from Starfish draws. Bottom: Sample spectrum of 2MASSJ00552554+4130184 (normalized). Vertical bands indicate SpeX filter regions. Key discrepancies occur in the Z, J, and H bands. With a systematic SpeX dataset in hand, we hypothesize the bulk of the correlated noise being driven by deficiencies in \texttt{SPHINX I} for characterizing these mid-to-late type M-dwarfs.}
    \label{cloud-free-starfish-lateM}
\end{figure}

\begin{figure}[!tbp]
    \includegraphics[width=\columnwidth]{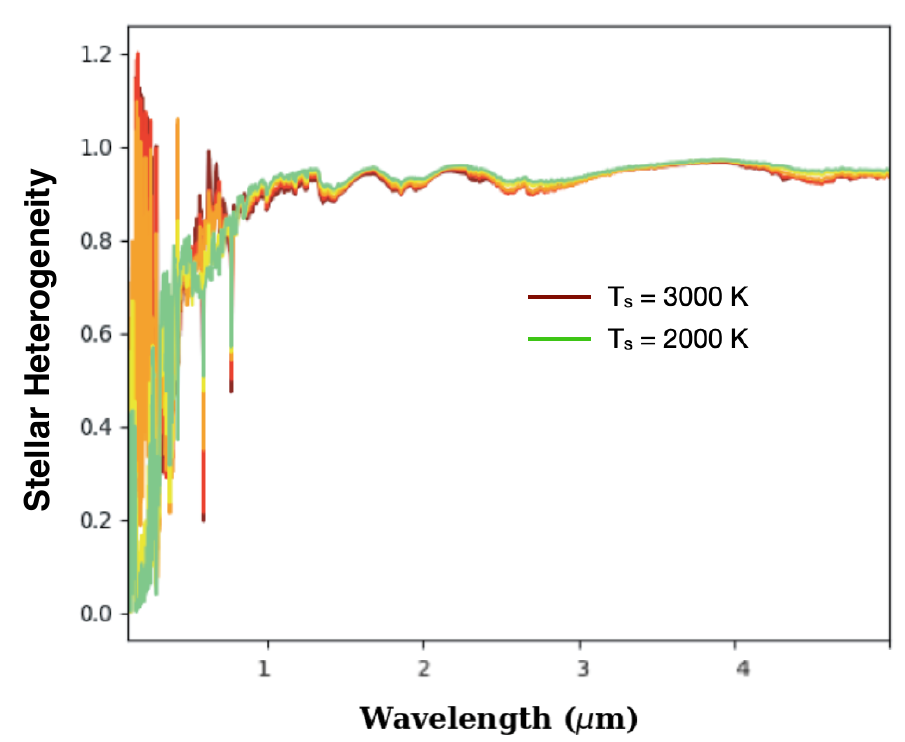}
    \caption{Simulated spectral differences in a 3000 K M-dwarf due to starspots with coverage of 20\% and spot temperatures from 2000–3000 K. Degeneracies arise in the same NIR regions impacted by clouds and convection.Value of 1 would indicate a quiescent photosphere spectrum.}
    \label{stellaractivityinlateM}
\end{figure}

There is also the question of correlated noise sources stemming either from model deficiencies or data systematics. To probe these systematics further, we ran a grid of cloud-free, $\alpha=1$ \texttt{SPHINX I} models using the GP-based $Starfish$ likelihood approach as done in Paper I \citep{iyer2023sphinx}. The goal of this exercise is to understand the dominant source of residuals when working with this cohesive unformly reduced Spex Prism Library data. Stacked residuals (Figure \ref{cloud-free-starfish-lateM}) confirm that the dominant model-data mismatches arise in the Z, J, and H bands—wavelengths sensitive to FeH, TiO/VO, and cloud opacity. Residuals peak below 0.8 $\mu$m and in narrow spectral intervals between 1.0–1.3 and 1.6–1.8 $\mu$m, consistent with expected telluric or opacity limitations. This shows that although we have a uniform, systematically reduced dataset for this analysis, model deficiencies from \texttt{SPHINX I} when applied to this sample of cooler M-dwarfs, are likely dominating the residuals. While the mean residual working with a uniform dataset has shown some marginal improvement relative to the ``stitched'' multi-instrument dataset used in Paper I \cite{iyer2023sphinx}, they certainly do not eliminate them entirely, underscoring the need for missing cloud-convection physics and high fidelity datasets. In fact, these stellar spectra are additionally only a snapshot in time of the star and it is worth noting that these results could also be influenced by stellar activity. Figure \ref{stellaractivityinlateM} for instance shows how starspots in late-Ms contribute wavelength-dependent slopes and feature variations, particularly in the Z and J bands—adding further degeneracies to metallicity or cloud interpretations within the same bandpass.

\begin{figure}[!tbp]
    \includegraphics[width=\columnwidth]{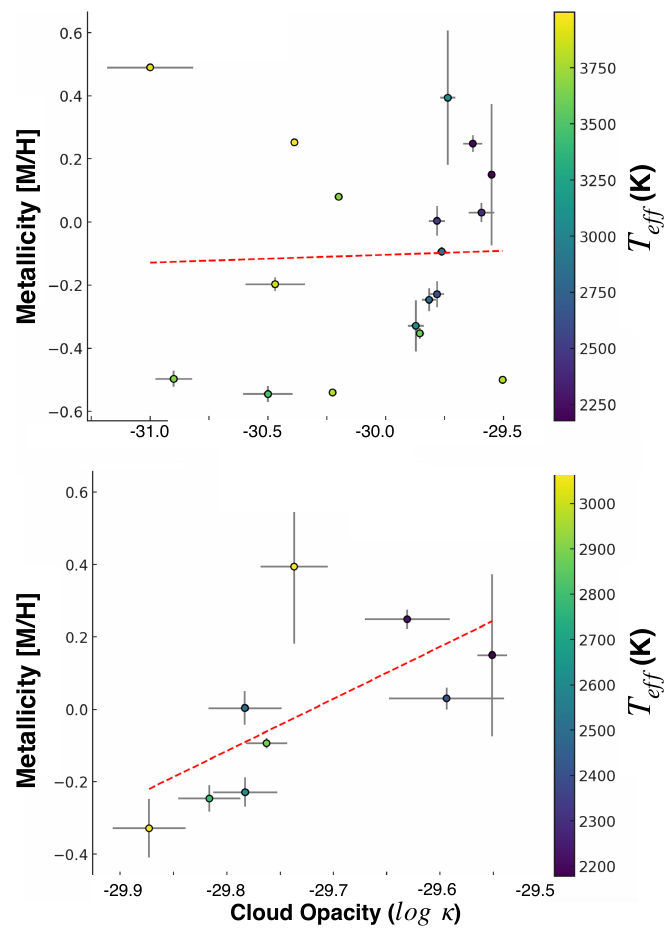}
    \caption{\textbf{(Top)} Metallicity vs. cloud opacity for all targets from \citep{mann2014prospecting}; no significant trend found (p = 0.89). \textbf{(Bottom)} Same plot restricted to $T_{\rm eff} < 3000$ K. A marginal trend appears (p = 0.06), suggesting cloud opacity may correlate with metallicity in cooler M-dwarfs.}
    \label{fig:met_vs_cloudopacity_main}
\end{figure}

Tackling degeneracies such as these motivate a need to explore further connections to 3D processes and improvements within 1D parameterizations, which are valuable to retain computational efficiency when performing grid-trievals. Some 1D non-gray cloud models and 3D Cloud-Convection simulations of brown dwarf atmospheres by \citet{Morley2014water} and \citep{lefevre2022cloud} demonstrate that radiatively active clouds, particularly MgSiO$_3$, can substantially reshape the vertical thermal profile. They find that silicate clouds lead to secondary convective zones and strong heating at cloud bases—features that are not adequately reproduced simply via gray cloud prescriptions. Our own Figure \ref{cloudplot} shows the effective consequence of cloud opacity: muted spectral features and reddened slopes. Prior models \citep{Morley2014water,lefevre2022cloud} also show that metallicity and particle number influence convection depth and thermal perturbations. This trend is echoed in our results, where we show in figure \ref{fig:met_vs_cloudopacity_main} (top) that there is no strong correlation between cloud opacity and metallicity in the full sample, however in the $T_{\rm eff} < 3000$ K subset (bottom), a marginal trend emerges—suggesting a potential metallicity-cloud link at cooler temperatures. 

Condensate cloud opacity in cool dwarf atmospheres is expected to correlate with metallicity because higher [M/H] increases the partial pressures of refractory species (e.g., Mg, Si, Fe), enhancing supersaturation and cloud mass loading. This metallicity–cloud link has been discussed in the context of ultracool dwarfs and giant-planet atmospheres \citep{ackerman2001precipitating,helling2008comparison,Stephens2009,Helling2014,marley2014cool}. The cloud-metallicity trend among the coolest stars (T$_{eff}$ $<$ 3000K) are consistent with the expectation that condensates become more influential as temperatures decrease. Further exploration with larger samples and models incorporating magnetic activity and detailed cloud microphysics can help refine this relationship. With respect to influence of convective mixing-length on the cloud-metallicity trend, lower $\alpha_{MLT}$ values generally reduce the need for higher metallicity to reproduce near-IR slopes \citep{iyer2023sphinx}, therefore exploring $\alpha_{MLT}$ $<$ 0.5 in future work may also further test the robustness of this trend.

\begin{figure}[!tbp]
    \centering
    \includegraphics[width=0.9\columnwidth]{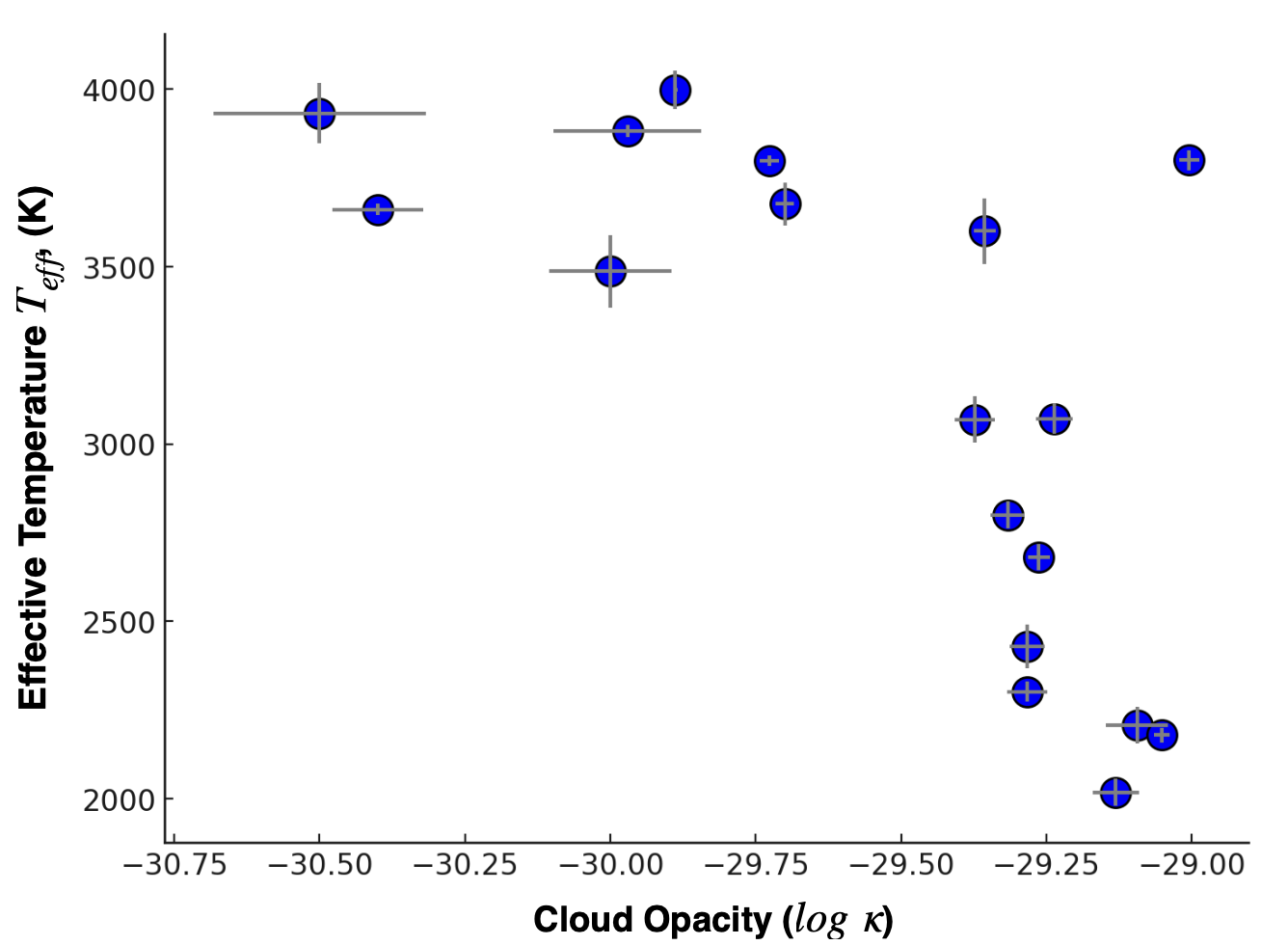}
    \caption{Retrieved cloud opacity versus effective temperature for 10 mid-to-late type M-dwarfs spanning the full T$_{\mathrm{eff}}$ range of the SpeX sample, along with 8 early-type M-dwarfs analyzed in \citep{iyer2023sphinx} taken from \citep{mann2015constrain}. 
    Cooler stars favor more optically thick clouds, consistent with dust formation physics and cloud--convection feedback. 
    For the warmer, less cloudy targets, the grid-trieval compensates by driving T$_{\mathrm{eff}}$ toward the upper edge of the priors when using cloud-free models, effectively biasing the fits toward artificially higher temperatures. However, this underscores the trend of incorporating clouds for the cooler sample.}
    \label{fig:cloud_vs_teff_alone}
\end{figure}

\begin{figure}[!tbp]
    \includegraphics[width=\columnwidth]{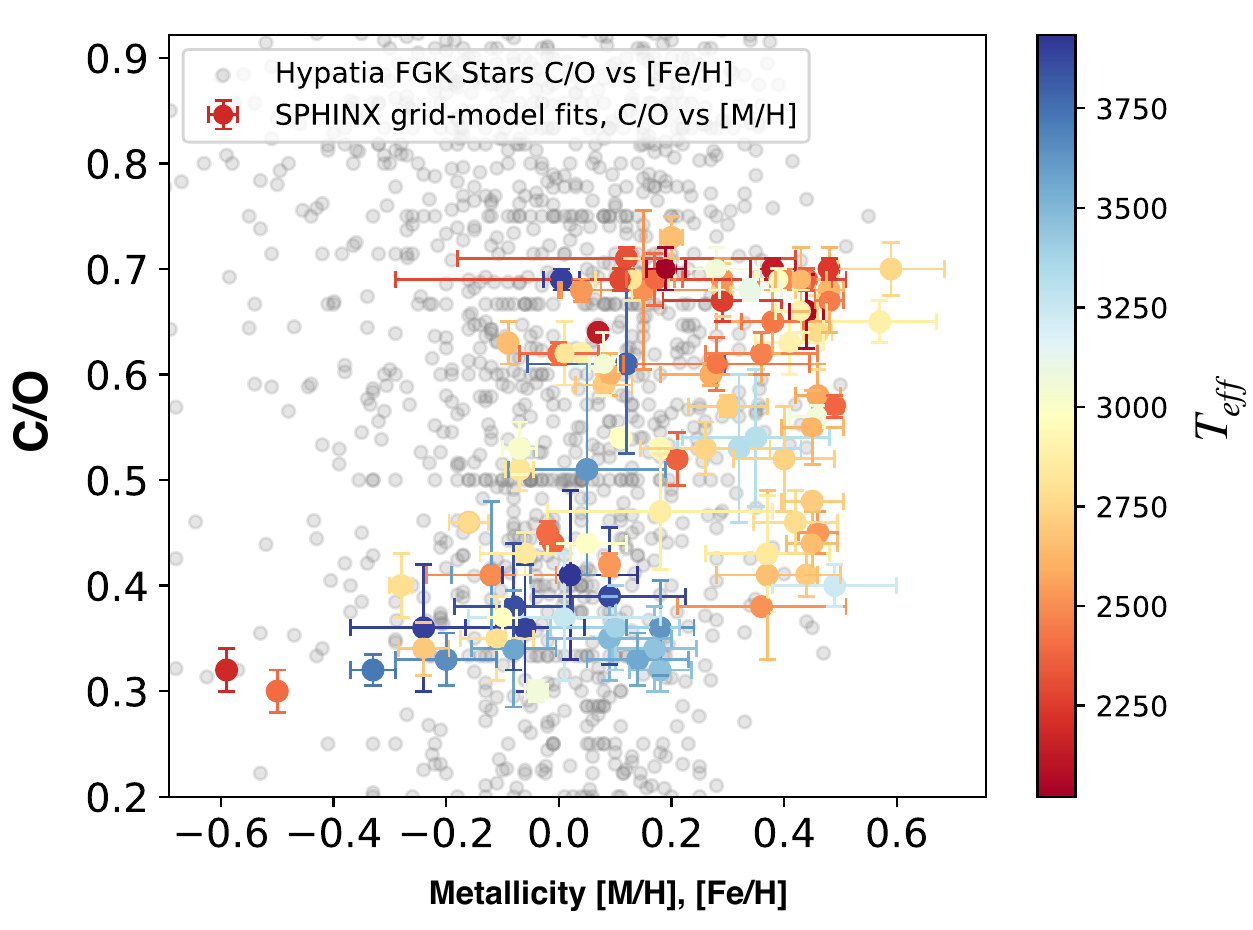}
    \caption{Inferred [M/H] vs. C/O for our full sample from this work. Early-type M-dwarfs (from Part I \citet{iyer2023sphinx}) cluster near solar C/O, while late-Ms span a broader range. FGK-type stars from the Hypatia Catalog shown in gray. This figure shows that even the FGKM+M binaries have consistent metallicities when fitting with the minimal cloud + $\alpha_{mlt} = 0.5$ model. In addition to the right panel of figure \ref{fig:met_vs_ctoo_beforeandafter}, this includes benchmark binary mid-to-late Ms from \citep{mann2013metal}.}
    \label{fig:met_vs_ctoo_ALL}
\end{figure}

We also find a clear anticorrelation between effective temperature and retrieved cloud opacity (Figure \ref{fig:cloud_vs_teff_alone}), consistent with increasing condensate formation as temperatures fall. We examine metallicity–C/O trends across our full sample. Figure \ref{fig:met_vs_ctoo_ALL} shows that \texttt{SPHINX II} produces consistent [M/H] and C/O values with empirical FGK benchmarks (Hypatia catalog), though mid-to-late M-dwarfs span a wider range in C/O than early types. The mean scatter is $\sim$13\%, providing unprecedented precision in low-resolution C/O constraints. To assess whether the apparent concentrations in Figure \ref{fig:met_vs_ctoo_ALL} reflect true chemical substructure, we computed a 2D kernel–density estimate (KDE) and Gaussian-mixture model comparison in the C/O–[M/H] plane. The distribution is smooth and continuous, with only mild density peaks, and the statistical tests do not support a significant bimodal split. We therefore interpret the pattern as consistent with normal Solar-neighborhood abundance variations and small-sample structure rather than distinct chemical populations.

Finally, a note on the lower C/O values: at late-M dwarf temperatures, nearly all carbon is locked in CO, so lowering C/O (i.e., increasing oxygen relative to carbon) enhances H$_2$O abundances while leaving CO largely unchanged (e.g., \cite{lodders2003,Burrows1999}. We verified this behavior in our grid calculations: spectra with lower C/O show stronger H$_2$O bands in the near-IR while the CO bandhead at 2.3 $\mu$m remains insensitive, consistent with carbon-limited equilibrium chemistry (provided as a supplementary figure).

\subsection{Putting Results in Context with Recent JWST Results}\label{jwstsection}
Significant time has been dedicated to observing and characterizing exoplanets and brown dwarf companions around M-dwarf hosts via JWST with Cycles 1-4 \citep{nikolov2022trexolists}. However, there are relatively fewer efforts dedicated to directly characterize stellar hosts. For transit transmission spectra with JWST, the standard approach to tackle the Transit Light Source Effect (TLSE) \citep{rackham2018} has been to correct for the contribution of unocculted stellar surface spots on the transmission spectrum and then marginalizing jointly over the stellar (photosphere-spot/faculae fractional coverage in area and temperature contrasts) and planetary atmospheric parameters (Equilibrium temperatures and planetary chemical compositions)\citep{zhang2018,pinhas2018,iyer2020influence}. However, a key result from \cite{iyer2020influence} warns us that for configurations of small planets orbiting a heterogenous M-dwarf with spot coverage above 1$\%$; we will continue to incur biases in inferred planetary atmospheric properties from transmission spectra---if we do not understand the stellar disk integrated spectrum adequately. In fact, this will be a persistent problem regardless of joint star-planet retrievals, appropriate TLSE corrections and even JWST-quality data precision \citep{iyer2020influence}. 

Another point to note is that significant efforts are being spent in measuring UV/Optical activity indicators that are valuable in understanding photochemical processes on the planetary atmosphere, however; there is no straightforward way yet to connect these parameters to understand the level of photospheric activity in order to infer the actual wavelength-dependent contribution to exoplanet transmission spectra. Stellar variability in the form of rotational modulation at best provides us with a lower limit of the level of activity \citep{jeffries2012}. An additional avenue for model-driven errors could be due to the simplistic assumption that the spot/faculae regions of the photosphere have the same spectra as the rest of the quiescent photosphere, except only varying in effective temperatures. The wavelength dependent behavior  (Figure \ref{stellaractivityinlateM}) described by this assumption is only part of the entire picture.  In fact, several works (e.g. \cite{witzke2022can,norris2023spectral}) show that there is a spectral shape dependence due to photospheric heterogeneity conditions, which are 3D, non-trivial, and heavily influenced by magnetic effects. 

Modeling M-dwarf host stars remains challenging due to several coupled effects.
First, surface heterogeneity (e.g., spots and faculae) and uncertainties in limb-darkening treatments can bias inferred stellar and planetary parameters \citep{patel2022}. Second, one-dimensional LTE spectral synthesis can miss key physics, including dust opacity effects and three-dimensional processes such as turbulent convection and granulation \citep{wende2009}. Together, these factors introduce degeneracies and systematics that limit the accuracy and precision of retrieved fundamental stellar properties, which in turn propagate directly into uncertainties in exoplanet characterization. Not understanding fundamental atmospheric properties of the host star leaves us at a disadvantage when it comes to refining our understanding of the formation mechanisms for substellar mass objects. Additionally, such model deficiencies have far-reaching implications not just for small planets around M-dwarfs, but also for giant exoplanets such as from the GEMS survey \citep{kanodia2024} or M-dwarfs with brown dwarf companions such as ROSS 19B \citep{schneider2021ross}.

With these issues in perspective, we recommend a multi-pronged approach involving observational, theoretical, and statistical inference tools to refine our overall understanding of M-dwarfs. On the observational side, we recommend significant time and efforts be devoted to building high fidelity stellar spectral databases both from space and ground based observations \citep{kesseli2017empirical,jing2024half}. Particularly, gathering observations of M-dwarf binaries and moving group members to provide the necessary grounding and validation of models. We also advocate for multi-band long-baseline monitoring observations of these targets to understand both spectral and time-dependent nature of variability \citep{quintana2021pandora,mori2024}. On the theoretical side, we recommend continuing model development balancing computational tradeoffs with 3D Radiative Transfer / Magneto Hydrodyanamic parameterizations \citep{witzke2022can} into 1D models, and robust statistical tools such as Gaussian Process based Inference methods within a retrieval framework \citep{czekala2015starfish,zhang2021,iyer2023sphinx}.

\subsection{The Case of Trappist-1}\label{Trappist}

\begin{figure}[!tbp]
    \centering
    \includegraphics[width=\columnwidth]{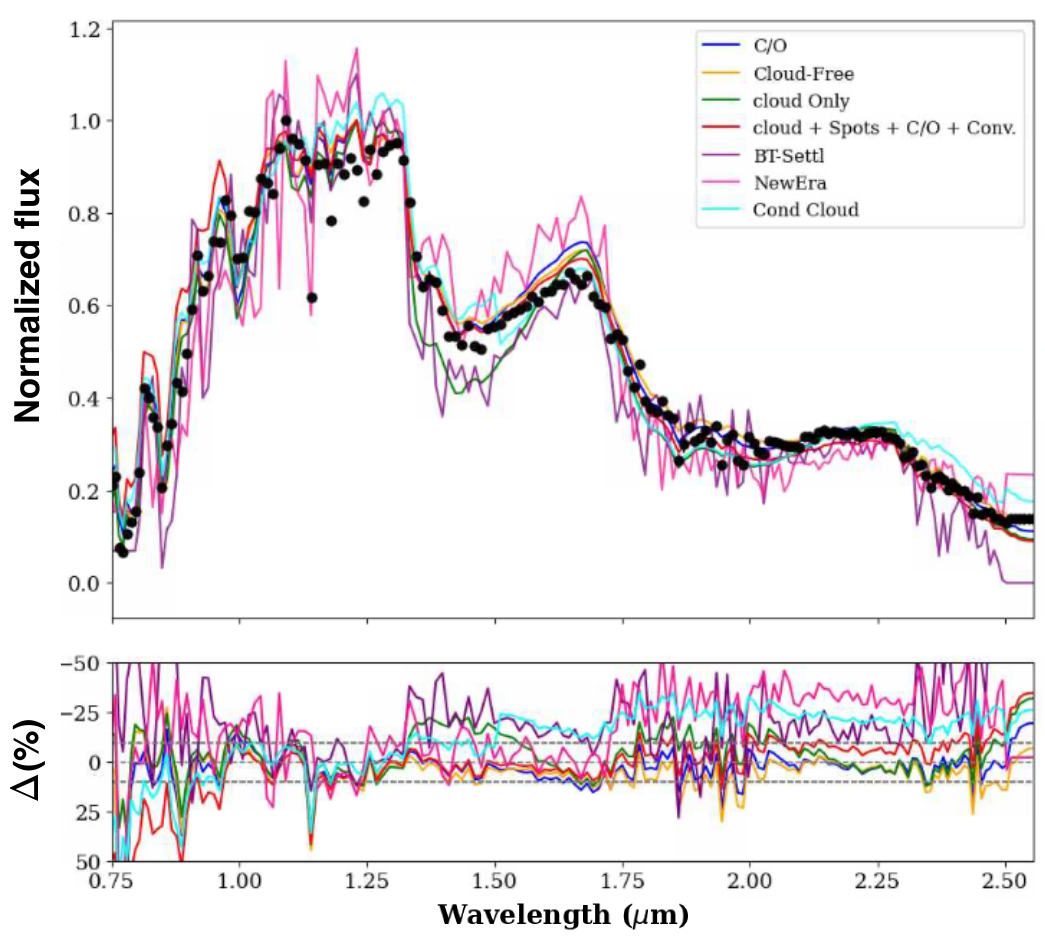}
    \caption{
    Comparison of observed SpeX IRTF spectrum of Trappist-1 (black points) with forward models under different assumptions. Models include: C/O–varying (blue), cloud-free (yellow), gray clouds-only (green), gray cloud+spots+convection (red), BT-Settl \citep{allard2013bt} (purple), NewEra \citep{hauschildt2025newera} (pink), and Cloud condensate + convection feedback model (cyan). The residual panel below shows the percent deviation relative to the data. The gray cloud + $\alpha_{mlt}=0.5$ convection \texttt{SPHINX II} model (red) best reproduces the continuum suppression around 1~$\mu$m and the overall spectral shape across 0.8–2.5~$\mu$m. Simpler models such as the cloud-free \texttt{SPHINX I} case \citep{iyer2023sphinx} (yellow) are statistically disfavored, highlighting the necessity of including cloud opacity and convective feedback in late-M dwarfs.}
    
    \label{fig:Trappistspec}
\end{figure}

\begin{figure}[!tbp]
    \centering
    \includegraphics[width=\columnwidth]{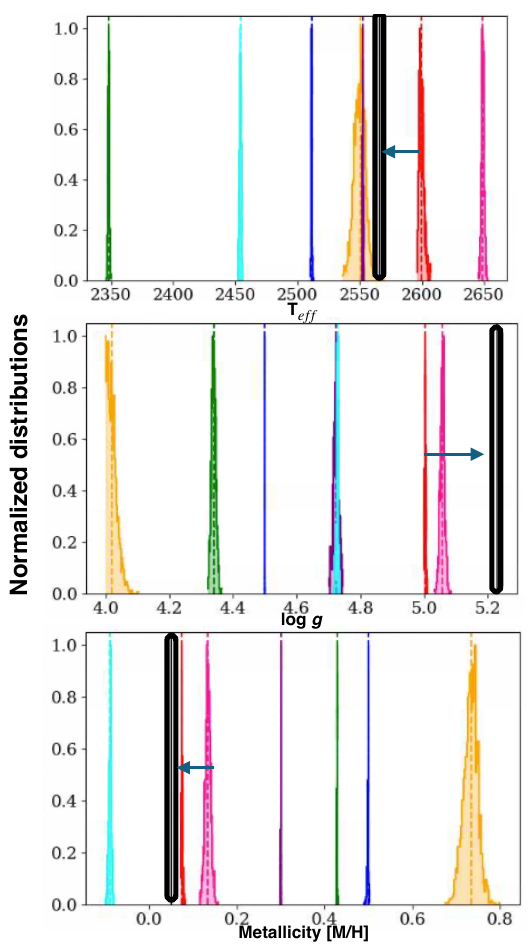}
    \caption{Posterior distributions for Trappist-1 from grid-trievals with different stellar atmosphere models.  
    Shown are marginalized distributions in effective temperature \textbf{(top)}, surface gravity \textbf{(middle)}, and metallicity \textbf{(bottom)}.  
    Colors correspond to different model families: cloud-free \texttt{SPHINX I} \citep{iyer2023sphinx} (yellow), gray cloud-only (green), C/O variation (blue), 
    gray cloud + $\alpha_{mlt} = 0.5$ convection + spots model (red), BT-Settl \citep{allard2013bt} (purple), NewEra \citep{hauschildt2025newera} (pink) and Cloud condensate + convection feedback model (cyan).  
    Vertical dashed lines mark the median values for each posterior.  
    Black boxes and arrows highlight observational values \citep{agol2021,ducrot2020}. The gray cloud + $\alpha_{mlt}$ = 0.5 \texttt{SPHINX II} model and the NewEra grid, 
    both converge toward higher $\log g$ and modest sub-solar metallicities compared to cloud-free solutions.  
    The clustering of solutions around $T_{\mathrm{eff}} \sim 2550$–2600 K, $\log g \sim 5.0$–5.2, and 
    [M/H] $\sim 0.0$–0.2 strongly favors models including a gray cloud opacity, lower mixing length in convection and photospheric heterogeneities, 
    reinforcing the evidence of complexity in Trappist-1 atmosphere.}
    \label{fig:Trappisthist}
\end{figure}

A grid-trieval analysis on low-resolution SpeX IRTF spectra \citep{rayner2003} of Trappist-1 demonstrates (see Figures \ref{fig:Trappistspec} and \ref{fig:Trappisthist}) the challenge in modeling the late-type M-dwarf regime, where model assumptions strongly drive the stellar fits. Bayesian model comparison (via BIC) consistently favors more complex models that include stellar clouds, convection, and surface inhomogeneities over purely gas-opacity models. For context, we also performed grid-trievals with BT-Settl \citep{allard2013bt} and the NewEra model grid \citep{hauschildt2025newera}, which carry their own internal assumptions. Since the bulk of this paper is focused on SpeX IRTF data, we use the same dataset for Trappist-1. To understand the effect of clouds alone, we see that it falls short in describing Trappist-1, when comparing fits with the two versions of \texttt{SPHINX-II} gray cloud only model (green) versus condensate cloud + convection feedback only model (cyan)--that there are no significant improvements in the constraints of all three fundamental parameters shown in Figure \ref{fig:Trappisthist}. Across all frameworks, the models that best reproduce the observations are those that incorporate degeneracies between spots, condensates, and convective feedback, specifically the gray cloud + $\alpha_{mlt} = 0.5$ convection \texttt{SPHINX II} model including spot/faculae parameterizations for the grid-trievals.

\begin{figure*}[!tbp]
    \includegraphics[width=\textwidth]{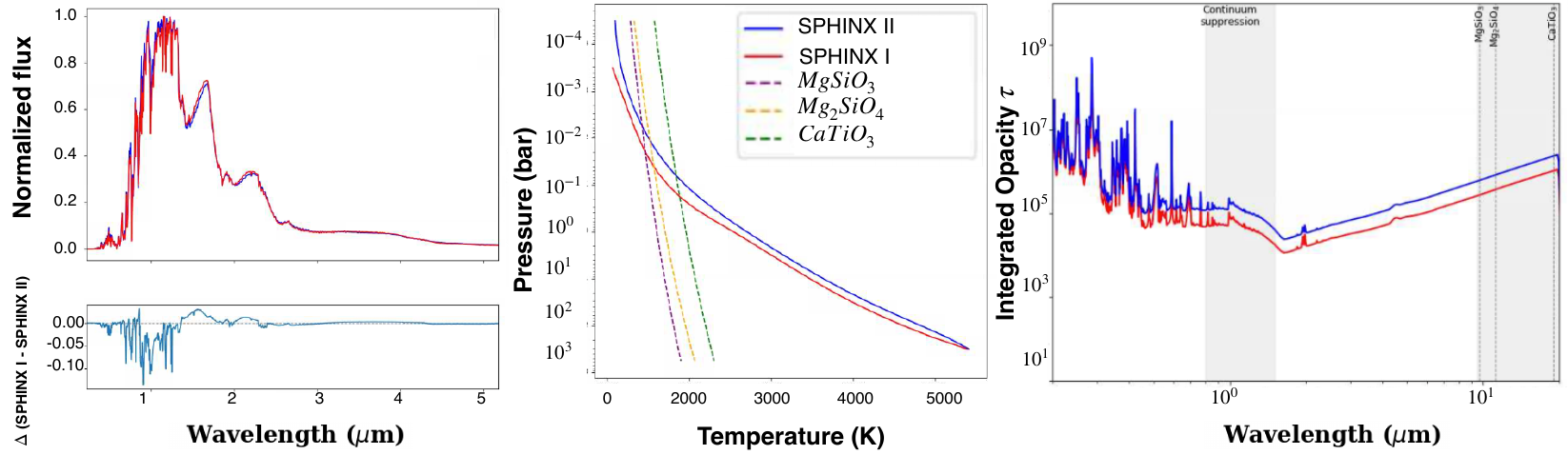}
    \caption{
Comparison of cloud‐free and cloudy forward models for Trappist-1. 
(\textbf{Left}) The condensate‐cloud + convective feedback \texttt{SPHINX II} model (blue) yields a hotter TP structure than the cloud‐free \texttt{SPHINX I} model (red), leading to muted near‐IR flux in the residuals. 
(\textbf{Middle}) Both TP profiles are overlaid with theoretical condensation curves for MgSiO$_3$, Mg$_2$SiO$_4$, and CaTiO$_3$ \citep{lodders2003,visscher2010,marley2013}, showing that the modeled photosphere intersects the silicate condensation boundaries near $10^{-2}$~bar. 
(\textbf{Right}) Integrated optical depth: inclusion of condensates raises $\tau$ across 0.8–2~$\mu$m and enhances opacity through $\sim$8–10~$\mu$m, consistent with broad silicate vibrational behavior \citep{henning1996dust,jaeger1998,min2007}. 
Together these panels illustrate how condensates reshape both the emergent spectrum and the thermal structure of TRAPPIST-1.}
    \label{Trappist1modcompare}
\end{figure*}

To dig deeper into understanding the nature of clouds and their relevance for Trappist-1 atmosphere, we now move on from the gray cloud model and explore the effect of condensates in the Trappist-1 atmosphere. We perform an exercise comparing two forward models--in Figure~\ref{Trappist1modcompare}, assuming $T_{\rm eff}=2566$ K, $\log g = 5.24$ \citep{agol2021}, [Fe/H] $=+0.05$ \citep{ducrot2020A}, and C/O $=0.5$ with cloud-free \texttt{SPHINX I} versus \texttt{SPHINX II} including condensate cloud + convection feedback prescription. The middle panel shows that the \texttt{SPHINX II} (blue) produces a hotter TP structure relative to the cloud-free \texttt{SPHINX I} model (red), which directly correlates with suppressed near-IR flux in the residuals bottom left panel. The middle panel overlays both TP profiles with theoretical condensation curves for MgSiO$_3$, Mg$_2$SiO$_4$, and CaTiO$_3$ \citep{lodders2003,visscher2010,marley2013}, highlighting that the modeled photosphere indeed crosses the silicate condensation boundaries near $10^{-2}$ bar. The right panel shows the integrated optical depth, where the inclusion of condensates substantially elevates $\tau$ around 1 $\mu$m and increases mid-IR opacity, consistent with known vibrational features of silicates at between 10 to 20 $\mu$m \citep{henning1996dust,jaeger1998,min2007}. The effect is a smooth rise in opacity across the 0.8–2 $\mu$m window, mimicking a broad haze-like continuum and a redenning of the overall slope at longer wavelengths. The \texttt{SPHINX II} Condensate cloud+Convection model remains elevated relative to \texttt{SPHINX I} due to the added continuum contribution of condensates, as the former reproduces the overall expected silicate resonance behavior \citep{Henning1997,jaeger1998,draine2003}.

While silicate vibrational modes formally contribute near 10–20~$\mu$m, the predicted condensate column is too low to yield a distinct emission or absorption feature. Instead, these submicron-scale grains act as a smooth gray opacity source, broadly suppressing the continuum without producing a recognizable silicate band—consistent with the lack of such features in late-M dwarfs and Trappist-1’s observed spectrum \citep{Cushing2006,Stephens2009,marley2013}. The $\tau \approx 1$ photosphere occurs near $10^{-1}$–$10^{-2}$~bar, coincident with the upper edge of the modeled condensation zone (see Figure~\ref{Trappist1modcompare}). At longer wavelengths, the added continuum opacity from these submicron silicate grains produces a mild reddening of the spectral energy distribution (SED), slightly flattening the mid-infrared slope—an effect analogous to that seen in L-dwarf transition objects where thin, optically tenuous silicate layers modulate the 3–8~$\mu$m flux distribution \citep{Stephens2009,saumon2012,marley2013}. This behavior is directly reflected in the integrated opacity spectrum for Trappist-1 (Figure~\ref{Trappist1modcompare}, right panel): the condensate contribution raises the continuum opacity across $3$–$8~\mu$m—flattening the mid-IR SED slope, as seen by the muted molecular features (CO) around 2 and 5$\mu$m. The result is a smooth, gray-like suppression of the near-IR and a mild reddening through the mid-IR, without a distinct 10–20~$\mu$m silicate feature.

The condensate behavior is also shown in appendix Figure~\ref{Trappist1condensate}. The left panel presents the vertical distribution of the condensate mass fraction ($q_c$) for Mg$_2$SiO$_4$, the dominant silicate species. $q_c$ peaks at $\sim10^{-9}$–$10^{-8}$ near $10^{-2}$ bar—precisely where the TP profile intersects the condensation curve—before declining sharply aloft due to the balance of vertical mixing and sedimentation \citep{ackerman2001precipitating}. The right panel shows the effective particle radius ($r_{\rm eff}$) of Mg$_2$SiO$_4$, which increases monotonically with depth. However, $r_{\rm eff}$ remains extremely small, submicron, reaching nanometer scales at the cloud top ($<0.2$ nm) and increasing gradually with depth (to $\sim1.2\times10^{-8}\,\mu$m). This indicates that while condensation is triggered, particle growth is strongly mass-limited, producing only incipient, sub-micron (nanometer level) grains rather than optically thick silicate clouds.

The extremely small particle radii reflect the limited vertical mixing within the radiative zone above the convective boundary. The choice of the small vertical mixing coefficient ($K_{zz} = 1$ cm$^2$ s$^{-1}$) used in the \citet{ackerman2001precipitating} cloud routine could likely play a role here. In Trappist-1, the silicate condensation zone lies well within the radiative regime above the convective boundary (see Figure \ref{Trappist1condensate}), where large-scale mixing is expected to be weak \citep[e.g., overshoot velocities and mixing coefficients decline sharply above the convective zone in 3D RHD models;][]{freytag2010,allard2012,tremblin2015cloud}. Increasing $K_{zz}$ by orders of magnitude would raise particle lofting and effective particle radii, but we do not expect it to substantially alter the cloud base pressure because the local temperature gradient is radiatively controlled. In this regime, low $K_{zz}$ values are physically plausible and reproduce the observed near-IR flux suppression without invoking optically thick clouds. A more detailed exploration of non-convective mixing efficiencies (range of K$_{zz}$ values) and their impact on $r_{\rm eff}$ will be pursued in future work. Such modeling will help clarify whether small-scale mixing could explain the subtle continuum reddening and flux suppression observed in the latest-type M dwarfs.

Despite their diminutive size, these grains measurably impact Trappist-1’s emergent spectrum. Their continuum opacity raises the near-IR $\tau$, shifts the radiative–convective balance, and drives the flux suppression around 1 $\mu$m relative to the cloud-free case (Figure~\ref{Trappist1modcompare}, left). This is a key result: we are able to predict condensates in Trappist-1, even if these are mass-limited hazes rather than thick silicate decks. The impact of modeling them on the stellar SED is non-negligible, and their inclusion is essential for accurately modeling late-M dwarfs.

\subsection{Condensate Opacity in M-dwarfs}\label{condensates}

We perform a simple exercise to understand the condensation behavior of forsterite (Mg$_2$SiO$_4$) alone, across the effective temperature range characteristic of mid-to-late type M-dwarfs. Forsterite is one of the primary silicate condensates expected to form in these atmospheres \citep[e.g.,][]{Lodders2006, marley2013}, and its opacity contribution is sensitive to both the mass fraction of condensed material ($q_c$), the effective particle radius ($r_\mathrm{eff}$), and the efficiency of vertical mixing, which regulates condensate lofting. Equilibrium chemistry studies predict that Mg-silicates first condense as forsterite before converting toward enstatite (MgSiO$_3$) in oxygen-rich conditions \citep{lodders2003,Gail2014}. Other condensates such as CaTiO$_3$ and Fe-bearing silicates may also form depending on the elemental mixture and local thermodynamic conditions; and will be analyzed in detail in future.

\begin{figure}[ht]
\centering
\includegraphics[width=1.05\columnwidth]{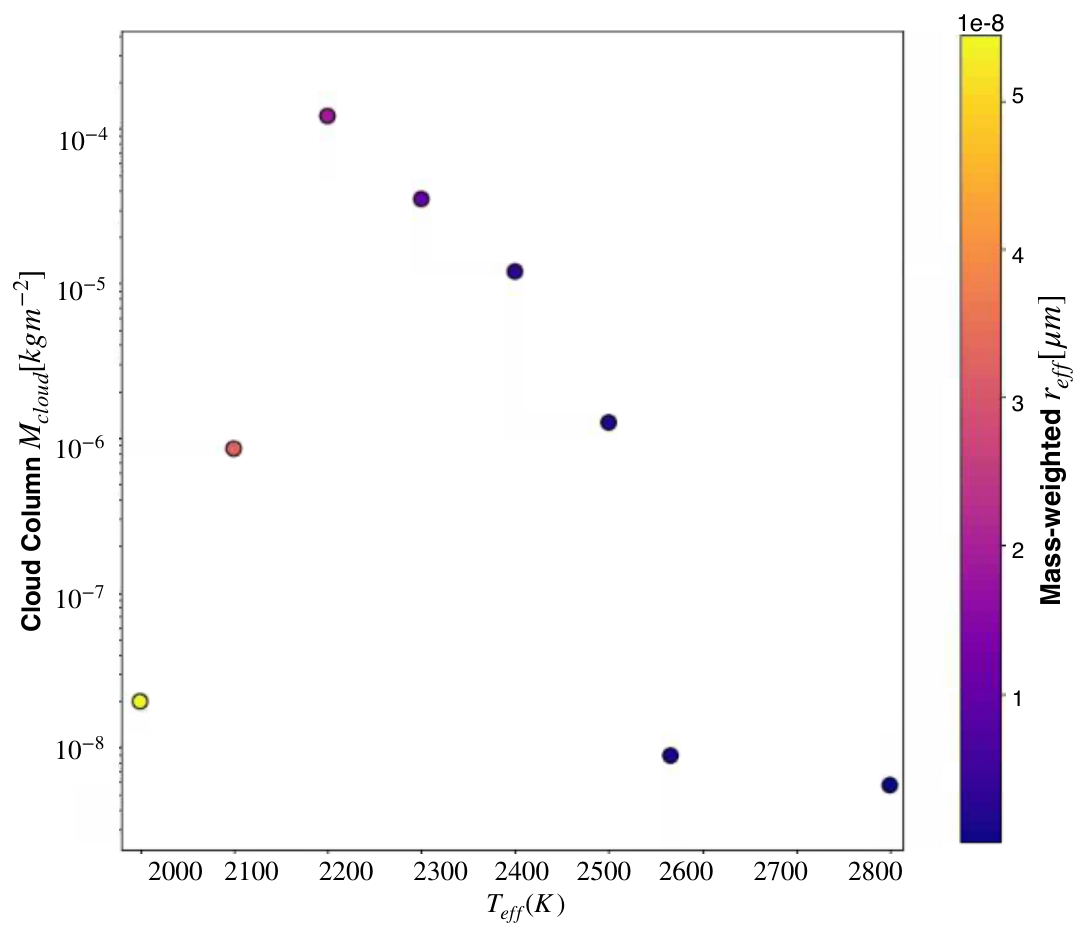}
\caption{
Mass‐weighted cloud column ($M_{\mathrm{cloud}}$) as a function of effective temperature for the \texttt{SPHINX II} grid. 
Silicate clouds are most abundant near 2100–2400~K, with declining mass columns at higher $T_{\mathrm{eff}}$ as condensate formation becomes inefficient. 
The color bar denotes the mass‐weighted mean particle radius ($r_{\mathrm{eff}}$), which remains submicron and decreases with increasing $T_{\mathrm{eff}}$ as the available condensate supply diminishes. 
These trends indicate that mid‐type M dwarfs develop the most optically significant silicate cloud decks, directly influencing the degree of continuum flattening and MIR reddening described in Section~\ref{condensates}.}

\label{fig:cloud_column}
\end{figure}

Table~\ref{tab:mg2sio4_clouds} summarizes the peak condensate mass fraction, corresponding effective radius, and pressure levels ($P_\mathrm{peak}$) for a grid of condensate cloud models spanning $T_\mathrm{eff} = 2000$--2900~K, with fixed log$g$ = 5.0, [M/H] = 0.0, and C/O = 0.5. The results demonstrate that condensation is not strictly monotonic with temperature: models near 2200--2400~K yield the largest $q_c$ peaks and cloud mass columns, while cooler ($\sim2000$~K) models form clouds but with smaller peak abundances. This non-monotonic behavior (see Figure \ref{fig:cloud_column}) arises from how the atmospheric $T$--$P$ profile intersects the forsterite condensation curve. At intermediate 2200 K $<$ $T_\mathrm{eff}$ $<$ 2400 K, both the local temperature gradient and vapor supply favor supersaturation, maximizing condensation efficiency, whereas cooler models ($<$ 2200 K) condense deeper where less material remains available in the upper atmosphere. In these cooler cases, the condensation zone lies below the photosphere—forming a deep cloud deck / buried cloud that contributes little to the emergent flux \citep[e.g.,][]{tsuji2002water,Helling2014}. The precise condensation chemistry may also vary with the stellar Mg/Si ratio, which governs whether forsterite, enstatite, or quartz becomes the dominant silicate species \citep[e.g.,][]{Calamari2024,Lodders2006}. It is important to note that clouds do not disappear in this regime; rather, in late-M dwarfs the photosphere moves above the cloud base as $T_{\rm eff}$ decreases, reducing the visible cloud signature. In contrast, true L-dwarfs retain their cloud deck within the line-forming region, producing the characteristic dusty L-dwarf spectra. Our models therefore capture the onset of photospheric condensation in late-M dwarfs, not the fully cloudy L-dwarf regime. At higher temperatures ($\gtrsim2600$~K), condensation is increasingly suppressed, with only weak residual cloud signatures by $T_\mathrm{eff} \sim 2800$~K as refractory species remain in the gas phase at these higher temperatures, delaying condensation until the $T$--$P$ curve no longer intersects the silicate stability regime. Effective particle sizes remain in the sub-micron regime throughout, reinforcing that silicate clouds in M-dwarfs are dominated by small grains \citep{Ackerman2001, Helling2008}. 

A higher vertical mixing efficiency (larger K$_{zz}$) could, in principle, keep condensate aloft at higher altitudes in the $<$ 2200K models and increase photospheric opacity. However, it will not shift the cloud base (set by the T-P–condensation intersection) nor remove the overall non-monotonic trend with T$_{eff}$; the deepest, buried clouds still contribute less efficiently to the emergent flux of the 2000 K regime than the mid-T$_{eff}$ cases. A full exploration of enhanced radiative-zone K$_{zz}$ is deferred to future work.

The precise condensation chemistry may also vary with the stellar Mg/Si ratio, which governs whether forsterite, enstatite, or quartz becomes the dominant silicate species \citep[e.g.,][]{Calamari2024,Lodders2006}. At higher temperatures ($\gtrsim2600$~K), condensation is increasingly suppressed, with only weak residual cloud signatures by $T_\mathrm{eff} \sim 2800$~K as refractory species remain in the gas phase at these higher temperatures, delaying condensation until the $T$--$P$ curve no longer intersects the silicate stability regime. Effective particle sizes remain in the sub-micron regime throughout, reinforcing that silicate clouds in M-dwarfs are dominated by small grains \citep{Ackerman2001, Helling2008}. A higher vertical mixing efficiency (larger K$_{zz}$) could, in principle, keep condensate aloft at higher altitudes in the $<$ 2200~K models and increase photospheric opacity. However, it will not shift the cloud base (set by the $T$--$P$–condensation intersection) nor remove the overall non-monotonic trend with $T_{\rm eff}$; the deepest, buried clouds still contribute less efficiently to the emergent flux of the 2000~K regime than the mid-$T_\mathrm{eff}$ cases. A full exploration of enhanced radiative-zone K$_{zz}$ is deferred to future work.

The trends in condensate behavior can also be visualized by considering the total cloud column mass, $M_\mathrm{cloud}$, as a function of effective temperature. Figure~\ref{fig:cloud_column} shows $M_\mathrm{cloud}$ across the same grid of models, color-coded by the mass-weighted mean particle size. Consistent with Table~\ref{tab:mg2sio4_clouds}, the maximum condensate burden occurs near 2100--2400~K, where both the pressure and temperature conditions favor efficient silicate condensation. Cooler models around 2000~K form clouds but with significantly lower total mass columns, while hotter models ($\gtrsim2600$~K) show a sharp decline in $M_\mathrm{cloud}$ as condensation is confined to increasingly low-pressure layers. The colorbar trend highlights that particle radii remain in the sub-micron regime throughout, decreasing steadily with increasing $T_\mathrm{eff}$ as the reduced condensate supply yields smaller grains.

\begin{table*}[ht]
\centering
\small
\begin{tabular}{c c c c c}
\hline
\textbf{Teff [K]} & 
\textbf{$P_\mathrm{peak}$ [bar]} & 
\textbf{$q_c$ peak [ppm]} & 
\textbf{$M_\mathrm{cloud}$ [kg m$^{-2}$]} & 
\textbf{$\langle r_\mathrm{eff} \rangle_\mathrm{mass}$ [$\mu$m]} \\
\hline
2000 & $4.4 \times 10^{-1}$ & $1.3 \times 10^{-3}$ & $2.0 \times 10^{-8}$ & $5.5 \times 10^{-8}$ \\
2100 & $2.5 \times 10^{-1}$ & $9.9 \times 10^{-2}$ & $8.6 \times 10^{-7}$ & $3.3 \times 10^{-8}$ \\
2200 & $1.4 \times 10^{-1}$ & $2.5 \times 10^{1}$  & $1.2 \times 10^{-4}$ & $1.9 \times 10^{-8}$ \\
2300 & $7.8 \times 10^{-2}$ & $1.3 \times 10^{1}$  & $3.5 \times 10^{-5}$ & $1.1 \times 10^{-8}$ \\
2400 & $2.5 \times 10^{-2}$ & $1.4 \times 10^{1}$  & $1.2 \times 10^{-5}$ & $3.4 \times 10^{-9}$ \\
2500 & $1.4 \times 10^{-2}$ & $2.6 \times 10^{0}$  & $1.3 \times 10^{-6}$ & $1.9 \times 10^{-9}$ \\
2566 & $7.8 \times 10^{-3}$ & $5.6 \times 10^{2}$  & $8.9 \times 10^{-9}$ & $1.1 \times 10^{-9}$ \\
2600 & $2.5 \times 10^{-3}$ & $6.3 \times 10^{-3}$ & $5.4 \times 10^{-10}$ & $3.7 \times 10^{-10}$ \\
2800 & $4.4 \times 10^{-5}$ & $3.7 \times 10^{0}$  & $5.7 \times 10^{-9}$ & $6.8 \times 10^{-12}$ \\
\hline
\end{tabular}
\caption{Summary of Mg$_2$SiO$_4$ condensate cloud properties across effective temperature. 
$P_\mathrm{peak}$ is the pressure at which the condensate mass fraction peaks, 
$q_c$ peak is the local maximum condensate mass fraction, 
$M_\mathrm{cloud}$ is the total mass column of condensates, 
and $\langle r_\mathrm{eff} \rangle_\mathrm{mass}$ is the mass-weighted mean particle radius.}
\label{tab:mg2sio4_clouds}
\end{table*}

The trend in $P_\mathrm{peak}$ in Table \ref{tab:mg2sio4_clouds} illustrates how the condensation base of forsterite is pushed to progressively lower pressures (higher altitudes) as $T_\mathrm{eff}$ increases. This behavior arises naturally from the intersection of the atmospheric $T$--$P$ profile with the forsterite condensation curve \citep{Lodders2006, visscher2010}. At cooler effective temperatures (2000--2200~K), condensation occurs deeper in the atmosphere ($\sim$0.1--0.4 bar), whereas for hotter models (2600--2800~K) the crossing shifts upward to $\sim10^{-2}$--$10^{-5}$ bar. Such upward migration of the cloud base with increasing $T_\mathrm{eff}$ is a robust feature of equilibrium cloud models \citep{Ackerman2001, allard2001, helling2008comparison, Helling2014}. Physically, hotter atmospheres maintain refractory species such as Mg and Si in the gas phase at deeper levels, delaying condensation until much lower pressures are reached. By $\sim$2900~K, the $T$--$P$ profile no longer intersects the forsterite condensation curve, leading to negligible condensate formation, consistent with expectations for earlier-type M dwarfs \citep{tsuji1996dust}.

\begin{figure*}[ht]
\centering
\includegraphics[width=\textwidth]{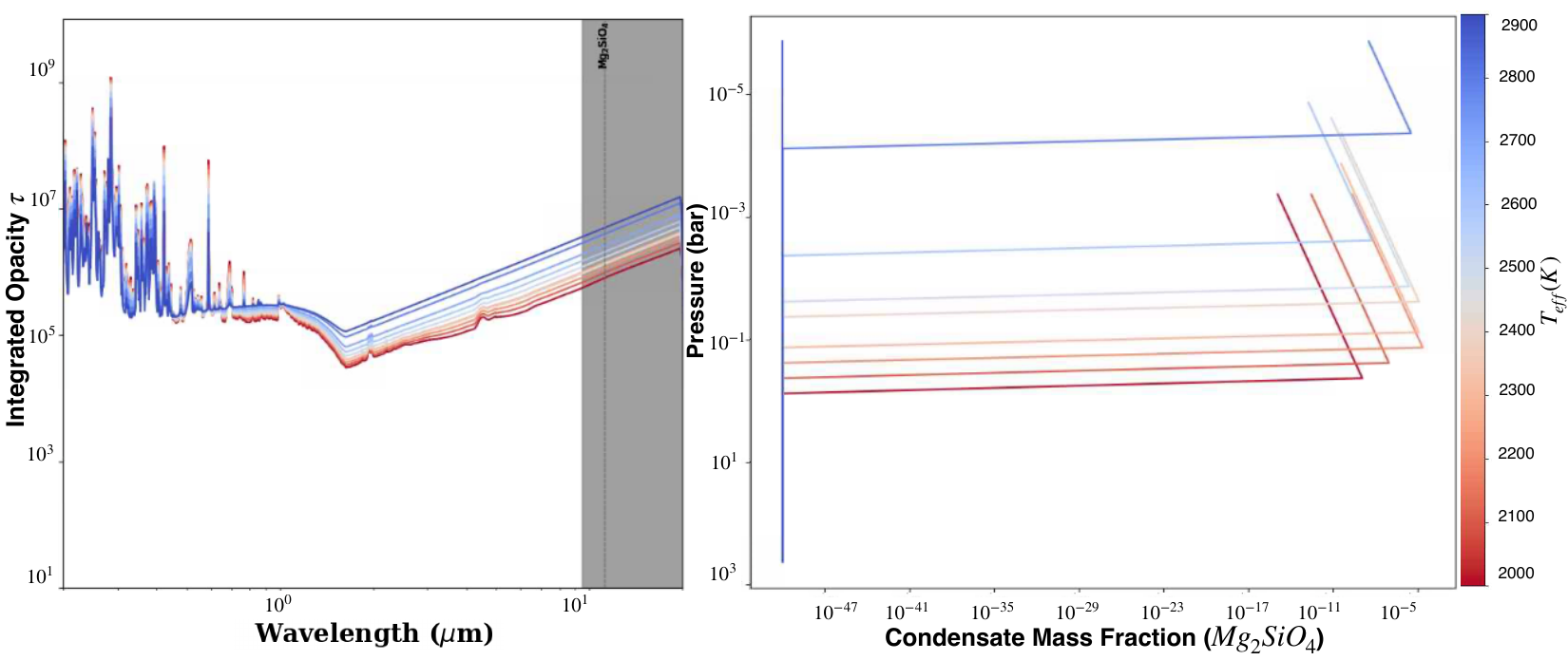}
\caption{
Condensate opacity behavior of forsterite (Mg$_2$SiO$_4$) across M‐dwarf effective temperatures. 
(\textbf{Left}) Integrated opacity spectra spanning $T_{\mathrm{eff}}=2000$–2900~K. 
Cooler models (red) exhibit stronger continuum flattening and muted near‐IR contrast, whereas hotter models (blue) approach cloud‐free slopes with negligible condensate contribution by $\sim$2900~K. 
(\textbf{Right}) Pressure‐dependent mass‐fraction profiles of Mg$_2$SiO$_4$ highlight the shift in condensation levels and declining cloud mass with increasing $T_{\mathrm{eff}}$. 
The combination of these panels demonstrates how condensate opacity progressively reddens the MIR continuum without producing a distinct 10–20~$\mu$m silicate feature and the upward migration of the cloud deck with increasing T$_{eff}$.}

\label{fig:forsterite_opacity}
\end{figure*}

From an observational perspective, Figures~\ref{fig:cloud_column} and \ref{fig:forsterite_opacity} together illustrate how condensates modify the mid-infrared (MIR) continuum. The cloud column trends in Figure~\ref{fig:cloud_column} show that silicate clouds are most abundant near 2100–2400 K, with smaller particle sizes and higher cloud decks at higher $T_\mathrm{eff}$, while cooler models develop deeper, optically thicker cloud bases. These same cooler models exhibit the strongest continuum flattening and reddening in the MIR. This behavior directly maps onto the integrated opacity spectra in Figure~\ref{fig:forsterite_opacity} (left), where the redder (cooler) curves lie below the bluer (hotter) ones in the near-infrared but show a noticeably steeper rise beyond $\sim$2–10 $\mu$m. This progressive slope change reflects increasing condensate burdens that mute molecular contrast in the near-infrared and produce a smooth, featureless opacity rise toward longer wavelengths. Conversely, at $T_\mathrm{eff} \gtrsim 2600$ K, reduced condensate abundance and elevated cloud bases yield atmospheres that are comparatively clearer, restoring steeper MIR slopes. These systematic shifts imply that cooler stars should appear progressively redder in the MIR, whereas those near 2800–2900 K approach the cloud-free limit. This continuum flattening without a distinct 10–20 $\mu$m silicate feature is consistent with \textit{Spitzer}/IRS observations across the M–L–T sequence, where silicate emission becomes apparent only in the early-L regime \citep{suarez2022,suarez2023ultracool,roellig2004spitzer}. These slope trends could be tested directly with \textit{JWST}/MIRI \citep{Greene2016} and complementary optical/NIR context from multi-band observations with future missions such as \textit{Pandora} \citep{quintana2021}. Importantly, understanding how condensates flatten and redden the emergent continuum provides a physical basis for interpreting spectral variability contrasts in the observable photosphere. By linking condensate-driven continuum behavior in the NIR/MIR to surface heterogeneity, we can better quantify and disentangle stellar contamination effects in exoplanet transmission and emission spectra \citep{rackham2018,barclay2025_PandoraMission}.

\section{Conclusions} \label{paper2summary}

Atmospheres of M-dwarfs have been non-trivial to characterize and our understanding of mid-to-late type M-dwarfs has been an even more daunting task. In this work, we expand on our work from \cite{iyer2023sphinx} and gather a more holistic understanding of low-mass stars. While it comes as no surprise--the manifestations of spectroscopic degeneracies due to varying atmospheric processes, the goal of this work was developing a self-consistent 1D synthesis model to acquire robust estimates of fundamental stellar properties of these stars. Models for ultracool dwarfs such as BT-Settl \citep{allard2012} and others include the chemical prescriptions and dynamics necessary to treat some of these challenges however they are missing a key ingredient: up-to-date molecular opacities. We address this gap by improving over the larger implications of not understanding M-dwarf atmospheres---particularly exoplanet characterization, Galactic and chemical evolution, and ultimately our understanding of the universe. With that, we present our upgraded synthesis model grid spectra and atmospheres for M-dwarfs; \texttt{SPHINX II}, by summarizing the following:
\begin{itemize}
    \item Treatment of clouds/dust become increasingly important for mid-to-late-type M-dwarfs, especially those with T$_{eff}$ $<$ 2900 K. We perform basic upgrades to the \texttt{SPHINX I} model grid from part I and include a minimal gray cloud opacity in the models varying the value ranging from a cloud-free scenario (log $\kappa$ = -32) to an opaque cloud scenario (log $\kappa$ = -28). We also incorporate a physically motivated condensate cloud treatment based on the \cite{ackerman2001precipitating} framework which parameterizes the vertical distribution of silicate condensates as a balance between upward mixing and downward rainout, coupling with convective feedback following \cite{lefevre2022cloud}.
    \item We show that assumptions regarding the convective mixing-length parameter also influence total convective heat flux in the adiabatic regions of the atmosphere (where convection is more efficient than radiation) causing differences in the spectral shape by 55$\%$ below 1 $\mu$m. 
    \item Mixed-model assumptions may be needed to describe a potentially young mid-type M-dwarf spectrum, for example with 2MASSJ00013044+1010146 SpeX Database \citep{burgasser2014spex} spectrum, we see the minimal gray cloud-only model and stellar surface heterogeneity models are equally favored, however the posterior probability distributions show that the cloud-only model might be more physically plausible. Moreover, these assumptions in conjunction give reasonable estimates of metallicities, consistent with neighborhood FGK-type stars.
    \item The combination of a gray cloud coupled with a lower convective efficiency improves metallicity constraints for all mid-to-late-type M-dwarf spectra for the sample used in this study (SpeX Database \citep{burgasser2014spex} and FGKM+companions \citep{mann2014prospecting}). The overall data-model relative residual differences using the cloudy models are $\sim$20$\%$ comparable to the cloud-free + $\alpha$=1 models, however the model upgrade results in significant improvements by yielding physically reasonable metallicities for these stars when compared to main-sequence FGK stars taken from the Hypatia catalog (\cite{hinkel2014} and see Figures \ref{fig:met_vs_ctoo_beforeandafter} and \ref{fig:met_vs_ctoo_ALL}).
    \item A case study of Trappist-1 shows that including a gray cloud, lower convective mixing length and photospheric heterogeneity substantially improves spectral fits compared to purely cloud-free models. Forward model comparisons of condensate cloud + convection feedback model versus cloud-free scenarios demonstrate that cloud opacity and convection jointly shape the emergent slopes, reinforcing the need to model these processes explicitly in late-type M-dwarfs.
    \item Using Mg$_2$SiO$4$ (forsterite) as a representative condensate, we find that silicate cloud opacity in M-dwarfs peaks near $T\mathrm{eff}$ $\sim$ 2100–2400 K and diminishes rapidly toward hotter temperatures around 2900K, marking a natural transition from cloud-dominated to effectively clear photospheres. The cooler temperature regime (below $<$ 2200K) marks another drop off in condensate column mass, this could be attributed to deeper/buried cloud bases in these temperature regimes. We also note an upward migration of the cloud base with increasing temperature, combined with sub-micron particle sizes, produces MIR continuum flattening and reddening at lower $T_\mathrm{eff}$, consistent with the observed slope evolution across the M–L boundary. These trends imply that condensate opacity—not molecular bands alone—governs the mid-IR spectral stellar SED slope in the coolest M dwarfs, a prediction testable with $JWST/MIRI$ and complementary optical/NIR context from missions like $Pandora$.
    \item The bulk of the correlated noise with our grid-trieval method coincides with spectral regions plagued with degeneracies between stellar heterogeneity, convection and clouds prompting to solve for these processes as free parameters--a step that is currently underway. In a future study (part III of this paper series), we will perform free Bayesian retrievals without invoking the self-consistent assumption as applied to grid models, and derive appropriate atmospheric structures using low-resolution M-dwarf data. This method will allow for arbitrary determinations of species abundances, and provide a framework to truly stress-test model assumptions that have repeatedly shown the potential to bias fundamental stellar properties.
\end{itemize}

In summary, we present an extended 1D radiative-convective thermochemical equilibrium chemistry model with atmospheres and spectra for M dwarfs that includes the latest molecular opacities, in addition to a basic gray cloud, condensate cloud sedimentation balance, and convective feedback prescription. We also emphasize the need to understand the level of biases induced from mixed-model parameterizations. Moving forward, these models and techniques will be essential for robust stellar characterization in support of exoplanet studies, particularly for complex systems such as Trappist-1. With access to facilities including \textit{JWST} and \textit{Pandora}, and future large missions, rigorous efforts to improve stellar atmosphere models are a necessity to disentangle stellar processes from planetary signals, advancing both stellar astrophysics and exoplanet science.

Supplementary Figures provided on \href{https://zenodo.org/records/17782633}{Zenodo} and model grid is available on \href{https://github.com/aishaiyer/SPHINX-II-1D-Stellar-Atmosphere-Models.git}{GitHub}.

\acknowledgments
\section*{Acknowledgments}
The authors would like to thank Ehsan [Sam] Gharib-Nezhad and the $EXOPLINES$ team \citep{gharib2021exoplines} for computing appropriate pressure broadening treatment for up-to-date molecular opacities used in this work. The authors would also like to thank Andrew Mann for all the low resolution SpeX IRTF M dwarf spectra used in this work. ARI would like to thank Mike Line, Jenny Patience, Evgenya Shkolnik, Joseph O'Rourke and Patrick Young for their support through the last months while undertaking part of this work for her dissertation \citep{iyer2023m}. ARI would also like to thank Ian Czekala, Miles Lucas, Michael Gully-Santiago, and Zhoujian (ZJ) Zhang for all their help and guidance in the working of $Starfish$. The authors would also like to credit NSF AAG Award, 2009592. ARI would like to acknowledge the NASA FINESST Grant 80NSSC21K1846. ARI would also like to acknowledge that this research was supported by an appointment to the NASA Postdoctoral Program at the Goddard Space Flight Center, administered by Oak Ridge Associated Universities under contract with NASA for successful completion post Ph.D. The authors acknowledge Research Computing at Arizona State University for providing {HPC, storage, etc.} resources that have contributed to the research results reported within this paper \citep{HPC:ASU23}. This work has made use of and benefited from: NumPy \citep{numpy}, SciPy \citep{2020SciPy-NMeth}, PyMultinest \citep{Buchner2014} and pygtc \citep{Bocquet2016}.
The research shown here acknowledges use of the Hypatia Catalog Database, an online compilation of stellar abundance data as described in \citealt{hinkel2014}, which was supported by NASA's Nexus for Exoplanet System Science (NExSS) research coordination network and the Vanderbilt Initiative in Data-Intensive Astrophysics (VIDA).This research has made use of the SIMBAD database,
operated at CDS, Strasbourg, France. "The SIMBAD astronomical database", \citealt{simbad2000}.This work has made use of data from the European Space Agency (ESA) mission {\it Gaia} (\url{https://www.cosmos.esa.int/gaia}), processed by the {\it Gaia} Data Processing and Analysis Consortium (DPAC, \url{https://www.cosmos.esa.int/web/gaia/dpac/consortium}). Funding for the DPAC has been provided by national institutions, in particular the institutions
participating in the {\it Gaia} Multilateral Agreement.

\vspace{5mm}
\facilities{SpeX SXD IRTF}
\appendix
\clearpage


\renewcommand{\thefigure}{A\arabic{figure}}
\setcounter{figure}{0}

\begin{figure}
    \noindent\includegraphics[width=0.85\textwidth]{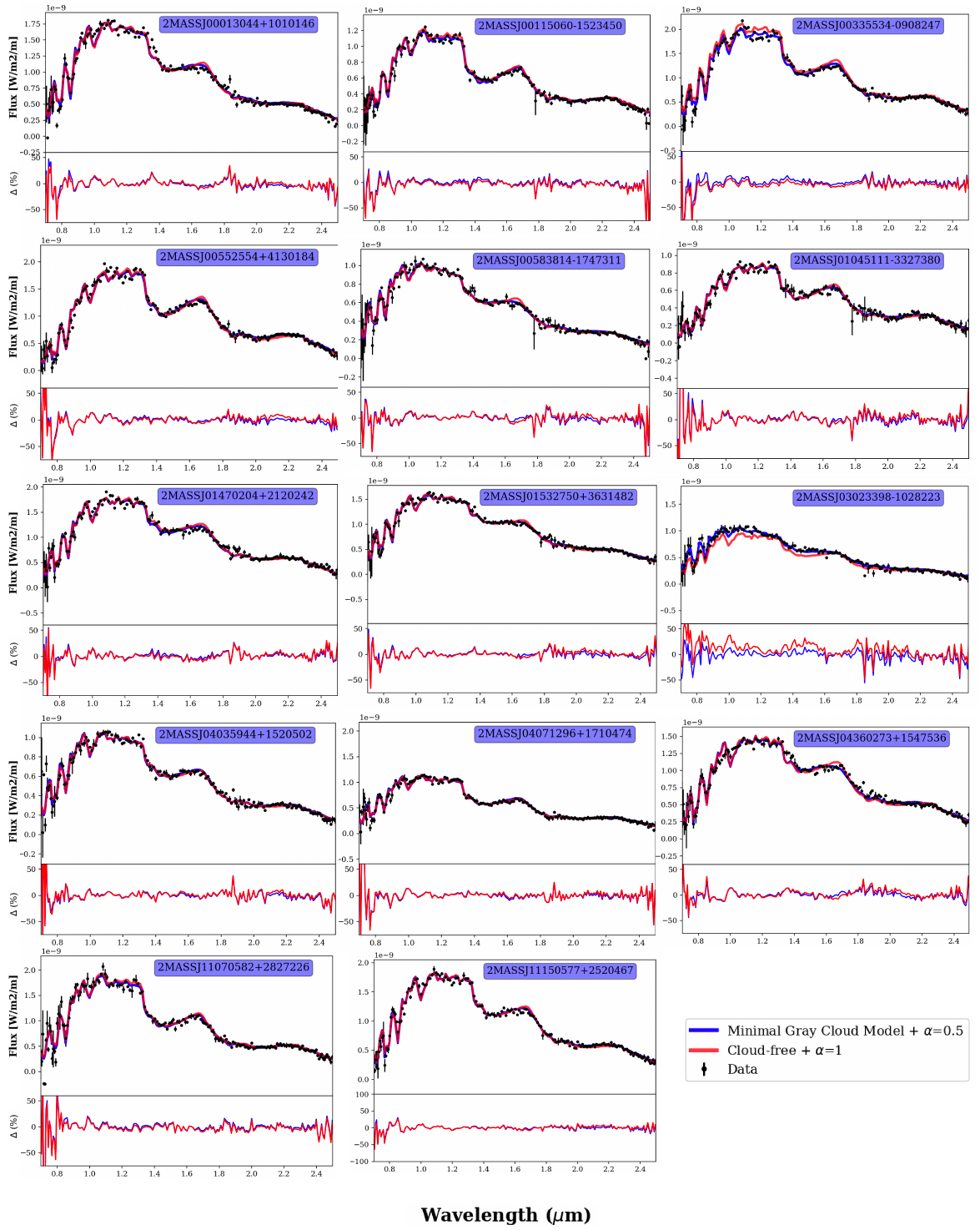}\qquad
     \centering
       \caption{Fitting SpeX IRTF NIR spectra (black points) taken from SpeX Prism Library Database with fiducial cloud-free + $\alpha$=1 \texttt{SPHINX I} model (red) vs. minimal cloudy + $\alpha=0.5$ \texttt{SPHINX II} model (blue).}
    \label{spexfits}
\end{figure}
   \begin{figure*}[!tbp]
\addtocounter{figure}{-1}
    \includegraphics[width=0.85\textwidth]{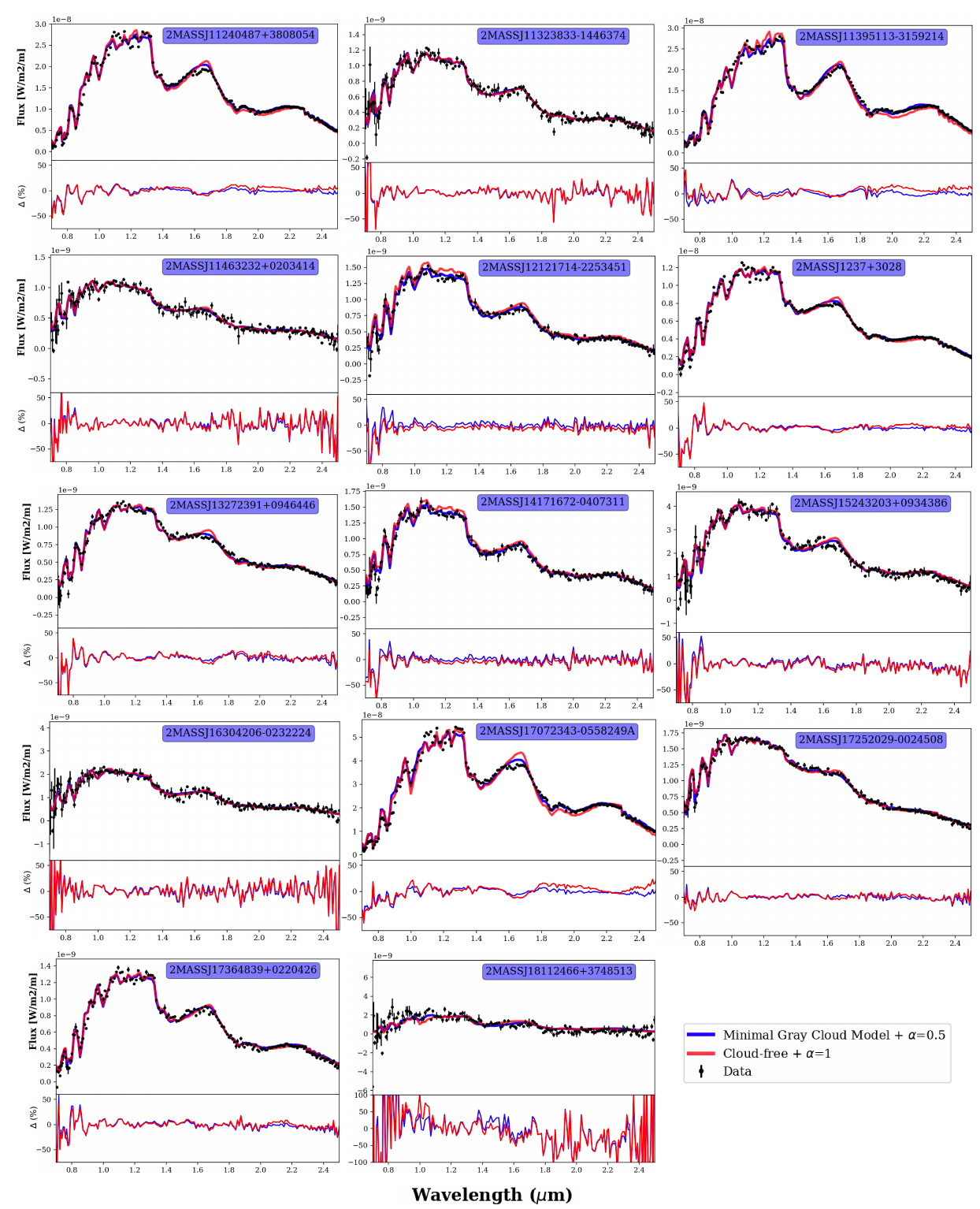}
    \centering
    \caption{(Continued) For most targets in this list (Table \ref{spexdatatable}), both models provide reasonable fits with a median relative error in about 20$\%$. The residual differences are worse at the edges of the bandpass, especially below 0.75 $\mu$m. For all these targets however, the model including a minimal gray cloud opacity along with a lower mixing-length convection value produces improved fits (also see Figure \ref{fig:met_vs_ctoo_beforeandafter})}.
    \end{figure*}
       \begin{figure}[!tbp]
\addtocounter{figure}{-1}
    \includegraphics[width=0.85\textwidth]{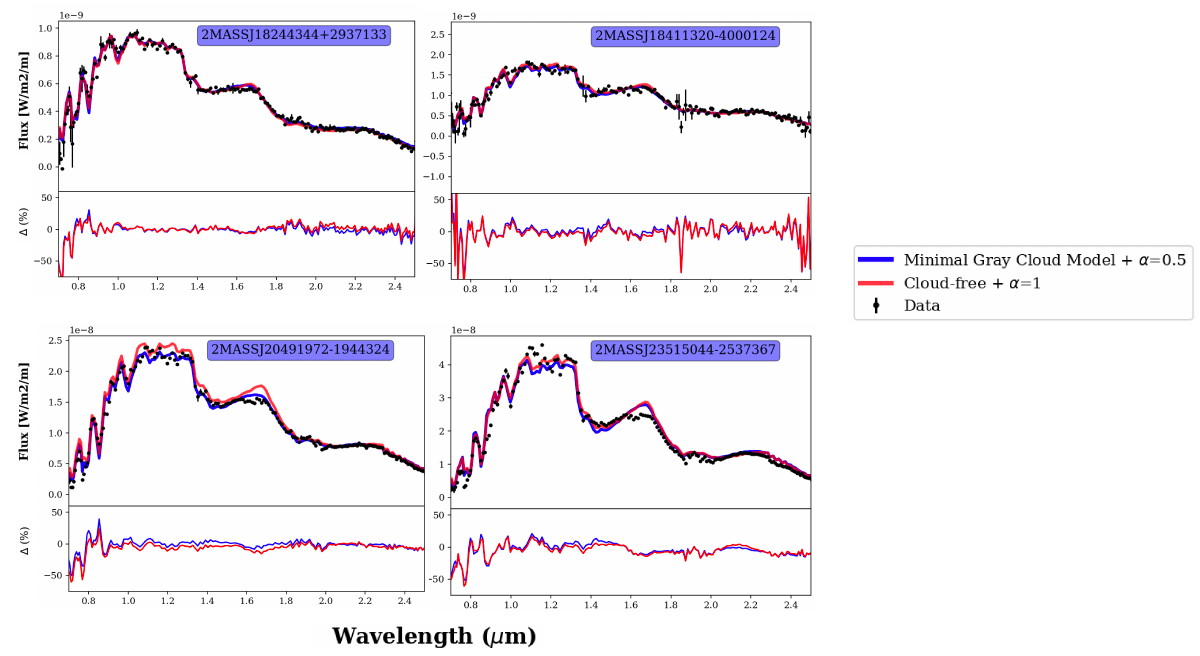}
    \centering
    \caption{(Continued) The relative difference is especially dramatic for a couple of targets: 2MASSJ18112466+3748513, where the residuals are above 50$\%$ throughout the NIR bandpass, and 2MASSJ23515044-2537367 where the residuals are comparable with other targets however both models fail to fit the spectra between 1.0--1.2$\mu$m and 1.6--1.8$\mu$m. Overall, for 90.6$\%$ (29 out of 32) of the targets in this list, we find that the Bayesian Information Criterion (BIC) value overwhelmingly favors the model including a minimal cloud and lower mixing length value. The three targets where the BIC favors both models equally are: 2MASSJ04071296+1710474, 2MASSJ11070582+2827226 and 2MASSJ16304206-0232224.}.
\end{figure}

\renewcommand{\thefigure}{A\arabic{figure}}
\begin{figure}[t]
    \centering
    \includegraphics[width=0.9\textwidth]{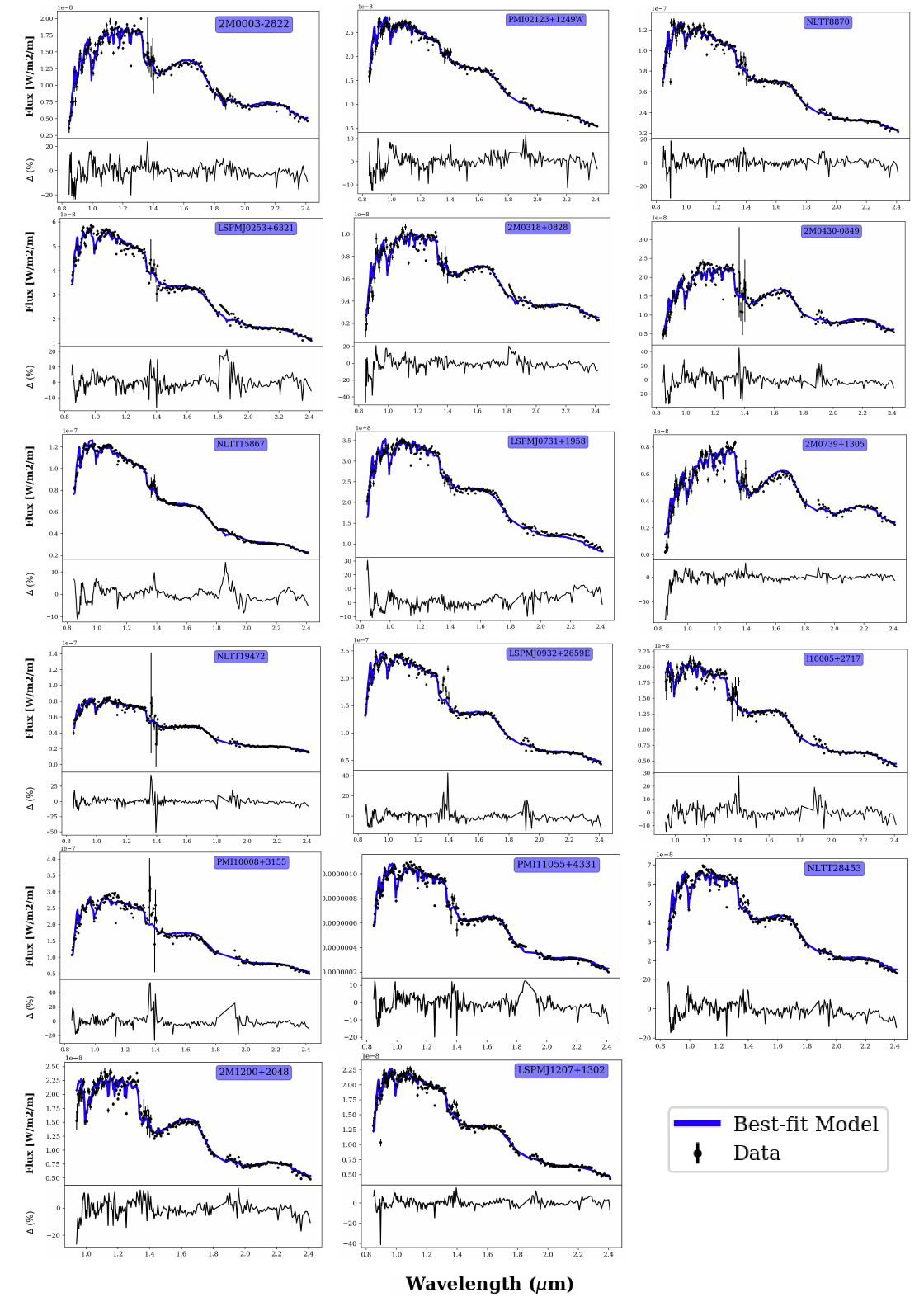}
    \caption{Fitting SpeX IRTF NIR spectra (black points) of FGKM+companion mid-to-late M-dwarfs from \cite{mann2014prospecting} with a minimal cloudy + $\alpha=0.5$ \texttt{SPHINX II} model.}
    \label{fig:mannfits_app}
\end{figure}
\addtocounter{figure}{-1}
\begin{figure*}[t]
    \centering
    \includegraphics[width=0.85\textwidth]{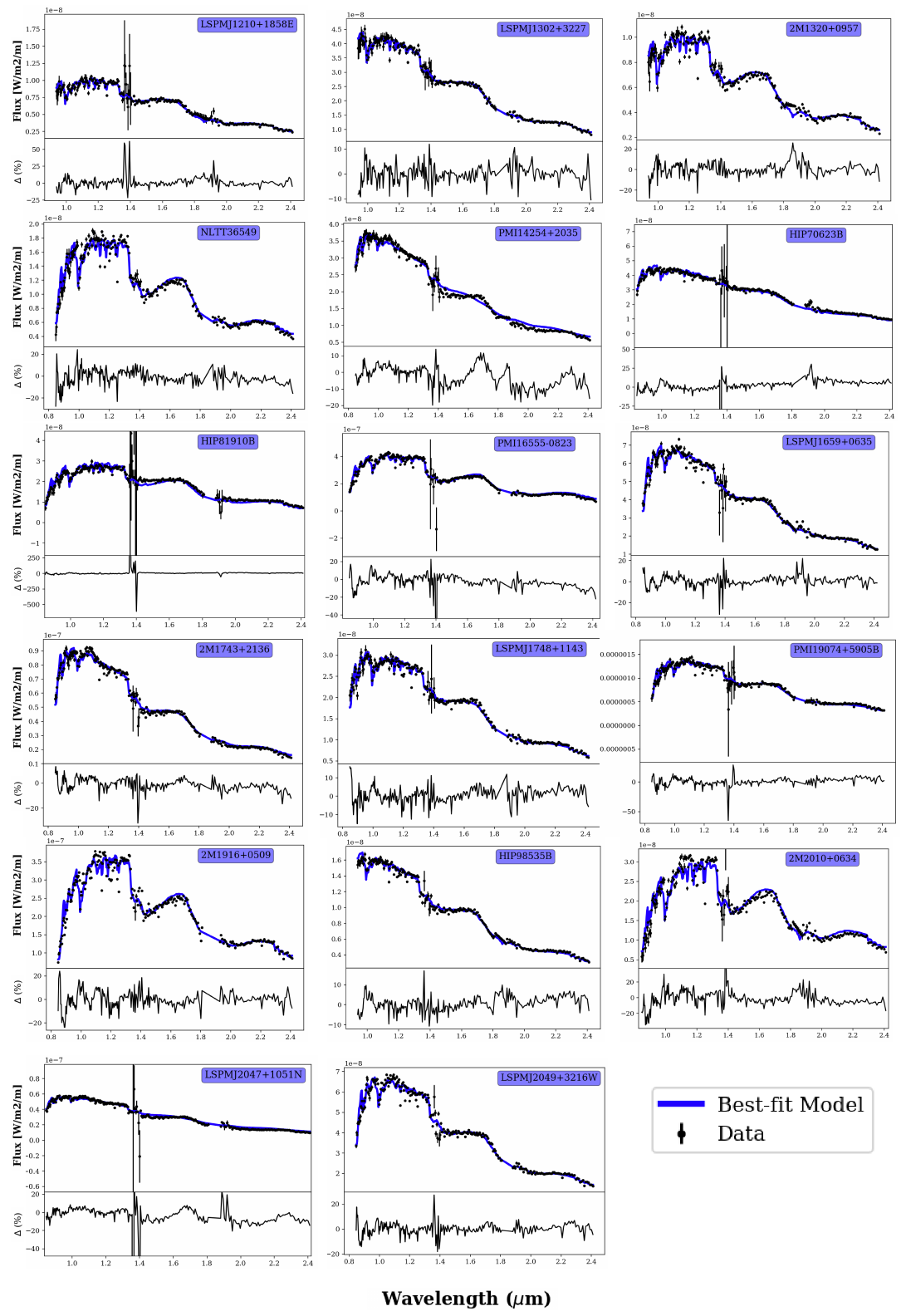}
    \caption{(Continued) For most targets in this list (Table \ref{manndatatable}), the best-fit models all achieve a median relative error under 20$\%$. Residuals are largest near regions susceptible to telluric noise, particularly around 1.4$\mu$m, but these are downweighted through the error inflation parameter (Equation \ref{errorinflate}).}
\end{figure*}
\addtocounter{figure}{-1}
\begin{figure}[t]
    \centering
    \includegraphics[width=0.9\textwidth]{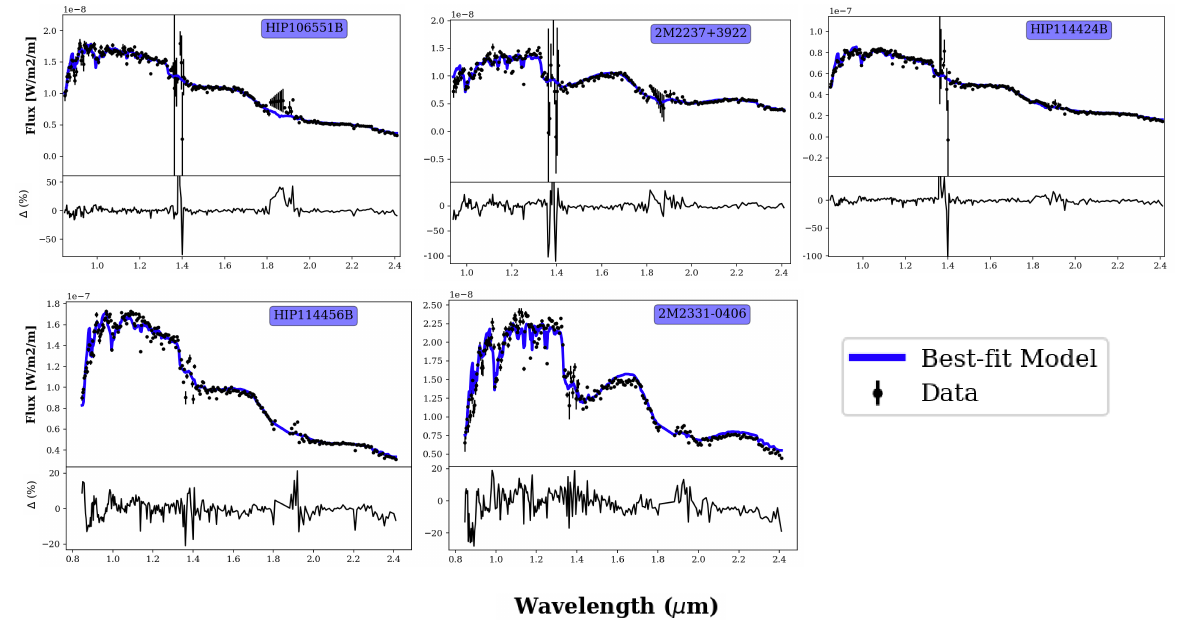}
    \caption{(Continued).}
\end{figure}

\renewcommand{\thefigure}{A\arabic{figure}}
\begin{figure*}[!tbp]
    \includegraphics[width=\textwidth]{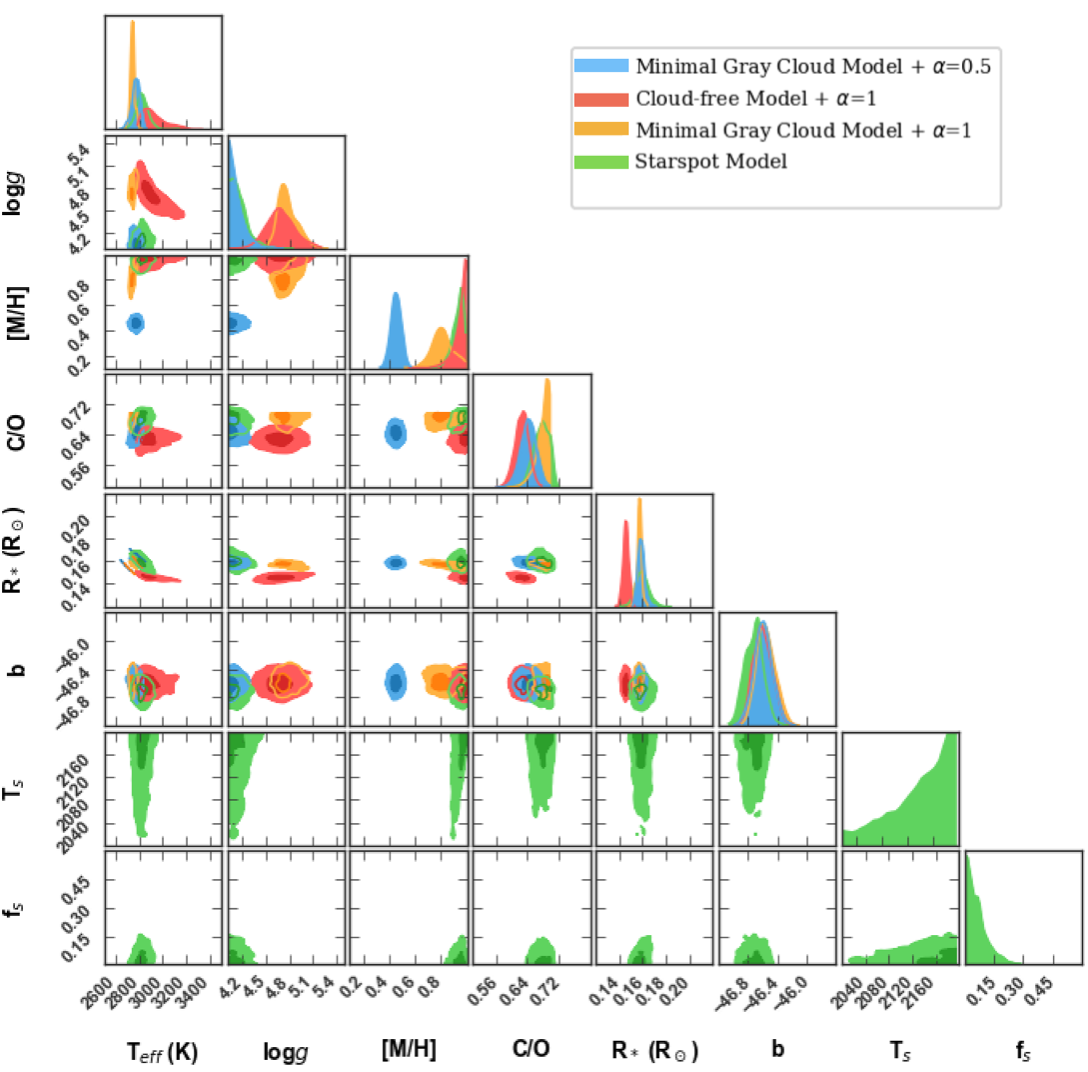}
    \centering
    \caption{Posterior histograms for all four models fitted to the target 2MASS J00013044+1010146 in Figure \ref{fig:allmodcompare_tospex}. We retrieve $T_{\rm eff}$, log $g$, [M/H], C/O, $R_*$, and error inflation exponent $b$. The starspot model (green) includes additional parameters for spot temperature $T_s$ and coverage fraction $f_s$. While both the blue and green models yield good fits, the blue (minimal cloud + low $\alpha$) is the only one that recovers metallicities consistent with FGK neighbors.}
\label{cornerplot}
\end{figure*}

\renewcommand{\thefigure}{A\arabic{figure}}
\begin{figure*}[!tbp]
    \includegraphics[width=\textwidth]{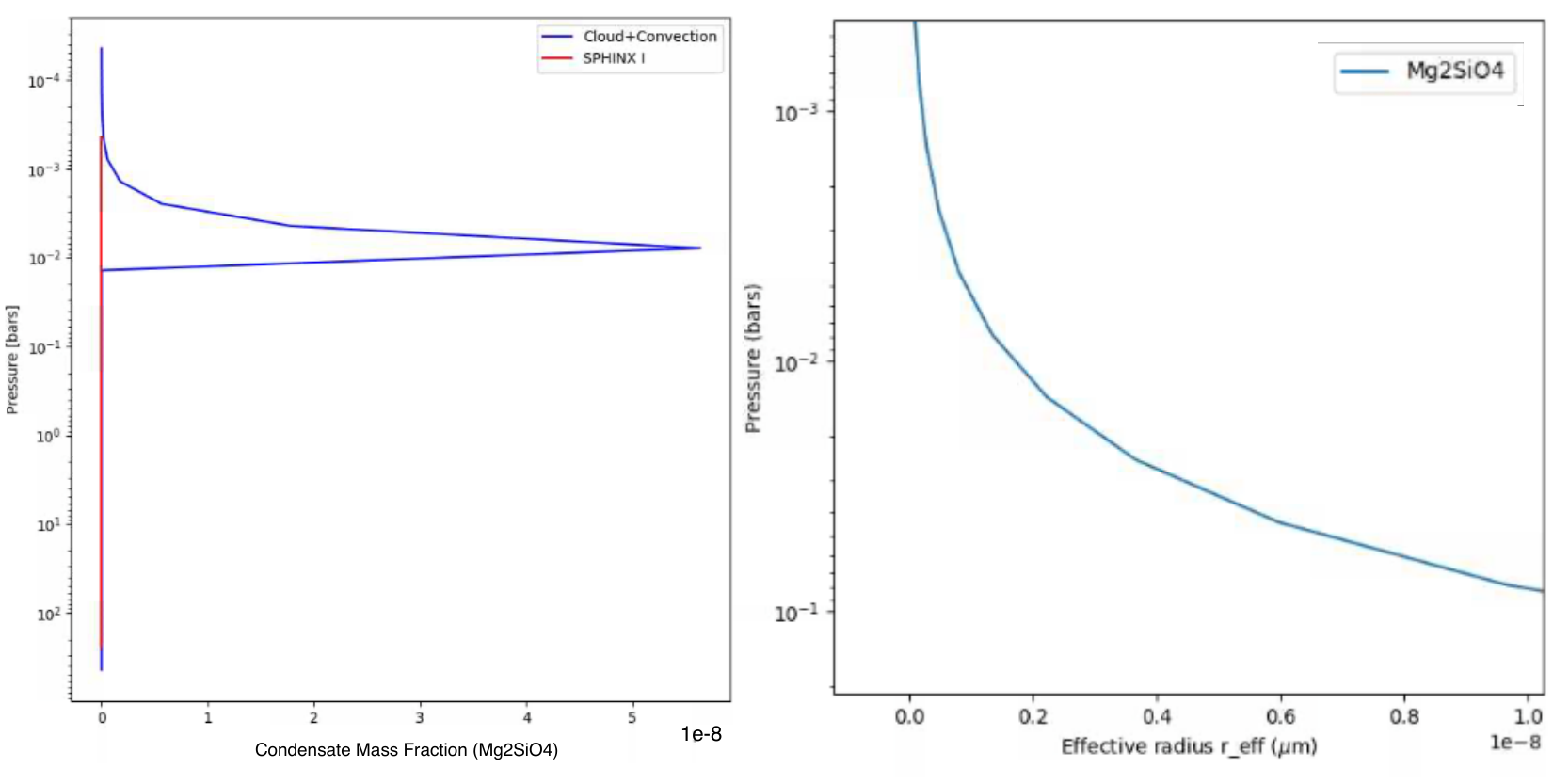}
    \caption{
    \textbf{Condensate properties for Trappist-1 Condensate cloud+Convection model.}
    \textbf{Left:} Vertical distribution of Mg$_2$SiO$_4$ condensate mass fraction ($q_c$), peaking at $\sim10^{-9}$–$10^{-8}$ near $10^{-2}$ bar where the TP profile crosses the condensation curve. $q_c$ declines sharply aloft due to mixing–sedimentation balance.  
    \textbf{Right:} Effective particle radius $r_{\rm eff}$ for Mg$_2$SiO$_4$ grows monotonically with depth but remains sub-nanometer throughout the cloud deck, indicating mass-limited growth.  
    Together, these panels show that while Trappist-1’s clouds are not optically thick, even trace condensation measurably alters the SED and radiative–convective structure.}
    \label{Trappist1condensate}
\end{figure*}

\begin{table*}[t]
\renewcommand{\thetable}{A\arabic{table}}
\setcounter{table}{0}
\centering
\caption{Targets used for this analysis from the SpeX Prism Spectral Libraries \citep{burgasser2014spex}. Listed here are also their respective H-band magnitudes, and distances.}\label{spexdatatable}
\resizebox{0.85\textwidth}{!}{
\begin{tabular}{|l | l | l |}
\hline
Target  & H-band Magnitude$^{a,b}$ & Distance$^c$ (pc) \\
\hline
2MASSJ00013044+1010146 & 15.132 $\pm$ 0.096 & 139.497 $\pm$ 11.6329\\
2MASSJ00115060-1523450 & 15.642 $\pm$ 0.151 & 121.819 $\pm$ 10.2766\\
2MASSJ00335534-0908247 & 15.001 $\pm$ 0.105 & 109.589 $\pm$ 6.7975\\
2MASSJ00552554+4130184 & 14.963 $\pm$ 0.077 & 78.169 $\pm$ 0.1960\\
2MASSJ00583814-1747311 & 15.742 $\pm$ 0.165 & 167.563 $\pm$ 12.9690\\
2MASSJ01045111-3327380 & 15.769 $\pm$ 0.160 & 95.000 $\pm$ 8.2805\\
2MASSJ01470204+2120242 & 15.028 $\pm$ 0.193 & 252.124 $\pm$ 42.274\\
2MASSJ01532750+3631482 & 15.153 $\pm$ 0.089 & 172.527 $\pm$ 12.3735\\
2MASSJ03023398-1028223 & 15.772 $\pm$ 0.145 & 193.810 $\pm$ 19.2206\\
2MASSJ04035944+1520502 & 15.683 $\pm$ 0.133 & 178.434 $\pm$ 23.134\\
2MASSJ04071296+1710474 & 15.69 $\pm$ 0.17 & 106.068 $\pm$ 9.8183\\
2MASSJ04360273+1547536 & 15.142 $\pm$ 0.083 & 223.284 $\pm$ 24.439\\
2MASSJ11070582+2827226 & 15.168 $\pm$ 0.097 & 108.696 $\pm$ 6.3126\\
2MASSJ11150577+2520467 & 15.027 $\pm$ 0.077 & 120.735 $\pm$ 6.8643\\
2MASSJ11240487+3808054 & 12.017 $\pm$ 0.027 & 18.412 $\pm$ 0.0506\\
2MASSJ11323833-1446374 & 15.621 $\pm$ 0.112 & 172.625 $\pm$ 13.7286\\
2MASSJ11395113-3159214 & 11.996 $\pm$ 0.022 & 49.709 $\pm$ 0.5112\\
2MASSJ11463232+0203414 & 15.681 $\pm$ 0.161 & 301.804 $\pm$ 36.780\\
2MASSJ12121714-2253451 & 15.404 $\pm$ 0.144 & 123.516 $\pm$ 6.3573\\
2MASSJ13272391+0946446 & 15.340 $\pm$ 0.107 & 145.841 $\pm$ 11.4260\\
2MASSJ15243203+0934386 & 14.261 $\pm$ 0.055 & 95.109 $\pm$ 2.3872\\
2MASSJ16304206-0232224 & 14.980 $\pm$ 0.077 & 128.223 $\pm$ 6.0027\\
2MASSJ17072343-0558249A & 11.260 $\pm$ 0.027 & 15.1 $\pm$ 1.9\\
2MASSJ17364839+0220426 & 15.358 $\pm$ 0.100 & 138.070 $\pm$ 11.8040\\
2MASSJ18112466+3748513 & 15.535 $\pm$ 0.046 & 78.7737 $\pm$ 1.1740\\
2MASSJ18244344+2937133 & 15.886 $\pm$ 0.079 & 140.129 $\pm$ 6.9708\\
2MASSJ18411320-4000124 & 15.022 $\pm$ 0.088 & 121.355 $\pm$ 7.0145\\
2MASSJ22120345+1641093 & 10.831 $\pm$ 0.025 & 37.2634 $\pm$ 0.0635\\
2MASSJ22341394+2359559 & 12.354 $\pm$ 0.023 & 18.4966 $\pm$ 0.0511\\
2MASSJ23515044-2537367 & 11.725 $\pm$ 0.026 & 20.3744 $\pm$ 0.1880\\
2MASSJ1237+3028 & 12.959 $\pm$ 0.024 & 36.263 $\pm$ 0.1822\\
2MASSJ12531308+2728028 & 12.365 $\pm$ 0.026 & 41.9891 $\pm$ 0.1880\\
\hline
\end{tabular}}
\begin{center}
\begin{minipage}{15.5cm}
$^a$ H-band magnitude of 2MASS All Sky Catalog of Point Sourcces \citealt{catri2003}\\
$^b$ GAIA EDR3 Collaboration \citealt{gaiadr3}\\
$^c$ Vizier Online Data Catalog \citealt{ducati2002}\\
\end{minipage}
\end{center}
\end{table*}

\begin{table*}[t]
\renewcommand{\thetable}{A\arabic{table}}
\centering
\caption{Benchmark targets used for this analysis: 39 M-dwarfs that are widely separated FGKM + mid-to-late type M binaries from IRTF SpeX observations done by \cite{mann2014prospecting}. Listed here are also their respective H-band magnitudes, and distances.}\label{manndatatable}
\resizebox{0.75\textwidth}{!}{
\begin{tabular}{|l | l | l |}
\hline
Target  & H-band Magnitude$^{a,b}$ & Distance$^c$ (pc) \\
\hline
2M0003-2822 &  12.376 $\pm$ 0.028 & 40.4029 $\pm$ 0.2657 \\
PMI02123+1249W & 12.09 $\pm$ 0.032 & 62.369 $\pm$ 0.2556\\
NLTT8870 & 10.59 $\pm$ 0.023 & 22 $\pm$ 0.2037\\
LSPMJ0253+6321 &  11.394 $\pm$ 0.031 & 30.059 $\pm$ 0.0336\\
2M0318+0828  &  13.087 $\pm$ 0.031 & 41.0423 $\pm$ 0.3407\\
2M0430-0849 & 12.204 $\pm$ 0.022 & 31.301 $\pm$ 0.2361\\
NLTT15867 &  10.637  $\pm$ 0.024 & 24.082 $\pm$ 0.0162\\
LSPMJ0731+1958 & 11.775 $\pm$ 0.025 & 48.773 $\pm$ 0.0162\\
2M0739+1305 & 13.29 $\pm$ 0.111 & 49.875 $\pm$ 0.6159\\
NLTT19472 & 10.988 $\pm$ 0.022 & 27.261 $\pm$ 0.2311\\
LSPMJ0932+2659E & 9.858 $\pm$ 0.024 & 18.806 $\pm$ 0.0233\\
I10005+2717 & 12.418 $\pm$ 0.024 & 36.071 $\pm$ 0.1345\\
PMI11055+4331 & 8.177 $\pm$ 0.024 & 14.941 $\pm$ 0.0850\\
NLTT28453 & 11.117 $\pm$ 0.027 & 19.347 $\pm$ 0.0199\\
2M1200+2048 & 12.261 $\pm$ 0.023 & 24.547 $\pm$ 0.0800\\
LSPMJ1207+1302 & 12.4 $\pm$ 0.023 & 43.576 $\pm$ 0.1722\\
LSPMJ1210+1858E & 13.052 $\pm$ 0.033 & 45.821 $\pm$ 0.2839\\
LSPMJ1302+3227 & 11.641 $\pm$ 0.02 & 31.674 $\pm$ 0.419\\
2M1320+0957 & 13.082 $\pm$ 0.031 & 36.182 $\pm$ 0.3190\\
NLTT36549 & 12.533 $\pm$ 0.023 & 26.824 $\pm$ 0.1060\\
PMI14254+2035 & 12.006 $\pm$ 0.03 & 43.759 $\pm$ 0.1722\\
HIP70623B & 11.506 $\pm$ 0.023 & 73.502 $\pm$ 0.1707\\
HIP81910B & 11.887 $\pm$ 0.026 & 47.112 $\pm$ 0.971\\
PMI16555-0823 & 9.201 $\pm$ 0.024 & 6.494 $\pm$ 0.0024\\
LSPMJ1659+0635 & 11.172 $\pm$ 0.028 & 48.288 $\pm$ 0.0921\\
2M1743+2136 & 11.016 $\pm$ 0.021 & 22.490 $\pm$ 0.0153\\
LSPMJ1748+1143 & 11.976 $\pm$ 0.028 & 36.007 $\pm$ 0.0731\\
PMI19074+5905B & 7.82 $\pm$ 0.033 & 22.443 $\pm$ 0.0079\\
2M1916+0509 & 9.226 $\pm$ 0.026 & 5.9187 $\pm$ 0.0023\\
HIP98535B & 12.724 $\pm$ 0.031 & 54.680 $\pm$ 0.1507\\
2M2010+0634 & 11.886 $\pm$ 0.023 & 16.168 $\pm$ 0.0379\\
LSPMJ2047+1051N & 11.494 $\pm$ 0.034 & 31.739 $\pm$ 0.0344\\
LSPMJ2049+3216W & 11.179 $\pm$ 0.021 & 23.343 $\pm$ 0.0194\\
HIP106551B & 12.606 $\pm$ 0.03 & 72.114 $\pm$ 0.3339\\
2M2237+3922 & 12.691 $\pm$ 0.021 & 20.988 $\pm$ 0.604\\
HIP114424B & 10.976 $\pm$ 0.022 & 36.752 $\pm$ 0.0789\\
HIP114456B & 10.227 $\pm$ 0.028 & 23.539 $\pm$ 0.1224\\
2M2331-0406 & 12.294 $\pm$ 0.026 & 25.965 $\pm$ 0.1058\\
PMI10008+3155 & 9.643 $\pm$ 0.016 & 14.9411 $\pm$ 0.0850\\
\hline
\end{tabular}}
\begin{center}
\begin{minipage}{15.5cm}
$^a$ H-band magnitude of 2MASS All Sky Catalog of Point Sourcces \citealt{catri2003}\\
$^b$ GAIA EDR3 Collaboration \citealt{gaiadr3}\\
$^c$ Vizier Online Data Catalog \citealt{ducati2002}\\
\end{minipage}
\end{center}
\end{table*}

\begin{table*}[t]
\renewcommand{\thetable}{A\arabic{table}}
\caption{Model-derived fundamental properties of mid-to-late type M-dwarfs from the Spex Prism Spectral Libraries \citep{burgasser2014spex}. For all best-fit values shown here, we used the upgraded \texttt{SPHINX II} model with a minimal gray cloud opacity (log $\kappa$ = -29) and lower convective mixing length assumption ($\alpha$ = 0.5).}\label{bestfitspexvalues}
\resizebox{0.95\textwidth}{!}{
\begin{tabular}{|l | l | l | l | l | l|}
\hline
Target  & T$_{eff} (K)$ & log$g$ (g/cm$^{-2}$)& [M/H] & C/O & R$_*$ (R$_{\odot}$)\\
\hline
2MASSJ00013044+1010146 & 2772.88$^{+30.24}_{-36.80}$ & 4.09$^{+0.12}_{-0.06}$ & $0.47^{+0.03}_{-0.03}$ & 0.64$^{+0.02}_{-0.02}$ & 0.16$^{+0.003}_{-0.003}$\\
2MASSJ00115060-1523450 & 2534.05$^{+31.00}_{-26.60}$ & 4.21$^{+0.14}_{-0.13}$ & 0.47$^{+0.04}_{-0.04}$ & 0.45$^{+0.02}_{-0.03}$ & 0.13$^{+0.003}_{-0.003}$\\
2MASSJ00335534-0908247 & 2575.13$^{+28.68}_{-29.06}$ & 4.16$^{+0.04}_{-0.04}$ & 0.46$^{+0.04}_{-0.04}$ & 0.58$^{+0.02}_{-0.02}$ & 0.15$^{+0.003}_{-0.003}$\\
2MASSJ00552554+4130184 & 2361.75$^{+18.91}_{-20.53}$ & 4.07$^{+0.08}_{-0.05}$ & 0.48$^{+0.02}_{-0.02}$ & 0.68$^{+0.01}_{-0.01}$ & 0.12$^{+0.002}_{-0.002}$\\
2MASSJ00583814-1747311 & 2901.11$^{+48.35}_{-53.53}$ & 4.11$^{+0.17}_{-0.07}$ & 0.41$^{+0.06}_{-0.06}$ & 0.63$^{+0.02}_{-0.02}$ & 0.13$^{+0.003}_{-0.003}$\\
2MASSJ01045111-3327380 & 2357.47$^{+27.12}_{-26.41}$ & 4.09$^{+0.11}_{-0.06}$ & 0.44$^{+0.05}_{-0.04}$ & 0.69$^{+0.01}_{-0.02}$ & 0.10$^{+0.003}_{-0.003}$\\
2MASSJ01470204+2120242 & 2590.60$^{+37.88}_{-52.16}$ & 4.66$^{+0.22}_{-0.22}$ & 0.38$^{+0.10}_{-0.09}$ & 0.65$^{+0.03}_{-0.04}$ & 0.32$^{+0.01}_{-0.01}$\\
2MASSJ01532750+3631482 & 2721.39$^{+57.44}_{-69.39}$ & 4.53$^{+0.25}_{-0.25}$ & 0.29$^{+0.12}_{-0.13}$ & 0.68$^{+0.01}_{-0.03}$ & 0.19$^{+0.01}_{-0.01}$\\
2MASSJ03023398-1028223 & 3060.58$^{+39.04}_{-61.49}$ & 4.08$^{+0.12}_{-0.06}$ & 0.46$^{+0.06}_{-0.06}$ & 0.56$^{+0.03}_{-0.03}$ & 0.15$^{+0.003}_{-0.003}$\\
2MASSJ04035944+1520502 & 2732.14$^{+41.82}_{-48.44}$ & 4.88$^{+0.27}_{-0.30}$ & 0.40$^{+0.09}_{-0.10}$ & 0.52$^{+0.04}_{-0.04}$ & 0.16$^{+0.01}_{-0.005}$\\
2MASSJ04071296+1710474 & 2703.64$^{+31.36}_{-37.43}$ & 4.73$^{+0.16}_{-0.17}$ & 0.44$^{+0.06}_{-0.06}$ & 0.41$^{+0.03}_{-0.03}$ & 0.10$^{+0.002}_{-0.002}$\\
2MASSJ04360273+1547536 & 2511.57$^{+64.84}_{-70.77}$ & 4.83$^{+0.23}_{-0.26}$ & 0.15$^{+0.15}_{-0.15}$ & 0.68$^{+0.01}_{-0.02}$ & 0.29$^{+0.02}_{-0.02}$\\
2MASSJ11070582+2827226 & 2654.12$^{+42.87}_{-39.26}$ & 4.07$^{+0.10}_{-0.05}$ & 0.45$^{+0.04}_{-0.05}$ & 0.44$^{+0.03}_{-0.03}$ & 0.13$^{+0.003}_{-0.003}$\\
2MASSJ11150577+2520467 & 2592.91$^{+32.41}_{-39.64}$ & 4.65$^{+0.15}_{-0.13}$ & 0.27$^{+0.09}_{-0.09}$ & 0.60$^{+0.03}_{-0.02}$ & 0.16$^{+0.01}_{-0.004}$\\
2MASSJ11240487+3808054 & 2243.31$^{+14.65}_{-14.80}$ & 4.06$^{+0.07}_{-0.04}$ & 0.49$^{+0.02}_{-0.02}$ & 0.70$^{+0.002}_{-0.01}$ & 0.12$^{+0.001}_{-0.001}$\\
2MASSJ11323833-1446374 & 2782.41$^{+58.60}_{-42.74}$ & 4.89$^{+0.27}_{-0.41}$ & 0.43$^{+0.08}_{-0.08}$ & 0.46$^{+0.05}_{-0.05}$ & 0.15$^{+0.004}_{-0.004}$\\
2MASSJ11395113-3159214 & 2065.50$^{+94.25}_{-20.85}$ & 4.06$^{+0.04}_{-0.03}$ & 0.44$^{+0.02}_{-0.04}$ & 0.66$^{+0.02}_{-0.02}$ & 0.38$^{+0.01}_{-0.04}$\\
2MASSJ11463232+0203414 & 2872.37$^{+123.14}_{-105.82}$ & 4.60$^{+0.39}_{-0.35}$ & 0.18$^{+0.18}_{-0.22}$ & 0.47$^{+0.07}_{-0.08}$ & 0.25$^{+0.02}_{-0.01}$\\
2MASSJ12121714-2253451 & 2711.11$^{+49.39}_{-51.24}$ & 4.14$^{+0.17}_{-0.09}$ & 0.44$^{+0.05}_{-0.05}$ & 0.48$^{+0.03}_{-0.03}$ & 0.48$^{+0.03}_{-0.03}$\\
2MASSJ13272391+0946446 & 2528.81$^{+42.23}_{-33.68}$ & 4.25$^{+0.16}_{-0.15}$ & 0.29$^{+0.09}_{-0.08}$ & 0.69$^{+0.01}_{-0.01}$ & 0.17$^{+0.01}_{-0.01}$\\
2MASSJ15243203+0934386 & 2619.88$^{+44.66}_{-41.76}$ & 4.29$^{+0.34}_{-0.21}$ & 0.45$^{+0.05}_{-0.05}$ & 0.55$^{+0.03}_{-0.04}$ & 0.17$^{+0.01}_{-0.01}$\\
2MASSJ16304206-0232224 & 2837.16$^{+73.30}_{-99.31}$ & 4.17$^{+0.22}_{-0.12}$ & 0.37$^{+0.11}_{-0.11}$ & 0.43$^{+0.05}_{-0.06}$ & 0.15$^{+0.01}_{-0.01}$\\
2MASSJ17072343-0558249A & 2133.92$^{+18.40}_{-17.19}$ & 4.27$^{+0.15}_{-0.09}$ & 0.38$^{+0.04}_{-0.04}$ & 0.70$^{+0.002}_{-0.002}$ & 0.16$^{+0.003}_{-0.003}$\\
2MASSJ17364839+0220426 & 2450.01$^{+23.58}_{-20.97}$ & 4.14$^{+0.15}_{-0.09}$ & 0.48$^{+0.03}_{-0.03}$ & 0.67$^{+0.02}_{-0.02}$ & 0.17$^{+0.003}_{-0.003}$\\
2MASSJ18112466+3748513 & 2518.22$^{+153.31}_{-118.45}$ & 4.54$^{+0.57}_{-0.38}$ & 0.36$^{+0.17}_{-0.17}$ & 0.38$^{+0.11}_{-0.05}$ & 0.11$^{+0.01}_{-0.01}$\\
2MASSJ18244344+2937133 & 2739.53$^{+62.81}_{-50.22}$ & 4.46$^{+0.28}_{-0.21}$ & 0.26$^{+0.12}_{-0.11}$ & 0.53$^{+0.03}_{-0.04}$ & 0.12$^{+0.004}_{-0.004}$\\
2MASSJ18411320-4000124 & 2436.09$^{+60.48}_{-84.41}$ & 4.46$^{+0.29}_{-0.25}$ & 0.28$^{+0.13}_{-0.20}$ & 0.61$^{+0.05}_{-0.06}$ & 0.17$^{+0.01}_{-0.01}$\\
2MASSJ22120345+1641093 & 3071.03$^{+138.08}_{-57.66}$ & 4.17$^{+0.20}_{-0.10}$ & 0.19$^{+0.20}_{-0.07}$ & 0.53$^{+0.02}_{-0.03}$ & 0.27$^{+0.01}_{-0.01}$\\
2MASSJ22341394+2359559 & 2035.99$^{+18.61}_{-18.35}$ & 4.34$^{+0.08}_{-0.10}$ & 0.18$^{+0.04}_{-0.05}$ & 0.70$^{+0.002}_{-0.003}$ & 0.13$^{+0.002}_{-0.002}$\\
2MASSJ23515044-2537367 & 2383.55$^{+19.91}_{-20.61}$ & 4.04$^{+0.05}_{-0.03}$ & 0.49$^{+0.01}_{-0.01}$ & 0.57$^{+0.02}_{-0.02}$ & 0.14$^{+0.002}_{-0.002}$\\
2MASSJ1237+3028 & 2439.01$^{+28.02}_{-23.17}$ & 4.35$^{+0.15}_{-0.14}$ & 0.38$^{+0.06}_{-0.05}$ & 0.65$^{+0.01}_{-0.02}$ & 0.14$^{+0.002}_{-0.002}$\\
2MASSJ12531308+2728028 & 2259.28$^{+19.76}_{-23.87}$ & 4.47$^{+0.10}_{-0.14}$ & 0.45$^{+0.05}_{-0.05}$ & 0.56$^{+0.02}_{-0.02}$ & 0.12$^{+0.002}_{-0.002}$\\
\hline
\end{tabular}}
\end{table*}

\begin{table*}[t]
\renewcommand{\thetable}{A\arabic{table}}
\centering
\caption{Model-derived fundamental properties of mid-to-late type M-dwarfs that are companions to FGKM stars \citep{mann2014prospecting}. For all best-fit values shown here, we used the upgraded \texttt{SPHINX II} model with a minimal gray cloud opacity (log $\kappa$ = -29) and lower convective mixing length assumption ($\alpha$ = 0.5).}\label{manndatabestfitvalues}
\resizebox{\textwidth}{!}{
\begin{tabular}{|l | l | l | l | l | l|}
\hline
Target  & T$_{eff} (K)$ & log$g$ (g/cm$^{-2}$)& [M/H] & C/O & R$_*$ (R$_{\odot}$)\\
\hline
2M0003-2822 & 2289.32$^{+27.90}_{-23.73}$ & 4.38$^{+0.09}_{-0.09}$ & 0.11$^{+0.04}_{-0.04}$ & 0.69$^{+0.001}_{-0.01}$ & 0.22$^{+0.002}_{-0.01}$\\
PMI02123+1249W & 3055.21$^{+64.96}_{-29.59}$ & 4.91$^{+0.08}_{-0.10}$ & 0.28$^{+0.07}_{-0.07}$ & 0.70$^{+0.001}_{-0.01}$ & 0.23$^{+0.004}_{-0.01}$\\
NLTT8870 & 2997.78$^{+48.34}_{-43.41}$ & 4.54$^{+0.12}_{-0.13}$ & 0.05$^{+0.07}_{-0.07}$ & 0.44$^{+0.02}_{-0.03}$ & 0.18$^{+0.004}_{-0.004}$\\
LSPMJ0253+6321 & 2784.17$^{+19.83}_{-19.14}$ & 4.48$^{+0.12}_{-0.10}$ & -0.28$^{+0.03}_{-0.03}$ & 0.40$^{+0.03}_{-0.02}$ & 0.19$^{+0.003}_{-0.003}$\\
2M0318+0828 & 2392.56$^{+27.27}_{-29.50}$ & 4.26$^{+0.11}_{-0.12}$ & 0.17$^{+0.05}_{-0.05}$ & 0.69$^{+0.01}_{-0.01}$ & 0.15$^{+0.004}_{-0.004}$\\
2M0430-0849 & 2254.03$^{+23.23}_{-21.89}$ & 4.32$^{+0.11}_{-0.14}$ & 0.29$^{+0.02}_{-0.02}$ & 0.67$^{+0.01}_{-0.02}$ & 0.19$^{+0.004}_{-0.004}$\\
NLTT15867 & 2982.10$^{+22.70}_{-21.05}$ & 4.28$^{+0.07}_{-0.09}$ & -0.10$^{+0.02}_{-0.03}$ & 0.37$^{+0.01}_{-0.02}$ & 0.20$^{+0.002}_{-0.002}$\\
LSPMJ0731+1958 & 2653.78$^{+13.10}_{-14.25}$ & 4.55$^{+0.09}_{-0.05}$ & -0.09$^{+0.01}_{-0.01}$ & 0.63$^{+0.02}_{-0.02}$ & 0.27$^{+0.003}_{-0.003}$\\
2M0739+1305 & 2021.28$^{+16.62}_{-13.40}$ & 4.15$^{+0.08}_{-0.07}$ & 0.19$^{+0.03}_{-0.04}$ & 0.70$^{+0.002}_{-0.002}$ & 0.23$^{+0.004}_{-0.004}$\\
NLTT19472 & 2787.77$^{+23.35}_{-24.13}$ & 4.86$^{+0.09}_{-0.12}$ & -0.11$^{+0.06}_{-0.05}$ & 0.35$^{+0.03}_{-0.03}$ & 0.20$^{+0.003}_{-0.003}$\\
LSPMJ0932+2659E & 2845.00$^{+70.98}_{-29.31}$ & 4.23$^{+0.12}_{-0.08}$ & -0.06$^{+0.11}_{-0.05}$ & 0.43$^{+0.03}_{-0.02}$ & 0.23$^{+0.005}_{-0.01}$\\
I10005+2717 & 2707.23$^{+26.91}_{-29.23}$ & 4.63$^{+0.13}_{-0.18}$ & 0.08$^{+0.05}_{-0.05}$ & 0.59$^{+0.02}_{-0.02}$ & 0.15$^{+0.003}_{-0.003}$\\
PMI11055+4331 & 2685.95$^{+19.66}_{-27.62}$ & 4.87$^{+0.08}_{-0.09}$ & -0.24$^{+0.04}_{-0.04}$ & 0.34$^{+0.02}_{-0.02}$ & 0.14$^{+0.002}_{-0.002}$\\
NLTT28453 & 2617.97$^{+12.93}_{-15.74}$ & 4.62$^{+0.08}_{-0.08}$ & 0.09$^{+0.02}_{-0.02}$ & 0.60$^{+0.02}_{-0.02}$ & 0.15$^{+0.002}_{-0.002}$\\
2M1200+2048 & 2496.16$^{+35.69}_{-23.41}$ & 5.07$^{+0.09}_{-0.08}$ & -0.12$^{+0.14}_{-0.09}$ & 0.41$^{+0.04}_{-0.04}$ & 0.12$^{+0.003}_{-0.003}$\\
LSPMJ1207+1302 & 2774.86$^{+21.21}_{-19.75}$ & 4.36$^{+0.11}_{-0.11}$ & -0.16$^{+0.03}_{-0.03}$ & 0.46$^{+0.02}_{-0.02}$ & 0.17$^{+0.003}_{-0.003}$\\
LSPMJ1210+1858E & 2532.61$^{+22.54}_{-20.16}$ & 4.85$^{+0.08}_{-0.08}$ & 0.03$^{+0.03}_{-0.03}$ & 0.68$^{+0.01}_{-0.01}$ & 0.15$^{+0.003}_{-0.003}$\\
LSPMJ1302+3227 & 2825.47$^{+17.61}_{-16.93}$ & 4.94$^{+0.08}_{-0.08}$ & -0.08$^{+0.03}_{-0.03}$ & 0.51$^{+0.02}_{-0.02}$ & 0.17$^{+0.003}_{-0.003}$\\
2M1320+0957 & 2395.00$^{+29.05}_{-32.89}$ & 4.61$^{+0.11}_{-0.14}$ & -0.01$^{+0.06}_{-0.06}$ & 0.62$^{+0.02}_{-0.03}$ & 0.13$^{+0.004}_{-0.004}$\\
NLTT36549 & 2398.26$^{+16.69}_{-18.28}$ & 4.72$^{+0.08}_{-0.09}$ & -0.02$^{+0.02}_{-0.02}$ & 0.45$^{+0.02}_{-0.02}$ & 0.13$^{+0.002}_{-0.002}$\\
PMI14254+2035 & 2961.38$^{+22.39}_{-18.57}$ & 4.06$^{+0.07}_{-0.04}$ & -0.50$^{+0.004}_{-0.004}$ & 0.30$^{+0.003}_{-0.003}$ & 0.20$^{+0.002}_{-0.002}$\\
HIP70623B & 3106.98$^{+12.42}_{-13.07}$ & 5.37$^{+0.05}_{-0.06}$ & 0.39$^{+0.05}_{-0.03}$ & 0.69$^{+0.004}_{-0.004}$ & 0.35$^{+0.003}_{-0.003}$\\
HIP81910B & 2530.34$^{+32.24}_{-32.02}$ & 4.92$^{+0.18}_{-0.18}$ & 0.34$^{+0.05}_{-0.05}$ & 0.68$^{+0.01}_{-0.01}$ & 0.26$^{+0.01}_{-0.01}$\\
PMI16555-0823 & 2469.25$^{+22.55}_{-18.94}$ & 4.53$^{+0.09}_{-0.10}$ & 0.09$^{+0.01}_{-0.01}$ & 0.42$^{+0.02}_{-0.02}$ & 0.14$^{+0.002}_{-0.002}$\\
LSPMJ1659+0635 & 3064.12$^{+35.42}_{-94.51}$ & 4.28$^{+0.11}_{-0.08}$ & 0.36$^{+0.04}_{-0.16}$ & 0.62$^{+0.01}_{-0.02}$ & 0.29$^{+0.01}_{-0.008}$\\
2M1743+2136 & 3014.89$^{+11.70}_{-9.36}$ & 4.08$^{+0.09}_{-0.05}$ & -0.04$^{+0.02}_{-0.02}$ & 0.30$^{+0.01}_{-0.005}$ & 0.16$^{+0.01}_{-0.008}$\\
LSPMJ1748+1143 & 2824.05$^{+18.66}_{-16.10}$ & 5.13$^{+0.07}_{-0.07}$ & -0.07$^{+0.04}_{-0.04}$ & 0.53$^{+0.02}_{-0.02}$ & 0.16$^{+0.002}_{-0.002}$\\
PMI19074+5905B & 2638.86$^{+17.84}_{-16.53}$ & 4.37$^{+0.08}_{-0.08}$ & 0.01$^{+0.02}_{-0.02}$ & 0.62$^{+0.02}_{-0.01}$ & 0.28$^{+0.01}_{-0.01}$\\
2M1916+0509 & 2366.17$^{+21.94}_{-24.31}$ & 4.28$^{+0.09}_{-0.11}$ & 0.43$^{+0.05}_{-0.05}$ & 0.69$^{+0.01}_{-0.01}$ & 0.12$^{+0.003}_{-0.003}$\\
HIP98535B & 3049.62$^{+22.46}_{-19.39}$ & 4.77$^{+0.12}_{-0.09}$ & 0.21$^{+0.02}_{-0.02}$ & 0.52$^{+0.01}_{-0.02}$ & 0.16$^{+0.002}_{-0.002}$\\
2M2010+0634 & 2180.09$^{+17.44}_{-16.20}$ & 4.62$^{+0.08}_{-0.06}$ & 0.08$^{+0.03}_{-0.03}$ & 0.61$^{+0.02}_{-0.02}$ & 0.12$^{+0.002}_{-0.002}$\\
LSPMJ2047+1051N & 2803.53$^{+6.32}_{-4.46}$ & 4.04$^{+0.01}_{-0.01}$ & -0.59$^{+0.003}_{-0.003}$ & 0.32$^{+0.01}_{-0.01}$ & 0.20$^{+0.001}_{-0.001}$\\
LSPMJ2049+3216W & 2708.68$^{+22.01}_{-19.11}$ & 4.52$^{+0.11}_{-0.11}$ & -0.17$^{+0.04}_{-0.04}$ & 0.46$^{+0.03}_{-0.02}$ & 0.17$^{+0.003}_{-0.003}$\\
HIP106551B & 2880.48$^{+30.19}_{-36.10}$ & 4.65$^{+0.08}_{-0.11}$ & 0.04$^{+0.05}_{-0.07}$ & 0.62$^{+0.02}_{-0.02}$ & 0.24$^{+0.01}_{-0.005}$\\
2M2237+3922 & 2130.74$^{+28.64}_{-36.16}$ & 4.39$^{+0.09}_{-0.14}$ & 0.07$^{+0.03}_{-0.03}$ & 0.64$^{+0.02}_{-0.02}$ & 0.11$^{+0.004}_{-0.004}$\\
HIP114424B & 2980.67$^{+24.63}_{-38.69}$ & 4.52$^{+0.10}_{-0.08}$ & 0.11$^{+0.05}_{-0.05}$ & 0.54$^{+0.02}_{-0.03}$ & 0.25$^{+0.004}_{-0.004}$\\
HIP114456B & 2933.97$^{+125.86}_{-53.34}$ & 4.27$^{+0.11}_{-0.11}$ & 0.18$^{+0.20}_{-0.08}$ & 0.53$^{+0.03}_{-0.03}$ & 0.24$^{+0.01}_{-0.01}$\\
2M2331-0406 & 2386.04$^{+17.94}_{-22.38}$ & 4.83$^{+0.10}_{-0.08}$ & -0.01$^{+0.02}_{-0.02}$ & 0.44$^{+0.02}_{-0.02}$ & 0.14$^{+0.002}_{-0.002}$\\
PMI10008+3155 & 2898.14$^{+41.22}_{-44.27}$ & 4.10$^{+0.15}_{-0.07}$ & 0.43$^{+0.05}_{-0.05}$ & 0.66$^{+0.02}_{-0.02}$ & 0.20$^{+0.004}_{-0.004}$\\
\hline
\end{tabular}}
\end{table*}

\clearpage

\bibliography{biblio}

@article{witzke2022can,
  title={Can 1D Radiative-equilibrium Models of Faculae Be Used for Calculating Contamination of Transmission Spectra?},
  author={Witzke, Veronika and Shapiro, Alexander I and Kostogryz, Nadiia M and Cameron, Robert and Rackham, Benjamin V and Seager, Sara and Solanki, Sami K and Unruh, Yvonne C},
  journal={The Astrophysical Journal Letters},
  volume={941},
  number={2},
  pages={L35},
  year={2022},
  publisher={IOP Publishing}
}

@phdthesis{iyer2023m,
  title={The M-dwarf Atmosphere Problem},
  author={Iyer, Aishwarya R},
  year={2023},
  school={Arizona State University}
}

@article{kanodia2024searching,
  title={Searching for Giant Exoplanets around M-dwarf Stars (GEMS) I: Survey Motivation},
  author={Kanodia, Shubham and Ca{\~n}as, Caleb I and Mahadevan, Suvrath and Ford, Eric B and Helled, Ravit and Anderson, Dana E and Boss, Alan and Cochran, William D and Delamer, Megan and Han, Te and others},
  journal={The Astronomical Journal},
  volume={167},
  number={4},
  pages={161},
  year={2024},
  publisher={IOP Publishing}
}

@ARTICLE{kanodia2024,
       author = {{Kanodia}, Shubham and {Ca{\~n}as}, Caleb I. and {Mahadevan}, Suvrath and {Ford}, Eric B. and {Helled}, Ravit and {Anderson}, Dana E. and {Boss}, Alan and {Cochran}, William D. and {Delamer}, Megan and {Han}, Te and {Libby-Roberts}, Jessica E. and {Lin}, Andrea S.~J. and {M{\"u}ller}, Simon and {Robertson}, Paul and {Stef{\'a}nsson}, Gumundur and {Teske}, Johanna},
        title = "{Searching for Giant Exoplanets around M-dwarf Stars (GEMS) I: Survey Motivation}",
      journal = {\aj},
         year = 2024,
        month = apr,
       volume = {167},
       number = {4},
          eid = {161},
        pages = {161},
          doi = {10.3847/1538-3881/ad27cb},
archivePrefix = {arXiv},
       eprint = {2402.04946},
 primaryClass = {astro-ph.EP},
       adsurl = {https://ui.adsabs.harvard.edu/abs/2024AJ....167..161K}
}

@ARTICLE{mori2024,
       author = {{Mori}, Mayuko and {Ikuta}, Kai and {Fukui}, Akihiko and {Narita}, Norio and {de Leon}, Jerome P. and {Livingston}, John H. and {Ikoma}, Masahiro and {Kawai}, Yugo and {Kawauchi}, Kiyoe and {Murgas}, Felipe and {Palle}, Enric and {Parviainen}, Hannu and {Fern{\'a}ndez Rodr{\'\i}guez}, Gareb and {Terada}, Yuka and {Watanabe}, Noriharu and {Tamura}, Motohide},
        title = "{Characterization of starspots on a young M-dwarf K2-25: multi-band observations of stellar photometric variability and planetary transits}",
      journal = {\mnras},
         year = 2024,
        month = mar,
          doi = {10.1093/mnras/stae841},
archivePrefix = {arXiv},
       eprint = {2403.13946},
 primaryClass = {astro-ph.EP},
       adsurl = {https://ui.adsabs.harvard.edu/abs/2024MNRAS.tmp..863M}
}

@ARTICLE{quintana2021,
       author = {{Quintana}, Elisa V. and {Col{\'o}n}, Knicole D. and {Mosby}, Gregory and {Schlieder}, Joshua E. and {Supsinskas}, Pete and {Karburn}, Jordan and {Dotson}, Jessie L. and {Greene}, Thomas P. and {Hedges}, Christina and {Apai}, D{\'a}niel and {Barclay}, Thomas and {Christiansen}, Jessie L. and {Espinoza}, N{\'e}stor and {Mullally}, Susan E. and {Gilbert}, Emily A. and {Hoffman}, Kelsey and {Kostov}, Veselin B. and {Lewis}, Nikole K. and {Foote}, Trevor O. and {Mason}, James and {Youngblood}, Allison and {Morris}, Brett M. and {Newton}, Elisabeth R. and {Pepper}, Joshua and {Rackham}, Benjamin V. and {Rowe}, Jason F. and {Stevenson}, Kevin},
        title = "{The Pandora SmallSat: Multiwavelength Characterization of Exoplanets and their Host Stars}",
      journal = {arXiv e-prints},
         year = 2021,
        month = aug,
          eid = {arXiv:2108.06438},
        pages = {arXiv:2108.06438},
          doi = {10.48550/arXiv.2108.06438},
archivePrefix = {arXiv},
       eprint = {2108.06438},
 primaryClass = {astro-ph.IM},
       adsurl = {https://ui.adsabs.harvard.edu/abs/2021arXiv210806438Q}
}

@ARTICLE{jeffries2012,
       author = {{Jackson}, R.~J. and {Jeffries}, R.~D.},
        title = "{Why do some young cool stars show spot modulation while others do not?}",
      journal = {\mnras},
         year = 2012,
        month = jul,
       volume = {423},
       number = {3},
        pages = {2966-2976},
          doi = {10.1111/j.1365-2966.2012.21119.x},
archivePrefix = {arXiv},
       eprint = {1204.4066},
 primaryClass = {astro-ph.SR},
       adsurl = {https://ui.adsabs.harvard.edu/abs/2012MNRAS.423.2966J}
}

@ARTICLE{wende2009,
       author = {{Wende}, S. and {Reiners}, A. and {Ludwig}, H. -G.},
        title = "{3D simulations of M star atmosphere velocities and their influence on molecular FeH lines}",
      journal = {\aap},
         year = 2009,
        month = dec,
       volume = {508},
       number = {3},
        pages = {1429-1442},
          doi = {10.1051/0004-6361/200913149},
archivePrefix = {arXiv},
       eprint = {0910.3493},
 primaryClass = {astro-ph.SR},
       adsurl = {https://ui.adsabs.harvard.edu/abs/2009A&A...508.1429W}
}

@ARTICLE{patel2022,
       author = {{Patel}, Jayshil A. and {Espinoza}, N{\'e}stor},
        title = "{Empirical Limb-darkening Coefficients and Transit Parameters of Known Exoplanets from TESS}",
      journal = {\aj},
         year = 2022,
        month = may,
       volume = {163},
       number = {5},
          eid = {228},
        pages = {228},
          doi = {10.3847/1538-3881/ac5f55},
archivePrefix = {arXiv},
       eprint = {2203.05661},
 primaryClass = {astro-ph.EP},
       adsurl = {https://ui.adsabs.harvard.edu/abs/2022AJ....163..228P}
}

@article{norris2023spectral,
  title={Spectral variability of photospheric radiation due to faculae--II. Facular contrasts for cool main-sequence stars},
  author={Norris, Charlotte M and Unruh, Yvonne C and Witzke, Veronika and Solanki, Sami K and Krivova, Natalie A and Shapiro, Alexander I and Yeo, Kok Leng and Cameron, Robert and Beeck, Benjamin},
  journal={Monthly Notices of the Royal Astronomical Society},
  volume={524},
  number={1},
  pages={1139--1155},
  year={2023},
  publisher={Oxford University Press}
}

@incollection{Lodders2006,
  author    = {Lodders, Katharina},
  title     = {Solar System Abundances and Condensation Temperatures of the Elements},
  booktitle = {Astrophysics Update 2},
  editor    = {Mason, John W.},
  publisher = {Springer},
  pages     = {1--28},
  year      = {2006},
  doi       = {10.1007/3-540-30313-8_1}
}

@article{Marley2013,
  author  = {Marley, Mark S. and Ackerman, Andrew S. and Cuzzi, Jeffrey N. and Kitzmann, Daniel},
  title   = {Clouds and Hazes in Exoplanet Atmospheres},
  journal = {Comparative Climatology of Terrestrial Planets},
  year    = {2013},
  pages   = {367--391},
  publisher = {University of Arizona Press},
  doi     = {10.2458/azu_uapress_9780816530595-ch16}
}

@article{Ackerman2001,
  author  = {Ackerman, Andrew S. and Marley, Mark S.},
  title   = {Precipitating Condensation Clouds in Substellar Atmospheres},
  journal = {The Astrophysical Journal},
  volume  = {556},
  number  = {2},
  pages   = {872--884},
  year    = {2001},
  doi     = {10.1086/321540}
}

@article{Helling2014,
  author  = {Helling, Christiane and Casewell, Sarah},
  title   = {Atmospheric Electrification in Brown Dwarfs and Exoplanets},
  journal = {The Astronomy and Astrophysics Review},
  volume  = {22},
  number  = {1},
  pages   = {80},
  year    = {2014},
  doi     = {10.1007/s00159-014-0080-0}
}

@article{Cushing2006,
  author  = {Cushing, Michael C. and Roellig, Thomas L. and Marley, Mark S. and Saumon, Didier and Leggett, S. K. and Kirkpatrick, J. Davy and Wilson, J. C. and Sloan, Gregory C. and Mainzer, Amy K. and Van Cleve, Jeffrey E. and Houck, James R.},
  title   = {A Spitzer Infrared Spectrograph Spectral Sequence of M, L, and T Dwarfs},
  journal = {The Astrophysical Journal},
  volume  = {648},
  number  = {1},
  pages   = {614--628},
  year    = {2006},
  doi     = {10.1086/505637}
}

@article{Stephens2009,
  author  = {Stephens, D. C. and Leggett, S. K. and Cushing, Michael C. and Marley, Mark S. and Saumon, Didier and Geballe, Thomas R. and Golimowski, David A. and Fan, Xiaohui and Noll, Keith S.},
  title   = {The 0.8--14.5 $\mu$m Spectra of Mid-L to Mid-T Dwarfs: Diagnostics of Effective Temperature, Grain Sedimentation, Gas Transport, and Surface Gravity},
  journal = {The Astrophysical Journal},
  volume  = {702},
  number  = {1},
  pages   = {154--170},
  year    = {2009},
  doi     = {10.1088/0004-637X/702/1/154}
}

@article{Greene2016,
  author  = {Greene, Thomas P. and Line, Michael R. and Montero, Camilo and Fortney, Jonathan J. and Lustig-Yaeger, Jacob and Luther, Kurt},
  title   = {Characterizing Transiting Exoplanet Atmospheres with JWST},
  journal = {The Astrophysical Journal},
  volume  = {817},
  number  = {1},
  pages   = {17},
  year    = {2016},
  doi     = {10.3847/0004-637X/817/1/17}
}

@article{Rackham2018,
  author  = {Rackham, Benjamin V. and Apai, D\'aniel and Giampapa, Mark S.},
  title   = {The Transit Light Source Effect: False Spectral Features and Incorrect Densities for M-dwarf Transiting Planets},
  journal = {The Astrophysical Journal},
  volume  = {853},
  number  = {2},
  pages   = {122},
  year    = {2018},
  doi     = {10.3847/1538-4357/aaa08c}
}

@article{hurt2024uniform,
  title={Uniform Forward-modeling Analysis of Ultracool Dwarfs. III. Late-M and L Dwarfs in Young Moving Groups, the Pleiades, and the Hyades},
  author={Hurt, Spencer A and Liu, Michael C and Zhang, Zhoujian and Phillips, Mark and Allers, Katelyn N and Deacon, Niall R and Aller, Kimberly M and Best, William MJ},
  journal={The Astrophysical Journal},
  volume={961},
  number={1},
  pages={121},
  year={2024},
  publisher={IOP Publishing}
}

@article{suarez2022,
  title={Ultracool dwarfs observed with the Spitzer infrared spectrograph--II. Emergence and sedimentation of silicate clouds in L dwarfs, and analysis of the full M5--T9 field dwarf spectroscopic sample},
  author={Su{\'a}rez, Genaro and Metchev, Stanimir},
  journal={Monthly Notices of the Royal Astronomical Society},
  volume={513},
  number={4},
  pages={5701--5726},
  year={2022},
  publisher={Oxford University Press}
}

@misc{diamondlowe2024hotrocks,
  author       = {Diamond-Lowe, Hannah and Mendonça, João and collaborators},
  title        = {The Hot Rocks Survey: Testing 9 Irradiated Terrestrial Exoplanets for Atmospheres},
  howpublished = {JWST General Observer Program 3730},
  year         = {2024},
  note         = {PI: H. Diamond-Lowe, Co-PI: J. Mendonça},
  url          = {https://www.stsci.edu/jwst/science-execution/approved-programs/general-observers/cycle-2-go}
}

@article{Barclay2025_PandoraMission,
  author  = {Barclay, Thomas and Quintana, Elisa V. and Col{\'o}n, Knicole D. and Hord, Benjamin J. and Mosby, Gregory and Schlieder, Joshua E. and Zellem, Robert T. and Karburn, Jordan and Heatwole, Peter F. and Kostov, Veselin B. and Gilbert, Emily A. and Lewis, Nikole K. and Rackham, Benjamin V. and Dotson, Jessie L. and Hedges, Christina},
  title   = {The Pandora SmallSat: A Low‐Cost, High Impact Mission to Study Exoplanets and Their Host Stars},
  journal = {arXiv e-prints},
  year    = {2025},
  eprint  = {2502.09730},
  primaryClass = {astro-ph.EP},
  url     = {https://arxiv.org/abs/2502.09730}
}

@article{Jaeger1998,
  author  = {Jaeger, C. and Molster, F. J. and Dorschner, J. and Henning, Th. and Mutschke, H. and Waters, L. B. F. M.},
  title   = {Steps toward Interstellar Silicate Mineralogy. IV. The Crystalline Revolution},
  journal = {Astronomy and Astrophysics},
  volume  = {339},
  pages   = {904--916},
  year    = {1998}
}

@article{Min2007,
  author  = {Min, M. and Waters, L. B. F. M. and de Koter, A. and Hovenier, J. W. and Keller, L. P. and Markwick-Kemper, F.},
  title   = {The 10 Micron Amorphous Silicate Feature of Protoplanetary Disks: A Tracer of Grain Growth?},
  journal = {Astronomy and Astrophysics},
  volume  = {462},
  number  = {2},
  pages   = {667--676},
  year    = {2007},
  doi     = {10.1051/0004-6361:20065243}
}

@article{Henning1997,
  author       = {Henning, Th. and Mutschke, H.},
  title        = {Low-temperature infrared properties of cosmic dust analogues},
  journal      = {Astronomy and Astrophysics},
  year         = {1997},
  volume       = {327},
  pages        = {743--754}
}

@article{Draine2003,
  author       = {Draine, B.T.},
  title        = {Interstellar Dust Grains},
  journal      = {Annual Review of Astronomy and Astrophysics},
  year         = {2003},
  volume       = {41},
  pages        = {241--289},
  doi          = {10.1146/annurev.astro.41.011802.094840}
}

@article{lodders2003,
  title={Solar system abundances and condensation temperatures of the elements},
  author={Lodders, Katharina},
  journal={The Astrophysical Journal},
  volume={591},
  number={2},
  pages={1220},
  year={2003},
  publisher={IOP Publishing}
}

@article{Visscher2010,
  author = {Visscher, Channon and Lodders, Katharina and Fegley, Bruce},
  title = {Atmospheric Chemistry in Giant Planets, Brown Dwarfs, and Low-Mass Dwarf Stars. II. Sulfur and Phosphorus},
  journal = {The Astrophysical Journal},
  volume = {716},
  number = {2},
  pages = {1060--1075},
  year = {2010},
  doi = {10.1088/0004-637X/716/2/1060}
}

@article{henning1996dust,
  title={Dust opacities for protoplanetary accretion disks: influence of dust aggregates.},
  author={Henning, Th and Stognienko, R},
  journal={Astronomy and Astrophysics, v. 311, p. 291-303},
  volume={311},
  pages={291--303},
  year={1996}
}

@ARTICLE{ducrot2020A,
       author = {{Ducrot}, E. and {Gillon}, M. and {Delrez}, L. and {Agol}, E. and {Rimmer}, P. and {Turbet}, M. and {G{\"u}nther}, M.~N. and {Demory}, B. -O. and {Triaud}, A.~H.~M.~J. and {Bolmont}, E. and {Burgasser}, A. and {Carey}, S.~J. and {Ingalls}, J.~G. and {Jehin}, E. and {Leconte}, J. and {Lederer}, S.~M. and {Queloz}, D. and {Raymond}, S.~N. and {Selsis}, F. and {Van Grootel}, V. and {de Wit}, J.},
        title = "{TRAPPIST-1: Global results of the Spitzer Exploration Science Program Red Worlds}",
      journal = {\aap},
         year = 2020,
        month = aug,
       volume = {640},
          eid = {A112},
        pages = {A112},
          doi = {10.1051/0004-6361/201937392},
archivePrefix = {arXiv},
       eprint = {2006.13826},
 primaryClass = {astro-ph.EP},
       adsurl = {https://ui.adsabs.harvard.edu/abs/2020A&A...640A.112D}
}

@article{Burrows1999,
  author       = {Burrows, Adam and Sharp, Christopher M.},
  title        = {Chemical Equilibrium Abundances in Brown Dwarf and Extrasolar Giant Planet Atmospheres},
  journal      = {The Astrophysical Journal},
  volume       = {512},
  number       = {2},
  pages        = {843--863},
  year         = {1999},
  doi          = {10.1086/306811}
}

@article{wakeford2015,
  title={Transmission spectral properties of clouds for hot Jupiter exoplanets},
  author={Wakeford, Hannah Ruth and Sing, David Kent},
  journal={Astronomy \& Astrophysics},
  volume={573},
  pages={A122},
  year={2015},
  publisher={EDP Sciences}
}

@article{Kitzmann2018,
  author  = {Kitzmann, Daniel and Heng, Kevin},
  title   = {A scattering greenhouse effect in Titan’s atmosphere},
  journal = {Monthly Notices of the Royal Astronomical Society},
  volume  = {475},
  number  = {1},
  pages   = {94--104},
  year    = {2018},
  doi     = {10.1093/mnras/stx3210}
}

@book{Gail2014,
  author       = {Gail, Hans-Peter and Sedlmayr, Erwin},
  title        = {Physics and Chemistry of Circumstellar Dust Shells},
  publisher    = {Cambridge University Press},
  year         = {2014},
  doi          = {10.1017/CBO9781139020732}
}

@article{Cox1981,
  author       = {Cox, A. N. and Shaviv, G. and Hodson, S. W.},
  title        = {On the Convection in Cool Main-Sequence Stars},
  journal      = {Astrophysical Journal Letters},
  volume       = {245},
  pages        = {L37--L40},
  year         = {1981},
  doi          = {10.1086/183527}
}

@article{Chabrier2007,
  author       = {Chabrier, Gilles and Gallardo, Jose and Baraffe, Isabelle},
  title        = {Evolution of low-mass star and brown dwarf eclipsing binaries},
  journal      = {Astronomy \& Astrophysics},
  volume       = {472},
  pages        = {L17--L20},
  year         = {2007},
  doi          = {10.1051/0004-6361:20077631}
}

@article{agol2021,
  title={Refining the transit-timing and photometric analysis of TRAPPIST-1: masses, radii, densities, dynamics, and ephemerides},
  author={Agol, Eric and Dorn, Caroline and Grimm, Simon L and Turbet, Martin and Ducrot, Elsa and Delrez, Laetitia and Gillon, Micha{\"e}l and Demory, Brice-Olivier and Burdanov, Artem and Barkaoui, Khalid and others},
  journal={The planetary science journal},
  volume={2},
  number={1},
  pages={1},
  year={2021},
  publisher={IOP Publishing}
}

@article{hauschildt2025newera,
  title={The NewEra model grid},
  author={Hauschildt, Peter H and Barman, T and Baron, E and Aufdenberg, JP and Schweitzer, A},
  journal={Astronomy \& Astrophysics},
  volume={698},
  pages={A47},
  year={2025},
  publisher={EDP Sciences}
}

@article{morley2012neglected,
  title={Neglected clouds in T and Y dwarf atmospheres},
  author={Morley, Caroline V and Fortney, Jonathan J and Marley, Mark S and Visscher, Channon and Saumon, Didier and Leggett, SK},
  journal={The Astrophysical Journal},
  volume={756},
  number={2},
  pages={172},
  year={2012},
  publisher={IOP Publishing}
}

@article{morley2024sonora,
  title={The Sonora substellar atmosphere models. III. Diamondback: atmospheric properties, spectra, and evolution for warm cloudy substellar objects},
  author={Morley, Caroline V and Mukherjee, Sagnick and Marley, Mark S and Fortney, Jonathan J and Visscher, Channon and Lupu, Roxana and Gharib-Nezhad, Ehsan and Thorngren, Daniel and Freedman, Richard and Batalha, Natasha},
  journal={The Astrophysical Journal},
  volume={975},
  number={1},
  pages={59},
  year={2024},
  publisher={IOP Publishing}
}

@article{lefevre2022cloud,
  title={Cloud-convection feedback in brown dwarf atmospheres},
  author={Lef{\`e}vre, Maxence and Tan, Xianyu and Lee, Elspeth KH and Pierrehumbert, RT},
  journal={The Astrophysical Journal},
  volume={929},
  number={2},
  pages={153},
  year={2022},
  publisher={IOP Publishing}
}

@ARTICLE{Arcangeli2018,
   author = {{Arcangeli}, J. and {D{\'e}sert}, J.-M. and {Line}, M.~R. and 
	{Bean}, J.~L. and {Parmentier}, V. and {Stevenson}, K.~B. and 
	{Kreidberg}, L. and {Fortney}, J.~J. and {Mansfield}, M. and 
	{Showman}, A.~P.},
    title = "{H$^{-}$ Opacity and Water Dissociation in the Dayside Atmosphere of the Very Hot Gas Giant WASP-18b}",
  journal = {\apjl},
archivePrefix = "arXiv",
   eprint = {1801.02489},
 primaryClass = "astro-ph.EP",
     year = 2018,
    month = mar,
   volume = 855,
      eid = {L30},
    pages = {L30},
      doi = {10.3847/2041-8213/aab272},
   adsurl = {http://adsabs.harvard.edu/abs/2018ApJ...855L..30A}
}

@ARTICLE{Kreidberg2018,
   author = {{Kreidberg}, L. and {Line}, M.~R. and {Parmentier}, V. and {Stevenson}, K.~B. and 
	{Louden}, T. and {Bonnefoy}, M. and {Faherty}, J.~K. and {Henry}, G.~W. and 
	{Williamson}, M.~H. and {Stassun}, K. and {Beatty}, T.~G. and 
	{Bean}, J.~L. and {Fortney}, J.~J. and {Showman}, A.~P. and 
	{D{\'e}sert}, J.-M. and {Arcangeli}, J.},
    title = "{Global Climate and Atmospheric Composition of the Ultra-hot Jupiter WASP-103b from HST and Spitzer Phase Curve Observations}",
  journal = {\aj},
archivePrefix = "arXiv",
   eprint = {1805.00029},
 primaryClass = "astro-ph.EP",
     year = 2018,
    month = jul,
   volume = 156,
      eid = {17},
    pages = {17},
      doi = {10.3847/1538-3881/aac3df},
   adsurl = {http://adsabs.harvard.edu/abs/2018AJ....156...17K}
}

@ARTICLE{pinhas2018,
       author = {{Pinhas}, Arazi and {Rackham}, Benjamin V. and {Madhusudhan}, Nikku and
         {Apai}, D{\'a}niel},
        title = "{Retrieval of planetary and stellar properties in transmission spectroscopy with AURA}",
      journal = {\mnras},
         year = "2018",
        month = "Nov",
       volume = {480},
       number = {4},
        pages = {5314-5331},
          doi = {10.1093/mnras/sty2209},
archivePrefix = {arXiv},
       eprint = {1808.10017},
 primaryClass = {astro-ph.EP},
       adsurl = {https://ui.adsabs.harvard.edu/abs/2018MNRAS.480.5314P}
}

@article{dupuy2010studying,
  title={Studying the physical diversity of late-M dwarfs with dynamical masses},
  author={Dupuy, Trent J and Liu, Michael C and Bowler, Brendan P and Cushing, Michael C and Helling, Christiane and Witte, Soeren and Hauschildt, Peter},
  journal={The Astrophysical Journal},
  volume={721},
  number={2},
  pages={1725},
  year={2010},
  publisher={IOP Publishing}
}

@article{zalesky2022uniform,
  title={A Uniform Retrieval Analysis of Ultra-cool Dwarfs. IV. A Statistical Census from 50 Late-T Dwarfs},
  author={Zalesky, J and Saboi, Kezman and Line, Michael R and Zhang, Zhoujian and Schneider, Adam C and Liu, Michael C and Best, William M. J. and Marley, Mark S.},
  journal={arXiv preprint arXiv:2206.01199},
  year={2022}
}

@article{helling2008comparison,
  title={A comparison of chemistry and dust cloud formation in ultracool dwarf model atmospheres},
  author={Helling, Ch and Ackerman, A and Allard, F and Dehn, M and Hauschildt, P and Homeier, D and Lodders, K and Marley, M and Rietmeijer, F and Tsuji, T and others},
  journal={Monthly Notices of the Royal Astronomical Society},
  volume={391},
  number={4},
  pages={1854--1873},
  year={2008},
  publisher={Blackwell Publishing Ltd Oxford, UK}
}

@article{ackerman2001precipitating,
  title={Precipitating condensation clouds in substellar atmospheres},
  author={Ackerman, Andrew S and Marley, Mark S},
  journal={The Astrophysical Journal},
  volume={556},
  number={2},
  pages={872},
  year={2001},
  publisher={IOP Publishing}
}

@article{fortney2012carbon,
  title={On the carbon-to-oxygen ratio measurement in nearby Sun-like stars: Implications for planet formation and the determination of stellar abundances},
  author={Fortney, Jonathan J},
  journal={The Astrophysical Journal Letters},
  volume={747},
  number={2},
  pages={L27},
  year={2012},
  publisher={IOP Publishing}
}

@article{allard2013bt,
  title={The BT-Settl model atmospheres for stars, brown dwarfs and planets},
  author={Allard, F},
  journal={Proceedings of the International Astronomical Union},
  volume={8},
  number={S299},
  pages={271--272},
  year={2013},
  publisher={Cambridge University Press}
}

@ARTICLE{bochanski2010,
       author = {{Bochanski}, John J. and {Hawley}, Suzanne L. and {Covey}, Kevin R. and {West}, Andrew A. and {Reid}, I. Neill and {Golimowski}, David A. and {Ivezi{\'c}}, {\v{Z}}eljko},
        title = "{The Luminosity and Mass Functions of Low-mass Stars in the Galactic Disk. II. The Field}",
      journal = {\aj},
         year = 2010,
        month = jun,
       volume = {139},
       number = {6},
        pages = {2679-2699},
          doi = {10.1088/0004-6256/139/6/2679},
archivePrefix = {arXiv},
       eprint = {1004.4002},
 primaryClass = {astro-ph.SR},
       adsurl = {https://ui.adsabs.harvard.edu/abs/2010AJ....139.2679B}
}

@article{schneider2021ross,
  title={Ross 19B: An Extremely Cold Companion Discovered via the Backyard Worlds: Planet 9 Citizen Science Project},
  author={Schneider, Adam C and Meisner, Aaron M and Gagn{\'e}, Jonathan and Faherty, Jacqueline K and Marocco, Federico and Burgasser, Adam J and Kirkpatrick, J Davy and Kuchner, Marc J and Gramaize, L{\'e}opold and Rothermich, Austin and others},
  journal={The Astrophysical Journal},
  volume={921},
  number={2},
  pages={140},
  year={2021},
  publisher={IOP Publishing}
}

@article{dressing2015occurrence,
  title={The occurrence of potentially habitable planets orbiting M dwarfs estimated from the full Kepler dataset and an empirical measurement of the detection sensitivity},
  author={Dressing, Courtney D and Charbonneau, David},
  journal={The Astrophysical Journal},
  volume={807},
  number={1},
  pages={45},
  year={2015},
  publisher={IOP Publishing}
}

@article{tsuji2002water,
  title={Water Observed in Red Giant and Supergiant Stars-Manifestation of a Novel Picture of the Stellar Atmosphere or else Evidence against the Classical Model Stellar Photosphere},
  author={Tsuji, Takashi},
  journal={arXiv preprint astro-ph/0209495},
  year={2002}
}

@article{czekala2015starfish,
  title={Starfish: Robust spectroscopic inference tools},
  author={Czekala, Ian and Andrews, Sean M and Mandel, Kaisey S and Hogg, David W and Green, Gregory M},
  journal={Astrophysics Source Code Library},
  pages={ascl--1505},
  year={2015}
}

@article{mann2013metal,
  title={Prospecting in late-type dwarfs: a calibration of infrared and visible spectroscopic metallicities of late K and M dwarfs spanning 1.5 dex},
  author={Mann, Andrew W and Brewer, John M and Gaidos, Eric and L{\'e}pine, S{\'e}bastien and Hilton, Eric J},
  journal={The Astronomical Journal},
  volume={145},
  number={2},
  pages={52},
  year={2013},
  publisher={IOP Publishing}
}

@article{burgasser2014spex,
  title={The SpeX Prism Library: 1000+ low-resolution, near-infrared spectra of ultracool M, L, T and Y dwarfs},
  author={Burgasser, Adam J},
  journal={arXiv preprint arXiv:1406.4887},
  year={2014}
}

@article{toon1989rapid,
  title={Rapid calculation of radiative heating rates and photodissociation rates in inhomogeneous multiple scattering atmospheres},
  author={Toon, Owen B and McKay, CP and Ackerman, TP and Santhanam, K},
  journal={Journal of Geophysical Research: Atmospheres},
  volume={94},
  number={D13},
  pages={16287--16301},
  year={1989},
  publisher={Wiley Online Library}
}

@article{hubeny2017model,
  title={Model atmospheres of sub-stellar mass objects},
  author={Hubeny, Ivan},
  journal={Monthly Notices of the Royal Astronomical Society},
  volume={469},
  number={1},
  pages={841--869},
  year={2017},
  publisher={Oxford University Press}
}

@article{gharib2019influence,
  title={The influence of h2o pressure broadening in high-metallicity exoplanet atmospheres},
  author={Gharib-Nezhad, Ehsan and Line, Michael R},
  journal={The Astrophysical Journal},
  volume={872},
  number={1},
  pages={27},
  year={2019},
  publisher={IOP Publishing}
}

@article{gharib2021exoplines,
  title={EXOPLINES: Molecular Absorption Cross-Section Database for Brown Dwarf and Giant Exoplanet Atmospheres},
  author={Gharib-Nezhad, Ehsan and Iyer, Aishwarya R and Line, Michael R and Freedman, Richard S and Marley, Mark S and Batalha, Natasha E},
  journal={The Astrophysical Journal Supplement Series},
  volume={254},
  number={2},
  pages={34},
  year={2021},
  publisher={IOP Publishing}
}

@article{bohm1958wasserstoffkonvektionszone,
  title={{\"U}ber die wasserstoffkonvektionszone in sternen verschiedener effektivtemperaturen und leuchtkr{\"a}fte. Mit 5 textabbildungen},
  author={B{\"o}hm-Vitense, Erika},
  journal={Zeitschrift fur Astrophysik},
  volume={46},
  pages={108},
  year={1958}
}

@ARTICLE{ducati2002,
       author = {{Ducati}, J.~R.},
        title = "{VizieR Online Data Catalog: Catalogue of Stellar Photometry in Johnson's 11-color system.}",
      journal = {VizieR Online Data Catalog},
         year = 2002,
        month = jan,
       adsurl = {https://ui.adsabs.harvard.edu/abs/2002yCat.2237....0D}
}

@article{mann2015constrain,
  title={How to constrain your M dwarf: measuring effective temperature, bolometric luminosity, mass, and radius},
  author={Mann, Andrew W and Feiden, Gregory A and Gaidos, Eric and Boyajian, Tabetha and Von Braun, Kaspar},
  journal={The Astrophysical Journal},
  volume={804},
  number={1},
  pages={64},
  year={2015},
  publisher={IOP Publishing}
}

@ARTICLE{catri2003,
       author = {{Cutri}, R.~M. and {Skrutskie}, M.~F. and {van Dyk}, S. and {Beichman}, C.~A. and {Carpenter}, J.~M. and {Chester}, T. and {Cambresy}, L. and {Evans}, T. and {Fowler}, J. and {Gizis}, J. and {Howard}, E. and {Huchra}, J. and {Jarrett}, T. and {Kopan}, E.~L. and {Kirkpatrick}, J.~D. and {Light}, R.~M. and {Marsh}, K.~A. and {McCallon}, H. and {Schneider}, S. and {Stiening}, R. and {Sykes}, M. and {Weinberg}, M. and {Wheaton}, W.~A. and {Wheelock}, S. and {Zacarias}, N.},
        title = "{VizieR Online Data Catalog: 2MASS All-Sky Catalog of Point Sources (Cutri+ 2003)}",
      journal = {VizieR Online Data Catalog},
         year = 2003,
        month = jun,
          eid = {II/246},
        pages = {II/246},
       adsurl = {https://ui.adsabs.harvard.edu/abs/2003yCat.2246....0C}
}

@ARTICLE{rayner2003,
       author = {{Rayner}, J.~T. and {Toomey}, D.~W. and {Onaka}, P.~M. and {Denault}, A.~J. and {Stahlberger}, W.~E. and {Vacca}, W.~D. and {Cushing}, M.~C. and {Wang}, S.},
        title = "{SpeX: A Medium-Resolution 0.8-5.5 Micron Spectrograph and Imager for the NASA Infrared Telescope Facility}",
      journal = {\pasp},
         year = 2003,
        month = mar,
       volume = {115},
       number = {805},
        pages = {362-382},
          doi = {10.1086/367745},
       adsurl = {https://ui.adsabs.harvard.edu/abs/2003PASP..115..362R}
}

@ARTICLE{zhang2021,
       author = {{Zhang}, Zhoujian and {Liu}, Michael C. and {Marley}, Mark S. and {Line}, Michael R. and {Best}, William M.~J.},
        title = "{Uniform Forward-modeling Analysis of Ultracool Dwarfs. I. Methodology and Benchmarking}",
      journal = {\apj},
         year = 2021,
        month = jul,
       volume = {916},
       number = {1},
          eid = {53},
        pages = {53},
          doi = {10.3847/1538-4357/abf8b2},
archivePrefix = {arXiv},
       eprint = {2011.12294},
 primaryClass = {astro-ph.SR},
       adsurl = {https://ui.adsabs.harvard.edu/abs/2021ApJ...916...53Z}
}

@ARTICLE{zalesky2019,
       author = {{Zalesky} and {Line}, Michael R. and {Schneider}, Adam C. and {Patience}, Jennifer},
        title = "{A Uniform Retrieval Analysis of Ultra-cool Dwarfs. III. Properties of Y Dwarfs}",
      journal = {\apj},
         year = 2019,
        month = may,
       volume = {877},
       number = {1},
          eid = {24},
        pages = {24},
          doi = {10.3847/1538-4357/ab16db},
archivePrefix = {arXiv},
       eprint = {1903.11658},
 primaryClass = {astro-ph.SR},
       adsurl = {https://ui.adsabs.harvard.edu/abs/2019ApJ...877...24Z}
}

@ARTICLE{line2017,
       author = {{Line} and {Marley}, Mark S. and {Liu}, Michael C. and {Burningham}, Ben and {Morley}, Caroline V. and {Hinkel}, Natalie R. and {Teske}, Johanna and {Fortney}, Jonathan J. and {Freedman}, Richard and {Lupu}, Roxana},
        title = "{Uniform Atmospheric Retrieval Analysis of Ultracool Dwarfs. II. Properties of 11 T dwarfs}",
      journal = {\apj},
         year = 2017,
        month = oct,
       volume = {848},
       number = {2},
          eid = {83},
        pages = {83},
          doi = {10.3847/1538-4357/aa7ff0},
archivePrefix = {arXiv},
       eprint = {1612.02809},
 primaryClass = {astro-ph.SR},
       adsurl = {https://ui.adsabs.harvard.edu/abs/2017ApJ...848...83L}
}

@article{line2015uniform,
  title={Uniform atmospheric retrieval analysis of ultracool dwarfs. I. Characterizing benchmarks, Gl 570D and HD 3651B},
  author={Line, Michael R and Teske, Johanna and Burningham, Ben and Fortney, Jonathan J and Marley, Mark S},
  journal={The Astrophysical Journal},
  volume={807},
  number={2},
  pages={183},
  year={2015},
  publisher={IOP Publishing}
}

@article{boyajian2012stellar,
  title={Stellar diameters and temperatures. II. Main-sequence K-and M-stars},
  author={Boyajian, Tabetha S and Von Braun, Kaspar and Van Belle, Gerard and McAlister, Harold A and Theo, A and Kane, Stephen R and Muirhead, Philip S and Jones, Jeremy and White, Russel and Schaefer, Gail and others},
  journal={The Astrophysical Journal},
  volume={757},
  number={2},
  pages={112},
  year={2012},
  publisher={IOP Publishing}
}

@article{iyer2020influence,
  title={The Influence of Stellar Contamination on the Interpretation of Near-infrared Transmission Spectra of Sub-Neptune Worlds around M-dwarfs},
  author={Iyer, Aishwarya R and Line, Michael R},
  journal={The Astrophysical Journal},
  volume={889},
  number={2},
  pages={78},
  year={2020},
  publisher={IOP Publishing}
}

@ARTICLE{hinkel2014,
       author = {{Hinkel}, Natalie R. and {Timmes}, F.~X. and {Young}, Patrick A. and {Pagano}, Michael D. and {Turnbull}, Margaret C.},
        title = "{Stellar Abundances in the Solar Neighborhood: The Hypatia Catalog}",
      journal = {\aj},
         year = 2014,
        month = sep,
       volume = {148},
       number = {3},
          eid = {54},
        pages = {54},
          doi = {10.1088/0004-6256/148/3/54},
archivePrefix = {arXiv},
       eprint = {1405.6719},
 primaryClass = {astro-ph.SR},
       adsurl = {https://ui.adsabs.harvard.edu/abs/2014AJ....148...54H}
}

@article{patience2012spectroscopy,
  title={Spectroscopy across the brown dwarf/planetary mass boundary-I. Near-infrared JHK spectra},
  author={Patience, Jennifer and King, RR and De Rosa, RJ and Vigan, Arthur and Witte, S and Rice, E and Helling, Ch and Hauschildt, P},
  journal={Astronomy \& Astrophysics},
  volume={540},
  pages={A85},
  year={2012},
  publisher={EDP Sciences}
}

@article{tsuji1996dust,
  title={Dust formation in stellar photospheres: a case of very low mass stars and a possible resolution on the effective temperature scale of M dwarfs.},
  author={Tsuji, T and Ohnaka, K and Aoki, W},
  journal={Astronomy and Astrophysics},
  volume={305},
  pages={L1},
  year={1996}
}

@ARTICLE{piskorz2018,
       author = {{Piskorz}, Danielle and {Buzard}, Cam and {Line}, Michael R. and {Knutson}, Heather A. and {Benneke}, Bj{\"o}rn and {Crockett}, Nathan R. and {Lockwood}, Alexandra C. and {Blake}, Geoffrey A. and {Barman}, Travis S. and {Bender}, Chad F. and {Deming}, Drake and {Johnson}, John A.},
        title = "{Ground- and Space-based Detection of the Thermal Emission Spectrum of the Transiting Hot Jupiter KELT-2Ab}",
      journal = {\aj},
         year = 2018,
        month = sep,
       volume = {156},
       number = {3},
          eid = {133},
        pages = {133},
          doi = {10.3847/1538-3881/aad781},
archivePrefix = {arXiv},
       eprint = {1809.05615},
 primaryClass = {astro-ph.EP},
       adsurl = {https://ui.adsabs.harvard.edu/abs/2018AJ....156..133P}
}

@ARTICLE{bonnefoy2018,
       author = {{Bonnefoy}, M. and {Perraut}, K. and {Lagrange}, A. -M. and {Delorme}, P. and {Vigan}, A. and {Line}, M. and {Rodet}, L. and {Ginski}, C. and {Mourard}, D. and {Marleau}, G. -D. and {Samland}, M. and {Tremblin}, P. and {Ligi}, R. and {Cantalloube}, F. and {Molli{\`e}re}, P. and {Charnay}, B. and {Kuzuhara}, M. and {Janson}, M. and {Morley}, C. and {Homeier}, D. and {D'Orazi}, V. and {Klahr}, H. and {Mordasini}, C. and {Lavie}, B. and {Baudino}, J. -L. and {Beust}, H. and {Peretti}, S. and {Musso Bartucci}, A. and {Mesa}, D. and {B{\'e}zard}, B. and {Boccaletti}, A. and {Galicher}, R. and {Hagelberg}, J. and {Desidera}, S. and {Biller}, B. and {Maire}, A. -L. and {Allard}, F. and {Borgniet}, S. and {Lannier}, J. and {Meunier}, N. and {Desort}, M. and {Alecian}, E. and {Chauvin}, G. and {Langlois}, M. and {Henning}, T. and {Mugnier}, L. and {Mouillet}, D. and {Gratton}, R. and {Brandt}, T. and {Mc Elwain}, M. and {Beuzit}, J. -L. and {Tamura}, M. and {Hori}, Y. and {Brandner}, W. and {Buenzli}, E. and {Cheetham}, A. and {Cudel}, M. and {Feldt}, M. and {Kasper}, M. and {Keppler}, M. and {Kopytova}, T. and {Meyer}, M. and {Perrot}, C. and {Rouan}, D. and {Salter}, G. and {Schmidt}, T. and {Sissa}, E. and {Zurlo}, A. and {Wildi}, F. and {Blanchard}, P. and {De Caprio}, V. and {Delboulb{\'e}}, A. and {Maurel}, D. and {Moulin}, T. and {Pavlov}, A. and {Rabou}, P. and {Ramos}, J. and {Roelfsema}, R. and {Rousset}, G. and {Stadler}, E. and {Rigal}, F. and {Weber}, L.},
        title = "{The GJ 504 system revisited. Combining interferometric, radial velocity, and high contrast imaging data}",
      journal = {\aap},
         year = 2018,
        month = oct,
       volume = {618},
          eid = {A63},
        pages = {A63},
          doi = {10.1051/0004-6361/201832942},
archivePrefix = {arXiv},
       eprint = {1807.00657},
 primaryClass = {astro-ph.EP},
       adsurl = {https://ui.adsabs.harvard.edu/abs/2018A&A...618A..63B}
}

@article{marley2014cool,
  title={On the cool side: modeling the atmospheres of brown dwarfs and giant planets},
  author={Marley, Mark S and Robinson, Tyler D},
  journal={arXiv preprint arXiv:1410.6512},
  year={2014}
}

@article{fischer2005planet,
  title={The planet-metallicity correlation},
  author={Fischer, Debra A and Valenti, Jeff},
  journal={The Astrophysical Journal},
  volume={622},
  number={2},
  pages={1102},
  year={2005},
  publisher={IOP Publishing}
}

@article{melendez2009peculiar,
  title={The peculiar solar composition and its possible relation to planet formation},
  author={Melendez, Jorge and Asplund, Martin and Gustafsson, Bengt and Yong, David},
  journal={The Astrophysical Journal},
  volume={704},
  number={1},
  pages={L66},
  year={2009},
  publisher={IOP Publishing}
}

@article{kopparapu2013habitable,
  title={Habitable zones around main-sequence stars: new estimates},
  author={Kopparapu, Ravi Kumar and Ramirez, Ramses and Kasting, James F and Eymet, Vincent and Robinson, Tyler D and Mahadevan, Suvrath and Terrien, Ryan C and Domagal-Goldman, Shawn and Meadows, Victoria and Deshpande, Rohit},
  journal={The Astrophysical Journal},
  volume={765},
  number={2},
  pages={131},
  year={2013},
  publisher={IOP Publishing}
}

@inproceedings{mahadevan2012habitable,
  title={The habitable-zone planet finder: a stabilized fiber-fed NIR spectrograph for the Hobby-Eberly Telescope},
  author={Mahadevan, Suvrath and Ramsey, Lawrence and Bender, Chad and Terrien, Ryan and Wright, Jason T and Halverson, Sam and Hearty, Fred and Nelson, Matt and Burton, Adam and Redman, Stephen and others},
  booktitle={Ground-based and airborne instrumentation for astronomy IV},
  volume={8446},
  pages={624--637},
  year={2012},
  organization={SPIE}
}

@article{iyer2023sphinx,
  title={The SPHINX M-dwarf Spectral Grid. I. Benchmarking New Model Atmospheres to Derive Fundamental M-dwarf Properties},
  author={Iyer, Aishwarya R and Line, Michael R and Muirhead, Philip S and Fortney, Jonathan J and Gharib-Nezhad, Ehsan},
  journal={The Astrophysical Journal},
  volume={944},
  number={1},
  pages={41},
  year={2023},
  publisher={IOP Publishing}
}

@article{schneider2018hazmat,
  title={HAZMAT. III. The UV evolution of mid-to late-M Stars with GALEX},
  author={Schneider, Adam C and Shkolnik, Evgenya L},
  journal={The Astronomical Journal},
  volume={155},
  number={3},
  pages={122},
  year={2018},
  publisher={IOP Publishing}
}

@article{faherty2016population,
  title={Population properties of brown dwarf analogs to exoplanets},
  author={Faherty, Jacqueline K and Riedel, Adric R and Cruz, Kelle L and Gagne, Jonathan and Filippazzo, Joseph C and Lambrides, Erini and Fica, Haley and Weinberger, Alycia and Thorstensen, John R and Tinney, CG and others},
  journal={The Astrophysical Journal Supplement Series},
  volume={225},
  number={1},
  pages={10},
  year={2016},
  publisher={IOP Publishing}
}

@article{delaunay1934bulletin,
  title={Bulletin de l’Academie des Sciences de l’URSS},
  author={Delaunay, B and Vide, S and Lam{\'e}moire, A and De Georges, V},
  journal={Classe des sciences math{\'e}matiques et naturelles},
  volume={6},
  pages={793--800},
  year={1934}
}

@article{rajpurohit2018exploring,
  title={Exploring the stellar properties of M dwarfs with high-resolution spectroscopy from the optical to the near-infrared},
  author={Rajpurohit, AS and Allard, F and Rajpurohit, S and Sharma, R and Teixeira, GDC and Mousis, O and Kamlesh, R},
  journal={Astronomy \& Astrophysics},
  volume={620},
  pages={A180},
  year={2018},
  publisher={EDP Sciences}
}

@article{rackham2023towards,
  title={Towards robust corrections for stellar contamination in JWST exoplanet transmission spectra},
  author={Rackham, Benjamin V and deWit, Julien},
  journal={arXiv preprint arXiv:2303.15418},
  year={2023}
}

@article{tsuji1996evolution,
  title={Evolution of dusty photospheres through red to brown dwarfs: how dust forms in very low mass objects.},
  author={Tsuji, T and Ohnaka, K and Aoki, W and Nakajima, T},
  journal={Astronomy and Astrophysics},
  volume={308},
  pages={L29--L32},
  year={1996}
}

@article{loyd2018hazmat,
  title={HAZMAT. IV. Flares and superflares on young M Stars in the far ultraviolet},
  author={Loyd, RO Parke and Shkolnik, Evgenya L and Schneider, Adam C and Barman, Travis S and Meadows, Victoria S and Pagano, Isabella and Peacock, Sarah},
  journal={The Astrophysical Journal},
  volume={867},
  number={1},
  pages={70},
  year={2018},
  publisher={IOP Publishing}
}

@article{shkolnik2014hazmat,
  title={HAZMAT. I. The evolution of far-UV and near-UV emission from early M stars},
  author={Shkolnik, Evgenya L and Barman, Travis S},
  journal={The Astronomical Journal},
  volume={148},
  number={4},
  pages={64},
  year={2014},
  publisher={IOP Publishing}
}

@article{muirhead2020magnetic,
  title={Magnetic inflation and stellar mass. V. Intensification and saturation of M-dwarf absorption lines with Rossby number},
  author={Muirhead, Philip S and Veyette, Mark J and Newton, Elisabeth R and Theissen, Christopher A and Mann, Andrew W},
  journal={The Astronomical Journal},
  volume={159},
  number={2},
  pages={52},
  year={2020},
  publisher={IOP Publishing}
}

@article{charbonneau2014solar,
  title={Solar dynamo theory},
  author={Charbonneau, Paul},
  journal={Annual Review of Astronomy and Astrophysics},
  volume={52},
  pages={251--290},
  year={2014},
  publisher={Annual Reviews}
}

@article{newton2013near,
  title={Near-infrared metallicities, radial velocities, and spectral types for 447 nearby M dwarfs},
  author={Newton, Elisabeth R and Charbonneau, David and Irwin, Jonathan and Berta-Thompson, Zachory K and Rojas-Ayala, Barbara and Covey, Kevin and Lloyd, James P},
  journal={The Astronomical Journal},
  volume={147},
  number={1},
  pages={20},
  year={2013},
  publisher={IOP Publishing}
}

@article{rojas2010metal,
  title={Metal-rich M-dwarf planet hosts: Metallicities with k-band spectra},
  author={Rojas-Ayala, B{\'a}rbara and Covey, Kevin R and Muirhead, Philip S and Lloyd, James P},
  journal={The Astrophysical Journal Letters},
  volume={720},
  number={1},
  pages={L113},
  year={2010},
  publisher={IOP Publishing}
}

@article{neves2012metallicity,
  title={Metallicity of M dwarfs-II. A comparative study of photometric metallicity scales},
  author={Neves, V and Bonfils, X and Santos, NC and Delfosse, X and Forveille, T and Allard, F and Nat{\'a}rio, C and Fernandes, CS and Udry, St{\'e}phane},
  journal={Astronomy \& Astrophysics},
  volume={538},
  pages={A25},
  year={2012},
  publisher={EDP Sciences}
}

@article{mann2014prospecting,
  title={Prospecting in ultracool dwarfs: measuring the metallicities of mid-and late-M dwarfs},
  author={Mann, Andrew W and Deacon, Niall R and Gaidos, Eric and Ansdell, Megan and Brewer, John M and Liu, Michael C and Magnier, Eugene A and Aller, Kimberly M},
  journal={The Astronomical Journal},
  volume={147},
  number={6},
  pages={160},
  year={2014},
  publisher={IOP Publishing}
}

@article{greene2019characterizing,
  title={Characterizing Transiting Exoplanets with JWST Guaranteed Time and ERS Observations},
  author={Greene, Thomas and Bean, Jacob and Beatty, Thomas and Bouwman, Jeroen and Fortney, Jonathan and Hasegawa, Yasuhiro and Henning, Thomas and Lafreniere, David and Lagage, Pierre-Olivier and Rieke, George and others},
  journal={arXiv preprint arXiv:1903.07152},
  year={2019}
}

@article{reiners2018carmenes,
  title={The CARMENES search for exoplanets around M dwarfs-High-resolution optical and near-infrared spectroscopy of 324 survey stars},
  author={Reiners, A and Zechmeister, M and Caballero, JA and Ribas, I and Morales, JC and Jeffers, SV and Sch{\"o}fer, P and Tal-Or, L and Quirrenbach, A and Amado, Pedro J and others},
  journal={Astronomy \& Astrophysics},
  volume={612},
  pages={A49},
  year={2018},
  publisher={Edp Sciences}
}

@article{woolf2020m,
  title={The M dwarf problem: Fe and Ti abundances in a volume-limited sample of M dwarf stars},
  author={Woolf, Vincent M and Wallerstein, George},
  journal={Monthly Notices of the Royal Astronomical Society},
  volume={494},
  number={2},
  pages={2718--2726},
  year={2020},
  publisher={Oxford University Press}
}

@article{tarter2007reappraisal,
  title={A reappraisal of the habitability of planets around M dwarf stars},
  author={Tarter, Jill C and Backus, Peter R and Mancinelli, Rocco L and Aurnou, Jonathan M and Backman, Dana E and Basri, Gibor S and Boss, Alan P and Clarke, Andrew and Deming, Drake and Doyle, Laurance R and others},
  journal={Astrobiology},
  volume={7},
  number={1},
  pages={30--65},
  year={2007},
  publisher={Mary Ann Liebert, Inc. 2 Madison Avenue Larchmont, NY 10538 USA}
}

@ARTICLE{gaiadr3,
       author = {{Gaia Collaboration}},
        title = "{VizieR Online Data Catalog: Gaia EDR3 (Gaia Collaboration, 2020)}",
      journal = {VizieR Online Data Catalog},
         year = 2020,
        month = nov,
          eid = {I/350},
        pages = {I/350},
       adsurl = {https://ui.adsabs.harvard.edu/abs/2020yCat.1350....0G}
}

@article{freytag2010role,
  title={The role of convection, overshoot, and gravity waves for the transport of dust in M dwarf and brown dwarf atmospheres},
  author={Freytag, Bernd and Allard, France and Ludwig, H-G and Homeier, Derek and Steffen, Matthias},
  journal={Astronomy \& Astrophysics},
  volume={513},
  pages={A19},
  year={2010},
  publisher={EDP Sciences}
}

@article{osborn2019,
    author = {Osborn, Ares and Bayliss, Daniel},
    title = "{Investigating the planet–metallicity correlation for hot Jupiters}",
    journal = {Monthly Notices of the Royal Astronomical Society},
    volume = {491},
    number = {3},
    pages = {4481-4487},
    year = {2019},
    month = {11},
    issn = {0035-8711},
    doi = {10.1093/mnras/stz3207},
    url = {https://doi.org/10.1093/mnras/stz3207},
    eprint = {https://academic.oup.com/mnras/article-pdf/491/3/4481/31559365/stz3207.pdf},
}

@ARTICLE{Buchner2014,
   author = {{Buchner}, J. and {Georgakakis}, A. and {Nandra}, K. and {Hsu}, L. and 
	{Rangel}, C. and {Brightman}, M. and {Merloni}, A. and {Salvato}, M. and 
	{Donley}, J. and {Kocevski}, D.},
    title = "{X-ray spectral modelling of the AGN obscuring region in the CDFS: Bayesian model selection and catalogue}",
  journal = {\aap},
archivePrefix = "arXiv",
   eprint = {1402.0004},
 primaryClass = "astro-ph.HE",
     year = 2014,
    month = apr,
   volume = 564,
      eid = {A125},
    pages = {A125},
      doi = {10.1051/0004-6361/201322971},
   adsurl = {http://adsabs.harvard.edu/abs/2014A%26A...564A.125B}
}

@article{mansfield2021unique,
  title={A unique hot Jupiter spectral sequence with evidence for compositional diversity},
  author={Mansfield, Megan and Line, Michael R and Bean, Jacob L and Fortney, Jonathan J and Parmentier, Vivien and Wiser, Lindsey and Kempton, Eliza M-R and Gharib-Nezhad, Ehsan and Sing, David K and L{\'o}pez-Morales, Mercedes and others},
  journal={Nature Astronomy},
  volume={5},
  number={12},
  pages={1224--1232},
  year={2021},
  publisher={Nature Publishing Group}
}

@article{Bocquet2016,
  doi = {10.21105/joss.00046},
  url = {http://dx.doi.org/10.21105/joss.00046},
  year  = {2016},
  month = {oct},
  publisher = {The Open Journal},
  volume = {1},
  number = {6},
  author = {Sebastian Bocquet and Faustin W. Carter},
  title = {pygtc: beautiful parameter covariance plots (aka. Giant Triangle Confusograms)},
  journal = {The Journal of Open Source Software}
}

@ARTICLE{2020SciPy-NMeth,
  author  = {Virtanen, Pauli and Gommers, Ralf and Oliphant, Travis E. and
            Haberland, Matt and Reddy, Tyler and Cournapeau, David and
            Burovski, Evgeni and Peterson, Pearu and Weckesser, Warren and
            Bright, Jonathan and {van der Walt}, St{\'e}fan J. and
            Brett, Matthew and Wilson, Joshua and Millman, K. Jarrod and
            Mayorov, Nikolay and Nelson, Andrew R. J. and Jones, Eric and
            Kern, Robert and Larson, Eric and Carey, C J and
            Polat, {\.I}lhan and Feng, Yu and Moore, Eric W. and
            {VanderPlas}, Jake and Laxalde, Denis and Perktold, Josef and
            Cimrman, Robert and Henriksen, Ian and Quintero, E. A. and
            Harris, Charles R. and Archibald, Anne M. and
            Ribeiro, Ant{\^o}nio H. and Pedregosa, Fabian and
            {van Mulbregt}, Paul and {SciPy 1.0 Contributors}},
  title   = {{{SciPy} 1.0: Fundamental Algorithms for Scientific
            Computing in Python}},
  journal = {Nature Methods},
  year    = {2020},
  volume  = {17},
  pages   = {261--272},
  adsurl  = {https://rdcu.be/b08Wh},
  doi     = {10.1038/s41592-019-0686-2},
}

@ARTICLE{allard1997,
       author = {{Allard}, France and {Hauschildt}, Peter H. and {Alexander}, David R. and {Starrfield}, Sumner},
        title = "{Model Atmospheres of Very Low Mass Stars and Brown Dwarfs}",
      journal = {\araa},
         year = 1997,
        month = jan,
       volume = {35},
        pages = {137-177},
          doi = {10.1146/annurev.astro.35.1.137},
       adsurl = {https://ui.adsabs.harvard.edu/abs/1997ARA&A..35..137A}
}

@ARTICLE{allard2001,
       author = {{Allard}, France and {Hauschildt}, Peter H. and {Alexander}, David R. and {Tamanai}, Akemi and {Schweitzer}, Andreas},
        title = "{The Limiting Effects of Dust in Brown Dwarf Model Atmospheres}",
      journal = {\apj},
         year = 2001,
        month = jul,
       volume = {556},
       number = {1},
        pages = {357-372},
          doi = {10.1086/321547},
archivePrefix = {arXiv},
       eprint = {astro-ph/0104256},
 primaryClass = {astro-ph},
       adsurl = {https://ui.adsabs.harvard.edu/abs/2001ApJ...556..357A}
}

@Article{numpy,
 title         = {Array programming with {NumPy}},
 author        = {Charles R. Harris and K. Jarrod Millman and St{\'{e}}fan J.
                 van der Walt and Ralf Gommers and Pauli Virtanen and David
                 Cournapeau and Eric Wieser and Julian Taylor and Sebastian
                 Berg and Nathaniel J. Smith and Robert Kern and Matti Picus
                 and Stephan Hoyer and Marten H. van Kerkwijk and Matthew
                 Brett and Allan Haldane and Jaime Fern{\'{a}}ndez del
                 R{\'{i}}o and Mark Wiebe and Pearu Peterson and Pierre
                 G{\'{e}}rard-Marchant and Kevin Sheppard and Tyler Reddy and
                 Warren Weckesser and Hameer Abbasi and Christoph Gohlke and
                 Travis E. Oliphant},
 year          = {2020},
 month         = sep,
 journal       = {Nature},
 volume        = {585},
 number        = {7825},
 pages         = {357--362},
 doi           = {10.1038/s41586-020-2649-2},
 publisher     = {Springer Science and Business Media {LLC}},
 url           = {https://doi.org/10.1038/s41586-020-2649-2}
}

@ARTICLE{simbad2000,
       author = {{Wenger}, M. and {Ochsenbein}, F. and {Egret}, D. and {Dubois}, P. and {Bonnarel}, F. and {Borde}, S. and {Genova}, F. and {Jasniewicz}, G. and {Lalo{\"e}}, S. and {Lesteven}, S. and {Monier}, R.},
        title = "{The SIMBAD astronomical database. The CDS reference database for astronomical objects}",
      journal = {\aaps},
         year = 2000,
        month = apr,
       volume = {143},
        pages = {9-22},
          doi = {10.1051/aas:2000332},
archivePrefix = {arXiv},
       eprint = {astro-ph/0002110},
 primaryClass = {astro-ph},
       adsurl = {https://ui.adsabs.harvard.edu/abs/2000A&AS..143....9W}
}

@ARTICLE{zhang2018,
       author = {{Zhang}, Zhanbo and {Zhou}, Yifan and {Rackham}, Benjamin V. and
         {Apai}, D{\'a}niel},
        title = "{The Near-infrared Transmission Spectra of TRAPPIST-1 Planets b, c, d, e, f, and g and Stellar Contamination in Multi-epoch Transit Spectra}",
      journal = {\aj},
         year = "2018",
        month = "Oct",
       volume = {156},
       number = {4},
          eid = {178},
        pages = {178},
          doi = {10.3847/1538-3881/aade4f},
archivePrefix = {arXiv},
       eprint = {1802.02086},
 primaryClass = {astro-ph.EP},
       adsurl = {https://ui.adsabs.harvard.edu/abs/2018AJ....156..178Z}
}

@article{rackham2024stellar,
  author = {Rackham, Benjamin V. and de Wit, Julien},
  title = {Stellar Contamination Is a Major Barrier to the Interpretation of Small Exoplanet Transmission Spectra},
  journal = {The Astronomical Journal},
  volume = {167},
  number = {3},
  pages = {117},
  year = {2024},
  doi = {10.3847/1538-3881/ad5833},
  publisher = {IOP Publishing}
}

@article{Chabrier2000,
  author       = {Chabrier, G. and Baraffe, I.},
  title        = {Theory of low-mass stars and substellar objects},
  journal      = {Annual Review of Astronomy and Astrophysics},
  volume       = {38},
  pages        = {337--377},
  year         = {2000},
  doi          = {10.1146/annurev.astro.38.1.337}
}

@article{Rajpurohit2013,
  author       = {Rajpurohit, A. S. and Reyl{\'e}, C. and Allard, F. and Homeier, D. and Schultheis, M. and Bessell, M. S. and Robin, A. C.},
  title        = {The effective temperature scale of M dwarfs. Comparison of synthetic spectra and photometry with observations},
  journal      = {Astronomy and Astrophysics},
  volume       = {556},
  pages        = {A15},
  year         = {2013},
  doi          = {10.1051/0004-6361/201220566}
}

@article{gagne2014banyan,
  title={BANYAN. II. Very low mass and substellar candidate members to nearby, young kinematic groups with previously known signs of youth},
  author={Gagn{\'e}, Jonathan and Lafreni{\`e}re, David and Doyon, Ren{\'e} and Malo, Lison and Artigau, {\'E}tienne},
  journal={The Astrophysical Journal},
  volume={783},
  number={2},
  pages={121},
  year={2014},
  publisher={IOP Publishing}
}

@ARTICLE{ducrot2020,
       author = {{Ducrot}, E. and {Gillon}, M. and {Delrez}, L. and {Agol}, E. and {Rimmer}, P. and {Turbet}, M. and {G{\"u}nther}, M.~N. and {Demory}, B. -O. and {Triaud}, A.~H.~M.~J. and {Bolmont}, E. and {Burgasser}, A. and {Carey}, S.~J. and {Ingalls}, J.~G. and {Jehin}, E. and {Leconte}, J. and {Lederer}, S.~M. and {Queloz}, D. and {Raymond}, S.~N. and {Selsis}, F. and {Van Grootel}, V. and {de Wit}, J.},
        title = "{TRAPPIST-1: Global results of the Spitzer Exploration Science Program Red Worlds}",
      journal = {\aap},
         year = 2020,
        month = aug,
       volume = {640},
          eid = {A112},
        pages = {A112},
          doi = {10.1051/0004-6361/201937392},
archivePrefix = {arXiv},
       eprint = {2006.13826},
 primaryClass = {astro-ph.EP},
       adsurl = {https://ui.adsabs.harvard.edu/abs/2020A&A...640A.112D}
}

@article{Magic2019,
  author       = {Magic, Z.},
  title        = {Calibrating mixing length from 3D stellar atmospheres},
  journal      = {Astronomy \& Astrophysics},
  volume       = {622},
  pages        = {A90},
  year         = {2019},
  doi          = {10.1051/0004-6361/201834507},
  url          = {https://doi.org/10.1051/0004-6361/201834507}
}

@article{nikolov2022trexolists,
  title={TrExoLiSTS: Transiting Exoplanets List of Space Telescope Spectroscopy},
  author={Nikolov, Nikolay K and Kovacs, Aiden and Martlin, Catherine},
  journal={Research Notes of the AAS},
  volume={6},
  number={12},
  pages={272},
  year={2022},
  publisher={The American Astronomical Society}
}

@article{feroz2009multinest,
  author       = {Feroz, F. and Hobson, M. P. and Bridges, M.},
  title        = {MULTINEST: an efficient and robust Bayesian inference tool for cosmology and particle physics},
  journal      = {Monthly Notices of the Royal Astronomical Society},
  volume       = {398},
  number       = {4},
  pages        = {1601--1614},
  year         = {2009},
  doi          = {10.1111/j.1365-2966.2009.14548.x}
}

@article{feroz2019multinest,
  author       = {Feroz, F. and Hobson, M. P. and Cameron, E. and Pettitt, A. N.},
  title        = {Importance Nested Sampling and the MultiNest Algorithm},
  journal      = {The Open Journal of Astrophysics},
  volume       = {2},
  pages        = {10},
  year         = {2019},
  doi          = {10.21105/astro.1306.2144}
}

@article{Saumon2012,
  author       = {Saumon, D. and Marley, M. S. and Abel, M. and Frommhold, L. and Freedman, R. S.},
  title        = {The Impact of Improved Collision-induced Absorption on the Modeling of Substellar Atmospheres},
  journal      = {The Astrophysical Journal},
  volume       = {750},
  number       = {1},
  pages        = {74},
  year         = {2012},
  doi          = {10.1088/0004-637X/750/1/74}
}

@article{roellig2004spitzer,
  title={Spitzer Infrared Spectrograph (IRS)* Observations of M, L, and T Dwarfs},
  author={Roellig, Thomas L and Van Cleve, Jeffrey E and Sloan, Gregory C and Wilson, John C and Saumon, Didier and Leggett, Sandy K and Marley, Mark S and Cushing, MC and Kirkpatrick, J Davy and Mainzer, Amanda K and others},
  journal={The Astrophysical Journal Supplement Series},
  volume={154},
  number={1},
  pages={418},
  year={2004},
  publisher={IOP Publishing}
}

@inproceedings{HPC:ASU23, 
  title  = "{The Sol Supercomputer at Arizona State University}", 
  doi    = {10.1145/3569951.3597573}, 
  year   = {2023}, 
  author = {
    Jennewein, Douglas M. and
    Lee, Johnathan and 
    Kurtz, Chris and
    Dizon, Will and 
    Shaeffer, Ian and 
    Chapman, Alan and 
    Chiquete, Alejandro and 
    Burks, Josh and 
    Carlson, Amber and 
    Mason, Natalie and 
    Kobwala, Arhat and 
    Jagadeesan, Thirugnanam and 
    Barghav, Praful and 
    Battelle, Torey and 
    Belshe, Rebecca and 
    McCaffrey, Debra and 
    Brazil, Marisa and 
    Inumella, Chaitanya and 
    Kuznia, Kirby and
    Buzinski, Jade and
    Dudley, Sean and 
    Shah, Dhruvil and 
    Speyer, Gil and 
    Yalim, Jason
  },
  isbn      = {9781450399852}, 
  month     = {Jul}, 
  pages     = {296--301}, 
  booktitle = {Practice and Experience in Advanced Research Computing}, 
  publisher = {Association for Computing Machinery}, 
  address   = {New York, NY, USA}, 
  numpages  = {6}, 
  location  = {Portland, OR, USA}, 
  series    = {PEARC '23}, 
}

@article{Calamari2024,
  author = {Calamari, M. and Helling, Ch. and Min, M. and Woitke, P. and Samra, D. and Lewis, N. K.},
  title = {The Condensate Chemistry of M-dwarf and Brown Dwarf Atmospheres: Dependence on Mg/Si Ratio and Cloud Composition},
  journal = {Astronomy \& Astrophysics},
  year = {2024},
  volume = {690},
  pages = {A105},
  doi = {10.1051/0004-6361/202347890}
}

@article{suarez2023ultracool,
  title={Ultracool dwarfs observed with the Spitzer Infrared Spectrograph--III. Dust grains in young L dwarf atmospheres are heavier},
  author={Su{\'a}rez, Genaro and Metchev, Stanimir},
  journal={Monthly Notices of the Royal Astronomical Society},
  volume={523},
  number={3},
  pages={4739--4747},
  year={2023},
  publisher={Oxford University Press}
}

@article{jing2024half,
  title={Half a Million Binary Stars from the low resolution spectra of LAMOST},
  author={Jing, Yingjie and Mao, Tian-Xiang and Wang, Jie and Liu, Chao and Chen, Xiaodian},
  journal={arXiv preprint arXiv:2411.03994},
  year={2024}
}

@article{kesseli2017empirical,
  title={An empirical template library of stellar spectra for a wide range of spectral classes, luminosity classes, and metallicities using SDSS BOSS spectra},
  author={Kesseli, Aurora Y and West, Andrew A and Veyette, Mark and Harrison, Brandon and Feldman, Dan and Bochanski, John J},
  journal={The Astrophysical Journal Supplement Series},
  volume={230},
  number={2},
  pages={16},
  year={2017},
  publisher={IOP Publishing}
}

@article{morley2014water,
  author       = {Morley, Caroline V. and Marley, Mark S. and Fortney, Jonathan J. and Lupu, Roxana E. and Saumon, Didier and Greene, Thomas and Lodders, Katharina},
  title        = {Water Clouds in Y Dwarfs and Exoplanets},
  journal      = {The Astrophysical Journal},
  year         = {2014},
  volume       = {787},
  number       = {1},
  pages        = {78},
  doi          = {10.1088/0004-637X/787/1/78}
}

@article{xuan2024validation,
  title={Validation of elemental and isotopic abundances in late-M spectral types with the benchmark HIP 55507 AB system},
  author={Xuan, Jerry W and Wang, Jason and Finnerty, Luke and Horstman, Katelyn and Grimm, Simon and Peck, Anne E and Nielsen, Eric and Knutson, Heather A and Mawet, Dimitri and Isaacson, Howard and others},
  journal={The Astrophysical Journal},
  volume={962},
  number={1},
  pages={10},
  year={2024},
  publisher={IOP Publishing}
}

@article{tremblin2015cloud,
  author       = {Tremblin, P. and Amundsen, D. S. and Mourier, P. and Baraffe, I. and Chabrier, G. and Drummond, B. and Homeier, D. and Venot, O. and Allard, F.},
  title        = {Cloud Dissipation and Radiative–Convective Transitions in Substellar Atmospheres},
  journal      = {The Astrophysical Journal},
  year         = {2015},
  volume       = {804},
  number       = {2},
  pages        = {L17},
  doi          = {10.1088/2041-8205/804/2/L17}
}

@article{Helling2008,
  author = {Helling, Ch. and Ackerman, A. S. and Allard, F. and Hauschildt, P. H. and Homeier, D. and Sudarsky, D. and Burrows, A.},
  year = {2008},
  title = {A comparison of chemistry and dust cloud formation in ultracool dwarf model atmospheres},
  journal = {Monthly Notices of the Royal Astronomical Society},
  volume = {391},
  pages = {1854--1873}
}

@article{Freytag2010,
  author = {Freytag, B. and Allard, F. and Ludwig, H.-G. and Homeier, D. and Steffen, M.},
  year = {2010},
  title = {The role of convection, overshoot, and gravity waves for the transport of dust in M and L dwarfs},
  journal = {Astronomy \& Astrophysics},
  volume = {513},
  pages = {A19},
  doi = {10.1051/0004-6361/200913354}}

@article{Mie1908,
  title={Beiträge zur Optik trüber Medien, speziell kolloidaler Metallösungen},
  author={Mie, Gustav},
  journal={Annalen der Physik},
  volume={330},
  number={3},
  pages={377--445},
  year={1908},
  doi={10.1002/andp.19083300302}
}

@article{Mukherjee2022,
  title        = {Probing the Extent of Vertical Mixing in Brown Dwarf Atmospheres with Disequilibrium Chemistry},
  author       = {Mukherjee, Sagnick and Fortney, Jonathan J. and Batalha, Natasha E. and Karalidi, Theodora and Marley, Mark S. and Visscher, Channon and Miles, Brittany E. and Skemer, Andrew J. I.},
  journal      = {\apj},
  volume       = {938},
  number       = {107},
  year         = {2022},
  doi          = {10.3847/1538-4357/ac8dfb},
  url          = {https://ui.adsabs.harvard.edu/abs/2022ApJ...938..107M}
}

@book{Bohren1983,
  title={Absorption and Scattering of Light by Small Particles},
  author={Bohren, Craig F and Huffman, Donald R},
  year={1983},
  publisher={Wiley},
  address={New York},
  isbn={978-0-471-29340-8}
}

@article{lim2023effects,
  title={Atmospheric reconnaissance of TRAPPIST-1 b with JWST/NIRISS: evidence for strong stellar contamination in the transmission spectra},
  author={Lim, Olivia and Benneke, Bj{\"o}rn and Doyon, Ren{\'e} and MacDonald, Ryan J and Piaulet, Caroline and Artigau, {\'E}tienne and Coulombe, Louis-Philippe and Radica, Michael and L’Heureux, Alexandrine and Albert, Lo{\"\i}c and others},
  journal={The Astrophysical Journal Letters},
  volume={955},
  number={1},
  pages={L22},
  year={2023},
  publisher={IOP Publishing}
}

@misc{fauchez2025impact,
  author = {Fauchez, Thomas J. and Lustig-Yaeger, Jacob and Lincowski, Andrew P. and Arney, Giada N. and Meadows, Victoria S.},
  title = {Impact of Stellar Model Deficiencies on Terrestrial Planet Transmission Spectra Orbiting M-dwarfs},
  year = {2025},
  eprint = {2502.19585},
  archivePrefix = {arXiv},
  primaryClass = {astro-ph.EP},
  note = {arXiv preprint},
  url = {https://arxiv.org/abs/2502.19585}
}

@misc{quintana2021pandora,
  author = {Quintana, Elisa V. and Mandell, Avi M. and Robinson, Tyler D. and Davis, Anne B. and Rauscher, Emily and Knutson, Heather and et al.},
  title = {The Pandora SmallSat Mission: Disentangling Stellar and Planetary Signals to Understand Exoplanet Atmospheres},
  year = {2021},
  eprint = {2108.06438},
  archivePrefix = {arXiv},
  primaryClass = {astro-ph.EP},
  note = {arXiv preprint},
  url = {https://arxiv.org/abs/2108.06438}
}

@ARTICLE{husser2013,
       author = {{Husser}, T. -O. and {Wende-von Berg}, S. and {Dreizler}, S. and
         {Homeier}, D. and {Reiners}, A. and {Barman}, T. and
         {Hauschildt}, P.~H.},
        title = "{A new extensive library of PHOENIX stellar atmospheres and synthetic spectra}",
      journal = {\aap},
         year = "2013",
        month = "May",
       volume = {553},
          eid = {A6},
        pages = {A6},
          doi = {10.1051/0004-6361/201219058},
archivePrefix = {arXiv},
       eprint = {1303.5632},
 primaryClass = {astro-ph.SR},
       adsurl = {https://ui.adsabs.harvard.edu/abs/2013A&A...553A...6H}
}

@article{allard2012,
  title={Models of very-low-mass stars, brown dwarfs and exoplanets},
  author={Allard, France and Homeier, Derek and Freytag, Bernd},
  journal={Philosophical Transactions of the Royal Society A: Mathematical, Physical and Engineering Sciences},
  volume={370},
  number={1968},
  pages={2765--2777},
  year={2012},
  publisher={The Royal Society Publishing}
}

@article{gordon1996nasa,
  title={NASA-Lewis Chemical Equilibrium Program CEA, Maritime Security},
  author={Gordon, S and McBride, B},
  journal={NASA Report},
  volume={1311},
  year={1996}
}
\bibliographystyle{aasjournal}

\end{document}